\documentclass{article}

\usepackage{microtype}
\usepackage{graphicx}
\usepackage{subfigure}
\usepackage{booktabs} 

\usepackage{amsmath}
\usepackage{amsfonts}

\usepackage{hyperref}

\def \v {\mathbf{v}}
\def \x {\mathbf{x}}
\def \w {\mathbf{w}}
\def \z {\mathbf{z}}

\def \V {\mathbf{V}}

\def \E {\mathbb{E}}
\def \I {\mathbb{I}}
\newtheorem{lemma}{Lemma}
\newtheorem{theorem}{Theorem}
\newtheorem{assumption}{Assumption}

\usepackage[accepted]{icml2020}

\icmltitlerunning{Communication-Efficient Distributed Stochastic AUC Maximization with Deep Neural Networks}

\begin{document}

\twocolumn[
\icmltitle{Communication-Efficient Distributed Stochastic AUC Maximization\\ with Deep Neural Networks}



\icmlsetsymbol{equal}{*}

\begin{icmlauthorlist}
\icmlauthor{Zhishuai Guo}{1}
\icmlauthor{Mingrui Liu}{1} 
\icmlauthor{Zhuoning Yuan}{1}
\icmlauthor{Li Shen}{2}
\icmlauthor{Wei Liu}{2}
\icmlauthor{Tianbao Yang}{1}
\end{icmlauthorlist}

\icmlaffiliation{1}{Department of Computer Science, The University of Iowa, Iowa City, IA 52242, USA}

\icmlaffiliation{2}{Tencent AI Lab, Shenzhen, China} 

\icmlcorrespondingauthor{Tianbao  Yang}{tianbao-yang@uiowa.edu} 
\icmlkeywords{Machine Learning, ICML}
\vskip 0.3in
]



\printAffiliationsAndNotice{}  

\begin{abstract}
In this paper, we study distributed algorithms for large-scale AUC maximization with a deep neural network as a predictive model.  
Although distributed learning techniques have been investigated extensively  in deep learning, they are not directly applicable to stochastic AUC maximization with deep neural networks due to its striking differences from standard loss minimization problems (e.g., cross-entropy).
Towards addressing this challenge,  we propose and analyze a communication-efficient distributed optimization algorithm  based on a {\it non-convex concave} reformulation of the AUC maximization, in which the communication of both the primal variable and the dual variable between each worker and the parameter server only occurs after multiple steps of gradient-based updates in each worker.  
Compared with the naive parallel version of an existing algorithm that computes stochastic gradients at individual machines and averages them for updating the model parameters, our algorithm requires a much less number of communication rounds and still achieves a linear speedup in theory. 
To the best of our knowledge, this is the \textbf{first} work that solves the {\it non-convex concave min-max} problem for  AUC maximization with deep neural networks in a communication-efficient distributed manner while still maintaining the linear speedup property in theory.  
Our experiments on several benchmark datasets show the effectiveness of our algorithm and also confirm our theory. 

\end{abstract}

%

\section{Introduction}
Large-scale distributed deep learning~\cite{dean2012large,li2014scaling} has achieved tremendous successes in various domains, including computer vision~\cite{goyal2017accurate}, natural language processing~\cite{devlin2018bert,yang2019xlnet}, generative modeling~\cite{brock2018large}, reinforcement learning~\cite{silver2016mastering,silver2017mastering}, etc. From the perspective of learning theory and optimization, most of them are trying to minimize a surrogate loss of a specific error measure using parallel minibatch stochastic gradient descent (SGD). For example, on the image classification task, the surrogate loss is usually the cross entropy between the estimated probability distribution according to the output of a certain neural network and the vector encoding ground-truth labels~\cite{krizhevsky2012imagenet,simonyan2014very,he2016deep}, which is a surrogate loss of the misclassification rate. Based on the surrogate  loss, parallel minibatch SGD~\cite{goyal2017accurate} is employed to update the model parameters.

However, when the data for classification is imbalanced, AUC (short for Area Under the ROC Curve) is a more suitable measure~\cite{elkan2001foundations}.   AUC is defined as the probability that a positive sample has a higher score than a negative sample~\cite{hanley1982meaning,hanley1983method}. Despite the tremendous applications of distributed deep learning in different fields, the study about optimizing AUC with distributed deep learning technologies is rare. The commonly used parallel mini-batch SGD for minimizing a surrogate loss of AUC will suffer from high communication costs in a distributed setting due to the non-decomposability nature of AUC measure. The reason is that  positive and negative data pairs that define a surrogate loss for AUC may sit on different machines.  To the best of our knowledge,~\citet{liu2019stochastic} is the only work trying to optimize a surrogate loss of {\it AUC with a deep neural network} that explicitly tackles the  non-decomposability of AUC measure. Nevertheless, their algorithms are designed only for the single-machine setting and hence are far from sufficient when encountering a huge amount of data. 
Although a naive parallel version of the stochastic algorithms proposed in~\cite{liu2019stochastic} can be used for distributed AUC maximization with a deep neural network, it would still suffer from high communication overhead due to a large number of communication rounds.

\begin{table*}[t]
	\caption{The summary of Iteration and Communication Complexities, where $K$ is the number of machines and $\mu\leq 1$. NP-PPD-SG denotes the naive parallel version of PPD-SG, which is also a special case of our algorithm, whose complexities can be derived by following our analysis. 
}\label{tab:1}
	\centering
	\label{tab:2}
	{\begin{tabular}{l|l|l|l}
			\toprule
			Algorithm & Setting&Iteration Complexity  &Communication Complexity\\
			\midrule
			PPD-SG~\cite{liu2019stochastic}& Single & $O(1/(\mu^2\epsilon))$& -  \\
			NP-PPD-SG&  Distributed & $O(1/(K\mu^2\epsilon))$& $O(1/(K\mu^2\epsilon))$  \\
			CoDA& Distributed & $O(1/(K\mu^2\epsilon))$& $O(1/(\mu^{3/2}\epsilon^{1/2}))$  \\
 		\bottomrule
	\end{tabular}}
	\vspace*{-0.15in}
\end{table*}
   

In this paper, we bridge the gap between stochastic AUC maximization and distributed deep learning by proposing  a communication-efficient distributed algorithm for stochastic AUC maximization with a deep neural network. {\bf The focus is to make the total number of communication rounds much less than the total number of iterative updates.}  %
We build our algorithm  upon the nonconvex-concave min-max reformulation of the original problem. The key ingredient is to design a communication-efficient distributed algorithm for solving the regularized min-max subproblems using multiple machines. %
Specifically, we follow the proximal primal-dual algorithmic framework proposed by~\cite{rafique2018non,liu2019stochastic}, i.e., by solving a sequence of quadratically regularized min-max saddle-point problems  with periodically updated regularizers successively. {\bf The key difference} is that the inner min-max problem solver is built on a distributed periodic  model averaging technique,  which  consists of a fixed number of stochastic primal-dual updates over individual machines and a small number of averagings of model parameters from multiple machines. %
This mechanism can greatly reduce the communication cost, which is similar to~\cite{zhou2017convergence,stich2018local,yu2019parallel}. %
However, their analysis cannot be applied to our case since their analysis only works for convex or non-convex {\it minimization} problems. In contrast,  our algorithm is designed for a particular {\it non-convex concave min-max} problem induced by the original AUC maximization problem. Our contributions are summarized as follows: 
\vspace*{-0.1in}\begin{itemize}
    \item We propose a communication-efficient distributed stochastic algorithm named CoDA for solving a nonconvex-concave min-max reformulation of AUC maximization with deep neural networks  by local primal-dual updating and periodically global variable averaging. To our knowledge, this is the first communication-efficient distributed stochastic algorithm for learning a deep neural network by AUC maximization. 
    
    \item We analyze the iteration complexity and communication complexity of the proposed algorithm under the commonly used  Polyak- \L ojasiewicz (PL) condition as in~\cite{liu2019stochastic}. Comparing with \cite{liu2019stochastic},  our theoretical result shows that the iteration complexity can be reduced by a factor of $K$ (the number of machines) in a certain region, while the communication complexity (the rounds of communication) is much lower than that of a naive distributed version of the stochastic algorithm proposed in~\cite{liu2019stochastic}. The summary of iteration and communication complexities is given in Table~\ref{tab:2}.

  \vspace*{-0.1in} \item We verify our theoretical claims by conducting experiments on several large-scale benchmark datasets. The experimental results show that our algorithm indeed exhibits good acceleration performance in practice.
\end{itemize} 

\section{Related Work}
\paragraph{Stochastic AUC Maximization.} It is challenging to directly solve the stochastic AUC maximization in the online learning setting since the objective function of AUC maximization depends on a sum of pairwise losses between samples from positive and negative classes. \citet{zhao2011online} addresses this problem by maintaining a buffer to store representative data samples, employing  the reservoir sampling technique to update the buffer, calculating gradient information based on the data in the buffer, and then performing a gradient-based update rule to update the classifier. \citet{gao2013one} does not maintain a buffer, while they instead maintained first-order and second-order statistics of the received data to update the classifier by gradient-based update. Both of them are infeasible in big data scenarios since~\citet{zhao2011online} suffers from a large amount of training data and~\citet{gao2013one} is not suitable for high dimensional data.~\citet{ying2016stochastic} addresses these issues by introducing a min-max reformulation of the original problem and solving it by a primal-dual stochastic gradient method~\cite{nemirovski2009robust}, in which no buffer is needed and the per-iteration complexity is the same magnitude as the dimension of the feature vector.~\citet{natole2018stochastic} improves the convergence rate by adding a strongly convex regularizer upon the original formulation. Based on the same saddle point formulation as in~\cite{ying2016stochastic}, ~\citet{liu2018fast} gets an  improved convergence rate by developing a multi-stage algorithm without adding the strongly convex regularizer. However, all of these studies focus on learning a linear model. Recently, \citet{liu2019stochastic} considers stochastic AUC maximization for learning a  deep non-linear model, in which they designed a proximal primal-dual gradient-based algorithm under the PL condition and established non-asymptotic convergence results. 
\vspace{-0.4cm} 
\paragraph{Communication Efficient Algorithms.} There are multiple approaches for reducing the communication cost in distributed optimization, including skipping communication and compression techniques. Due to limit of space, we mainly review the literature on skipping communication. For compression techniques, we refer the readers to~\cite{jiang2018linear,stich2018sparsified,basu2019qsparse,wangni2018gradient,bernstein2018signsgd} and references therein. Skipping communication is realized by doing multiple local gradient-based updates in each worker before aggregating the local model parameters. One special case is so-called one-shot averaging~\cite{zinkevich2010parallelized,mcdonald2010distributed,zhang2013communication}, where each machine solves a local problem and averages these solutions only at the last iterate. Some works  ~\cite{zhang2013communication,shamir2014distributed,godichon2017rates,jain2017parallelizing,DBLP:conf/icml/KoloskovaSJ19,koloskova*2020decentralized} consider one-shot averaging with one-pass of the entire data and establish statistical convergence, which is usually not able to  guarantee the convergence of the training error. 
The scheme of local SGD update at each worker with skipping communication is analyzed for convex~\cite{stich2018local,DBLP:conf/nips/JaggiSTTKHJ14} and nonconvex problems~\cite{zhou2017convergence,jiang2018linear,wang2018cooperative,lin2018don,wang2018adaptive,yu2019parallel,yu2019linear,basu2019qsparse,haddadpour2019local}. There are also several empirical studies~\cite{povey2014parallel,su2015experiments,mcmahan2016communication,chen2016scalable,mcmahan2016communication,lin2018don,kamp2018efficient} showing that this scheme exhibits good empirical performance for distributed deep learning. However, all of these works only consider minimization problems and do not apply to the nonconvex-concave min-max formulation as considered in this paper. 

\paragraph{Nonconvex Min-max Optimization.} Stochastic nonconvex min-max optimization has garnered increasing attention recently~\cite{rafique2018non,lin2018solving,sanjabi2018solving,lu2019hybrid,lin2019gradient,jin2019minmax,liu2020towards}. \citet{rafique2018non} considered  the case where the objective function is weakly-convex and concave and proposed an algorithm based on the spirit of proximal point method~\cite{rockafellar1976monotone}, in which a proximal subproblem with periodically updated reference points is approximately solved by an appropriate stochastic algorithm. They established the convergence to a nearly stationary point for the equivalent minimization problem. Under the same setting,~\citet{lu2019hybrid} designed a block-based algorithm and showed that it can converge to a solution with a small stationary gap, and~\citet{lin2019gradient} considered solving the problem using vanilla stochastic gradient descent ascent and established its convergence to a stationary point under the smoothness assumption.
There are also several papers~\cite{lin2018solving,sanjabi2018solving,liu2020towards} trying to solve non-convex non-concave min-max problems.~\citet{lin2018solving} proposed an inexact proximal point method for solving a class of weakly-convex weakly-concave problems, which was proven to converge to a nearly stationary point.~\citet{sanjabi2018solving} exploited the PL condition for the inner maximization problem and designed a multi-step alternating optimization algorithm which was able to converge to a stationary point.~\citet{liu2020towards} considered solving a class of nonconvex-nonconcave min-max problems by designing an adaptive gradient method and established an adaptive complexity for finding a stationary point. However, none of them is particularly designed for the distributed stochastic AUC maximization problem with a deep neural network. 
\vspace{-0.15in} 
\section{Preliminaries and Notations}
\vspace{-0.1in} 
The area under the ROC curve (AUC) on a population level for a scoring function $h: \mathcal{X}\rightarrow\mathbb{R}$ is defined as 
\vspace{-0.1in} 
\begin{equation} 
AUC(h) = \text{Pr}(h(\x) \geq h(\x') | y=1, y'=-1),
\end{equation}  
where $\z = (\x, y)$ and $\z'=(\x', y')$ are drawn independently from $\mathbb{P}$.
By employing the squared loss as the surrogate for the indicator function which is commonly used by previous studies~\cite{gao2013one,ying2016stochastic,liu2018fast,liu2019stochastic}, the deep AUC maximization problem can be formulated as 
\begin{equation*}
\min\limits_{\w\in\mathbb{R}^d} \E_{\z, \z'}\left[(1-h(\w;\x)+h(\w;\x'))^2| y=1, y'=-1\right],
\label{original_deep_auc}
\end{equation*}
where $h(\w; \x)$ denotes the prediction score for a data sample $\x$ made by a deep neural network parameterized by $\w$. 
It was shown in \cite{ying2016stochastic} that the above problem is equivalent to the following min-max problem: 
\begin{equation}
\label{min-max}
\min\limits_{\w\in \mathbb{R}^d \atop (a, b)\in \mathbb{R}^2} \max\limits_{\alpha\in \mathbb{R}} f(\w, a, b, \alpha)=\E_\z[F(\mathbf{w}, a, b, \alpha, \z)],
\end{equation}
where \begin{equation*}
\begin{split}
& F(\textbf{w}, a, b, \alpha; \z) = (1-p) (h(\w; \x)- a)^2 \mathbb{I}_{[y=1]} \\
&+ p(h(\w; \x) - b)^2\mathbb{I}_{[y=-1]} 
+ 2(1+\alpha)(p h(\w; \x)\mathbb{I}_{[y=-1]} \\
&- (1-p) h(\w, \x)\mathbb{I}_{[y=1]}) - p(1-p)\alpha^2,
\end{split}
\end{equation*}
where $p=\Pr(y=1)$ denotes the prior probability that an example belongs to the positive class, and $\mathbb I$ denotes an indicator function. The above min-max reformulation allows us to decompose the expectation over all data into the expectation over data on individual machines. 

In this paper, we consider the following distributed AUC maximization problem:
\begin{equation}
\label{opt:problem}
\min\limits_{\w\in \mathbb{R}^d \atop (a, b)\in \mathbb{R}^2} \max\limits_{\alpha\in \mathbb{R}} f(\w, a, b, \alpha)=\frac{1}{K}\sum\limits_{k=1}^{K} f_k(\mathbf{w}, a, b, \alpha),
\end{equation}
where $K$ is the total number of machines, $f_k(\w, a, b, \alpha)= \E_{\z^k}[F_k(\w, a,b,\alpha; \z^k)]$, $\z^k = (\x^k, y^k)\sim \mathbb{P}_k$, $\mathbb{P}_k$ is the data distribution on machine $k$, and $F_k(\w, a,b,\alpha; \z^k) = F(\w, a,b,\alpha; \z^k)$.
Our goal is to utilize $K$ machines to jointly solve the optimization problem~(\ref{opt:problem}). We emphasize that the $k$-th machine can only access data $\z^k\sim \mathbb P_k$ of its own. It is notable that our formulation includes both the batch-learning setting and the online learning setting. For the batch-learning setting, $\mathbb P_k$ represents the empirical distribution of data on the $k$-th machine and $p$ denotes the empirical positive ratio for all data. For the online learning setting, $\mathbb P_k= \mathbb P,\forall k$ represents the same population distribution of data and $p$ denotes the positive ratio in the population level. 

\paragraph{Notations.} We define the following notations:  
%
%
\begin{align*}
&\v = (\w^T, a, b)^T, \quad \phi(\v) = \max_{\alpha} f(\v, \alpha),\\
&  \phi_s (\v) = \phi(\v) + \frac{1}{2\gamma}\|\v-\v_{s-1}\|^2,\\
&\v^*_{\phi} = \arg\min\limits_{\v}\phi(\v),\quad \v^*_{\phi_s} = \arg\min\limits_{\v} \phi_s(\v).
\end{align*}
We make the following assumption throughout this paper.
\begin{assumption}~\\
\label{ass:1}
(i) There exist $\v_0, \Delta_0>0$ such that $\phi(\v_0) - \phi(\v^*_\phi)\leq \Delta_0$. 
(ii) For any $\x$, $\|\nabla h(\w; \x)\| \leq G_h$.\\
(iii) $\phi(\v)$ satisfies the $\mu$-PL condition, i.e., $\mu(\phi(\v)-\phi(\v_*))\leq \frac{1}{2}\|\nabla\phi(\v)\|^2$; $\phi(\v)$ is $L_1$-smooth, i.e., $\|\phi(\v_1) - \phi(\v_2)\| \leq L_1 \|\v_1 - \v_2\|$.\\
(iv) For any $\x$, $h(\w; \x)$ is $L_h$-smooth, and $h(\w; \x) \in [0, 1]$.
\label{assumption_1}
\end{assumption}
\vspace*{-0.15in}
\textbf{Remark:} Assumptions (i), (ii), (iii) and $h(\w; \x)\in [0, 1]$ of (iv) are also assumed in~\cite{liu2019stochastic}, which have been justified as well. 
$L$-smoothness of function $h$ is a standard assumption in the optimization literature. Finally, it should be noted that $\mu$ is usually much smaller than $1$~\cite{DBLP:conf/nips/Yuan0JY19}. This is important for us to understand our theoretical result later. 
\vspace{-0.1cm}

\section{Main Result and Theoretical Analysis}
In this section, we first describe our algorithm, and then present its convergence result followed by its analysis. 
For simplicity, we assume that the ratio $p$ of  data with the positive label is known. For the batch learning setting, $p$ is indeed the empirical ratio of positive examples. For the online learning setting with an unknown distribution, we can follow the online estimation technique in~\cite{liu2019stochastic} to do the parameter update.

 Algorithm~\ref{outer_loop} describes the proposed algorithm CoDA for optimizing AUC in a communication-efficient distributed manner.  CoDA shares the same algorithmic framework as proposed in~\cite{liu2019stochastic}. 
In particular, we employ a proximal-point algorithmic scheme that successively solves the following convex-concave problem approximately: 
\begin{equation}
    \min_{\v}\max_{\alpha}f(\v,\alpha)+\frac{1}{2\gamma}\|\v-\v_0\|^2,
\end{equation}
where $\gamma$ is an appropriate regularization parameter to make sure that the regularized  function is strongly-convex and strongly-concave. The reference point $\v_0$ is periodically updated after a number of iterations. At the $s$-th stage our algorithm invokes a communication-efficient algorithm for solving the above strongly-convex and strongly-concave subproblems. After obtaining a primal solution $\v_s$ at the $s$-th stage, we sample some data from individual machines to obtain an estimate of the corresponding dual variable $\alpha_s$. 
\begin{algorithm}[t]
\caption{CoDA}
\begin{algorithmic}[1]
\STATE{Initialization: $(\v_0 = \mathbf{0} \in \mathbb{R}^{\alpha + 2}, \alpha_0 = 0, \gamma)$.}
\FOR{$s=1, ..., S$}
\STATE{$\v_s = \text{DSG} (\v_{s-1}, \alpha_{s-1}, \eta_s, T_s, m_s, I_s, \gamma)$;}
\STATE{Each machine draws a minibatch $\{\z^k_1, ..., \z^k_{m_s}\}$ of size $m_s$ and does:}
\STATE{~~~~$h^k_{-} = \sum\limits_{i=1}^{m_s}h(\v_{s}; \x^k_i)\I_{y_i^k=-1}, N_{-}^{k} = \sum\limits_{i=1}^{m_s}\I_{y^k_i=-1}$,}
\STATE{~~~~$h^k_{+} = \sum\limits_{i=1}^{m_s}h(\v_{s}; \x^k_i)\I_{y^k_i=1}, N_{+}^{k} = \sum\limits_{i=1}^{m_s}\I_{y^k_i=1}$;}
\STATE{$\alpha_{s} = \frac{1}{K}\sum\limits_{k=1}^{K} \left[\frac{h_{-}^k}{N_{-}^k}
-  \frac{h_{+}^k}{N_{+}^k} \right]$; \hfill $\diamond$ communicate}
\ENDFOR 
\STATE{Return $\v_{S}$.} 
\end{algorithmic}
\label{outer_loop}
\end{algorithm}

Our new contribution is  the communication-efficient distributed algorithm  for solving the above strongly-convex and strongly-concave subproblems. The algorithm referred to as DSG is presented in  Algorithm~\ref{inner_loop}. Each machine makes a stochastic proximal-gradient update on the primal variable and a stochastic gradient update on the dual variable at each iteration. After every $I$ iterations, all the $K$ machines communicate to compute an average of local primal solutions $\v^k_t$ and local dual solutions $\alpha^k_t$. It is not difficult to show that when $I=1$, our algorithm reduces to the naive parallel version of the PPD-SG algorithm proposed in~\cite{liu2019stochastic}, i.e., by averaging individual primal and dual gradients and then updating the primal-dual variables according to the averaged gradient~\footnote{A tiny difference is that we use a proximal gradient update to handle the regularizer $\frac{1}{2\gamma}\|\v-\v_0\|^2$, while \citet{liu2019stochastic} used the gradient update. Using the proximal gradient update allows us to remove the assumption that $\|\v^k_t - \v_0\|$ is upper bounded.}. 
Our novel analysis allows us to use $I>1$ to skip communications, leading to a much less number of communications. The intuition behind this is that,  as long as the step size $\eta_s$ is sufficiently small we can control the distance between individual solutions $(\v^k_t, \alpha^k_t)$ to their global averages, which allows us to control the error term that is caused by the discrepancy between individual machines. We will provide more explanations as we present the analysis. 


Below, we present the main theoretical result of CoDA. Note that in the following presentation, $L_\v, H, B, \sigma_\v, \sigma_\alpha$ are appropriate constants, whose values are given in the proofs of Lemma~\ref{properties} and Lemma~\ref{lemma_one_stage} in the supplement.
\begin{theorem}
\label{thm:main}
Set $\gamma = \frac{1}{2L_{\v}}$, $c =\frac{\mu/L_{\v}}{5+\mu/L_{\v}}$, $\eta_s = \eta_0 K \exp(-(s-1)c)\leq O(1)$, 
$T_s = \frac{\max(8, 8G_h^2)}{L_{\v}\eta_0 K} \exp( (s-1)c)$, $I_s = \max(1, 1/\sqrt{K\eta_s})$ and $m_{s}=\max\left(\frac{1+C}{\eta_{s+1}^2 T_{s+1}  p^2(1-p)^2}, \frac{\log(K)}{\log(1/\Tilde{p})}\right)$, where $C = \frac{3\tilde{p}^{\frac{1}{\ln(1/\Tilde{p})}}}{2\ln(1/\Tilde{p})}$ and $\Tilde{p} = \max(p, 1-p)$. 
 To return $\v_S$ such that $\E[\phi(\v_S) - \phi(\v^*_{\phi})] \leq \epsilon$, it suffices to choose $S\geq \frac{5L_{\v}+\mu}{\mu} \max\bigg\{\log \left(\frac{2\Delta_0}{\epsilon}\right), \log S + \log\bigg[ \frac{2\eta_0}{\epsilon} \frac{6HB^2 + 12(\sigma_{\v}^2+\sigma_{\alpha}^2)}{5}\bigg]\bigg\}$.
 As a result, the number of iterations is at most $T=\widetilde{O}\bigg(\max\left(\frac{\Delta_0}{\mu \epsilon \eta_0 K}, \frac{L_{\v}}{\mu^2 K\epsilon}\right)\bigg)$ and the number of communications is at most $\widetilde{O}\bigg(\max\left(\frac{K}{\mu} + \frac{\Delta_0^{1/2}}{\mu(\eta_0 \epsilon)^{1/2}}, \frac{K}{\mu} +   \frac{L_{\v}^{1/2}}{\mu^{3/2}\epsilon^{1/2}}\right)\bigg)$, where $\widetilde{O}$ suppresses logarithmic factors, and $H, B, \sigma_\v, \sigma_\alpha$ are appropriate constants. 
\label{theorem_all_stage}
\end{theorem}
\begin{algorithm}[t]
\caption {DSG($\v_0, \alpha_0, \eta, T, I, \gamma$)}
\begin{algorithmic}
\STATE{Each machine does initialization: $\v_0^k = \v_0, \alpha_0^k = \alpha_0$,}
\FOR{$t=0, 1, ..., T-1$}
\STATE{Each machine $k$ updates its local solution in parallel:}
\STATE{~~~~$\v_{t+1}^k=\arg\min\limits_{\v}\bigg[\nabla_{\v} F_k(\v_t^k, \alpha_t^k; \z_t^k)^T \v$\\
~~~~~~~~~~~~~~~~~~~~~~~~~~~~~~
$+ \frac{1}{2\eta}\|\v - \v_t^k\|^2 + \frac{1}{2\gamma} \|\v - \v_0\|^2\bigg]$,}
\STATE{~~~~$\alpha_{t+1}^k = \alpha_t^k + \eta \nabla_{\alpha} F_k(\v_t^k, \alpha_t^k; \z_t^k)$,}
\IF{$t+1$ mod $I = 0$ }
\STATE{$\v^k_{t+1} = \frac{1}{K} \sum\limits_{k=1}^{K} \v_{t+1}^k$, \hfill $\diamond$ communicate }
\STATE{$\alpha^k_{t+1} = \frac{1}{K}\sum\limits_{k=1}^{K} \alpha_{t+1}^k$, \hfill $\diamond$ communicate}
\ENDIF
\ENDFOR
\STATE{Return $\tilde{\v} = \frac{1}{K}\sum\limits_{k=1}^{K}\frac{1}{T} \sum\limits_{t=1}^{T} \v_t^k$.}
\end{algorithmic}
\label{inner_loop}
\end{algorithm}
\vspace{-0.2cm}
We have the following remarks about Theorem~\ref{thm:main}.
\vspace{-0.2cm}
\begin{itemize}
\item  First, we can see that the step size $\eta_s$ is reduced geometrically in a stagewise manner. This is due to the PL condition. We note that a stagewise geometrically decreasing step size is usually used in practice in deep learning~\cite{DBLP:conf/nips/Yuan0JY19}.  Second, by setting $\eta_0= O(1/K)$ we have  $I_s =\Theta(\frac{1}{\sqrt{K}}\exp((s-1)c/2)$. It means two things: (i) the larger the number of machines the smaller value of $I_s$, i.e., more frequently the machines need to communicate. This is reasonable since more machines will create a larger discrepancy between data among different machines; (ii) the value $I_s$ can be increased geometrically across stages. This is because that the step size $\eta_s$ is reduced geometrically, which causes one step of primal and dual updates on individual machines to diverge less from their averaged solutions. As a result, more communications can be skipped. 
\vspace{-0.2cm}
\item Second, we can see that when $K\leq \Theta(1/\mu)$, we have the total iteration complexity given by $\widetilde O(\frac{1}{\mu^2K \epsilon})$. Compared with the iteration complexity of the PPD-SG algorithm proposed in~\cite{liu2019stochastic} that is $\widetilde O(\frac{1}{\mu^2 \epsilon})$, the proposed algorithm CoDA enjoys an iteration complexity that is reduced by a factor of $K$. This means that up to a certain large threshold  $\Theta(1/\mu)$ for the number $K$ of machines, CoDA enjoys a linear speedup. 
\vspace{-0.2cm}
\item Finally, let us compare CoDA with the naive parallel version of PPD-SG, which is CoDA by setting $I=1$.  In fact, our analysis of the iteration complexity for this case is still applicable, and it is not difficult to show that the iteration complexity of the naive parallel version of PPD-SG is given by  $\widetilde O(\frac{1}{\mu^2K \epsilon})$ when $K\leq 1/\mu$. As a result, its communication complexity is also  $\widetilde O(\frac{1}{\mu^2K \epsilon})$. In contrast, CoDA's communication complexity is  $\widetilde O(\frac{1}{\mu^{3/2}\epsilon^{1/2}})$ when $K\leq \frac{1}{\mu}\leq  \frac{1}{\mu^{1/2}\epsilon^{1/2}}\leq \frac{1}{\epsilon}$ according to Theorem~\ref{thm:main}~\footnote{Assume that $\epsilon$ is set to be smaller than $\mu$.}. Hence, our algorithm is more communication efficient, i.e., $\widetilde O(\frac{1}{\mu^{3/2}\epsilon^{1/2}})\leq \widetilde O(\frac{1}{\mu^2K \epsilon})$ when $K\leq \frac{1}{\mu}$~\footnote{Indeed, $K$ can be as large as $\frac{1}{\mu^{1/2}\epsilon^{1/2}}$ for CoDA to be more communication-efficient.}. This means that  up to a certain large threshold  $\Theta(1/\mu)$ for the number $K$ of machines, CoDA has a smaller communication complexity than the naive parallel version of PPD-SG. 
\end{itemize}




\vspace{-0.5cm}
\subsection{Analysis}
Below, we present a sketch of the proof of Theorem~\ref{thm:main} by providing some key lemmas. 
We first derive some useful properties regarding the random function $F_k(\v, \alpha, \z)$. 
\begin{lemma}
\label{properties}
Suppose that Assumption \ref{assumption_1} holds and $\eta \leq \min(\frac{1}{2p(1-p)}, \frac{1}{2(1-p)}, \frac{1}{2p})$.
Then there exist some constants $L_2, B_\alpha, B_\v, \sigma_{\v}, \sigma_{\alpha}$ such that
\begin{align*}
&\|\nabla_{\v} F_k(\v_1, \alpha; \z) - \nabla_{\v} F_k(\v_2, \alpha; \z)\| \leq L_2 \|\v_1-\v_2\|,\\
&\|\nabla_{\v} F_k(\v, \alpha; \z)\|^2 \leq B_{\v}^2, |\nabla_{\alpha} F_k(\v, \alpha; \z)|^2 \leq B_{\alpha}^2,\\ 
&\E[\|\nabla_{\v} f_k(\v, \alpha) - \nabla_{\v} F_k(\v, \alpha; \z)\|^2] \leq \sigma_{\v}^2,\\
&\E[|\nabla_{\alpha}f_k(\v, \alpha) - \nabla_{\alpha} F_k(\v, \alpha; \z)|] \leq \sigma^2_{\alpha}.
\end{align*}
\end{lemma}
\vspace*{-0.1in}
\textbf{Remark:} We include the proofs of these properties in the Appendix. In the following, we will denote $B^2 = \max(B_{\v}^2, B_{\alpha}^2)$ and $L_{\v} = \max(L_1, L_2)$.


Next, we introduce a key lemma,  which is of vital importance to establish the upper bound of the objective gap of the regularized subproblem. 
\begin{lemma}(One call of Algorithm \ref{inner_loop})
Let $\psi(\v) = \max\limits_{\alpha} f(\v, \alpha) + \frac{1}{2\gamma}\|\v - \v_0\|^2$, $\tilde{\v}$ be the output of Algorithm \ref{inner_loop}, and $\v^*_{\psi} = \arg\min \psi(\v)$, $\alpha^*(\tilde{\v}) = \arg\max\limits_{\alpha}f(\tilde{\v}, \alpha)+\frac{1}{2\gamma}\|\tilde{\v}-\v_0\|^2$. 
By running Algorithm \ref{inner_loop} with given input $\v_0, \alpha_0$ for $T$ iterations, $\gamma = \frac{1}{2L_{\v}}$, and $\eta \leq \min\{\frac{1}{L_{\v}+3G^2_{\alpha}/\mu_{\alpha}}, \frac{1}{L_{\alpha}+3G_{\v}^2/L_{\v}}, \frac{3}{2\mu_{\alpha}}, \frac{1}{2p(1-p)}, \frac{1}{2(1-p)}, \frac{1}{2p}\}$, we have \\
\begin{small}
\begin{align*}
\E[\psi(\tilde{\v}) - \min_{\v}\psi(\v)] \leq& \!\frac{2\|\v_0 - \v_{\psi}^*\|^2\! +\! \E[(\alpha_{0} - \alpha^*(\tilde{\v}))^2]}{\eta T} \\
&\!+\! H\!\eta^2\! I^2\! B^2 \I_{I>1}\! +\! \frac{\eta (2\sigma_{\v}^2\! +\! 3\sigma_{\alpha}^2)}{2K}, 
\end{align*}
\end{small}
where $\mu_{\alpha} = 2p(1-p)$, $L_{\alpha} = 2p(1-p)$, $G_{\alpha} = 2\max\{p, 1-p\}$,
$G_{\v} = 2\max\{p, 1-p\} G_h$,
and $H = \left(\frac{6G_{\v}^2}{\mu_{\alpha}} + 6L_{\v} + \frac{6G_{\alpha}^2}{L_{\v}} + \frac{6L_{\alpha}^2}{\mu_{\alpha}}\right)$.
\label{lemma_one_stage}  
\end{lemma} 
{\bf Remark: } The above result is similar to Lemma 2 in~\cite{liu2019stochastic}. The key difference lies in the second and third terms in the upper bound. The second term arises because of the discrepancy of updates between individual machines. The third term is due to the variance reduction by using multiple machines, which is the key to establish the linear speed-up. It is easy to see that by setting $I = \frac{1}{\sqrt{\eta K}}$, the second term and the third term have the same order. With the above lemma, the proof of Theorem~\ref{thm:main} follows similar analysis  to in~\cite{liu2019stochastic}. 

{\bf Sketch of the Proof of Lemma~\ref{lemma_one_stage}.} Below, we present a roadmap for the proof of the key Lemma~\ref{lemma_one_stage}. The main idea is to first bound the objective gap of the subproblem in Lemma~\ref{lem:objgap}. Then we further bound every term in the RHS in Lemma~\ref{lem:objgap}  appropriately, which is realized by Lemma~\ref{lem:var1}, Lemma~\ref{lem:var2} and Lemma~\ref{stich_lemma}. All the detailed proofs of Lemmas can be found in Appendix.



\begin{lemma}
\label{lem:objgap}
Define $\bar\v_t = \frac{1}{K}\sum_{k=1}^K\v^k_t, \bar{\alpha}_t =  \frac{1}{K}\sum_{k=1}^K\alpha^k_t$.  Suppose Assumption \ref{assumption_1} holds and by running Algorithm \ref{inner_loop}, we have
\begin{small}
\begin{align*}
&\psi(\tilde{\v}) - \min\limits_{\v} \psi(\v)\\
&\!\leq\! \frac{1}{T}\sum\limits_{t=1}^{T}\bigg[\! \underbrace{\!\langle \nabla_{\v}f(\bar{\v}_{t-1}, \bar{\alpha}_{t-1}),\! \bar{\v}_t\! -\! \v_{\psi}^*\rangle \!+\!2L_{\v}\! \langle \bar{\v}_t\!-
\!\v_0,\! \bar{\v}_t\!-\!\v_{\psi}^*\rangle}_{A_1} \\
& + \underbrace{ \langle \nabla_{\alpha} f(\bar{\v}_{t-1}, \bar{\alpha}_{t-1}),  \alpha^* - \bar{\alpha}_t \rangle }_{A_2}
\end{align*}
\begin{align*}
&+\!\underbrace{\frac{L_{\v} + 3G_{\alpha}^2/\mu_{\alpha}}{2}\|\bar{\v}_t - \bar{\v}_{t-1}\|^2  \!+\! \frac{L_{\alpha} + 3G_{\v}^2/L_{\v}}{2} (\bar{\alpha}_t - \bar{\alpha}_{t-1})^2}_{A_3} \\
&+ \frac{2L_{\v}}{3} \| \bar{\v}_{t-1} - \v_{\psi}^*\|^2 - L_{\v} \| \bar\v_t - \v_{\psi}^* \|^2   - \frac{\mu_{\alpha}}{3} (\bar{\alpha}_{t-1} - \alpha^*)^2\bigg].
\label{sum_v_alpha}
\end{align*}
\end{small}
\end{lemma}
Next, we will bound $A_1, A_2$ in Lemma~\ref{lem:var1} and Lemma~\ref{lem:var2}. 
$A_3$ can be cancelled with similar terms in the following two lemmas.
The remaining terms will be left to form a telescoping sum with other similar terms in the following two lemmas.  
\vspace{-0.1in} 
\begin{lemma}
\label{lem:var1}
Define \!$\hat{\v}_t\! =\! \arg\min\limits_{\v}\! \left(\! \frac{1}{K}\!\sum\limits_{k=1}^{K}\! \nabla_{\v}\! f\!(\!\v^k_{t-1}\!,\! \alpha^k_{t-1}\!)\!\right)^T\! \v\!$ $+ \frac{1}{2\eta} \|\v - \bar{\v}_{t-1}\|^2 + \frac{1}{2\gamma}\|\v - \v_{0}\|^2$.
 We have
\begin{small}
\begin{equation*}
\begin{split}
&A_1
\!\leq \! \frac{3G_{\alpha}^2}{2L_{\v}}\frac{1}{K} \sum\limits_{k=1}^{K} (\bar{\alpha}_{t-1}-\alpha_{t-1}^{k})^2 \!+\! \frac{3L_{\v}}{2}\frac{1}{K}\sum\limits_{k=1}^{K}\|\bar{\v}_{t-1}\!-\!\v_{t-1}^k\|^2 \\
&+ \eta \left\|\frac{1}{K}\sum\limits_{k=1}^{K}[\nabla_{\v}f_k(\v_{t-1}^k, \alpha_{t-1}^k) \!-\! \nabla_{\v}F_k(\v_{t-1}^{k}, \alpha_{t-1}^k; \z_{t-1}^k)]\right\|^2 \\
&\!+\! \frac{1}{K}\!\sum\limits_{k=1}^{K}\!\langle \!\nabla_{\v}f_k(\v_{t-1}^k,\! \alpha_{t-1}^k)\!-\!\nabla_{\v}F_k(\v_{t-1}^{k},\! \alpha_{t-1}^k;\z_{t-1}^k),\! \hat{\v}_t - \v_{\psi}^*\rangle\\
& \!+\!\frac{1}{2\eta} \!(\!\|\bar{\v}_{t-1}\!-\!\v_{\psi}^*\|^2 \!-\! \|\bar{\v}_{t-1} \!-\! \bar{\v}_t\|^2 \!-\! \|\bar{\v}_t \!-\! \v_{\psi}^*\|^2) 
\!+\! \frac{L_{\v}}{3}\|\bar{\v}_t\! -\! \v_{\psi}^*\|^2.
\end{split}
\end{equation*}
\end{small}
\end{lemma}
\vspace{-0.1in}
\begin{lemma}
\label{lem:var2}
Define $\hat{\alpha}_t\! = \!\bar{\alpha}_{t-1} + \frac{\eta}{K}\sum\limits_{k=1}^{K} \nabla_{\alpha} f_k(\v_{t-1}^{k}, \alpha_{t-1}^k)$, 
and 
\begin{small}
\begin{equation*}
\tilde{\alpha}_{t}\! =\! \tilde{\alpha}_{t-1}\! +\! \frac{\eta}{K}\sum\limits_{k=1}^{K}( 
 \nabla_{\alpha} F_k(\v_{t-1}^{k},\alpha_{t-1}^k; \z_{t-1}^k)  
\!-\!\nabla_{\alpha} f_k(\v_{t-1}^{k},\!\alpha_{t-1}^k) ). 
\end{equation*} 
\end{small}
\vspace{-0.1in} 
We have, 
\begin{small}
\begin{equation*}
\begin{split}
&A_2 \!\leq\! \frac{3G_{\v}^2}{2\mu_{\alpha}} \frac{1}{K}\sum\limits_{k=1}^{K}\|\bar{\v}_{t-1}-\v_{t-1}^k\|^2 \!+\! \frac{3L_{\alpha}^2}{2\mu_{\alpha}}\frac{1}{K}\sum\limits_{k=1}^{K} (\bar{\alpha}_{t-1} - \alpha_{t-1}^k)^2\\
&+\frac{3\eta}{2} ( \frac{1}{K} \sum\limits_{k=1}^{K}[ \nabla_{\alpha} f_k(\v_{t-1}^k, \alpha_{t-1}^k) -  \nabla_{\alpha} F_k (\v_{t-1}^k, \alpha_{t-1}^k; \z_{t-1})])^2\\ 
&+\!\frac{1}{K}\sum\limits_{k=1}^{K}\!\langle\! \nabla_{\alpha}\!  f_k(\v_{t-1}^k,\!\alpha_{t-1}^k)\!- \!  \nabla_{\alpha}\!F_k(\v_{t-1}^k,\!\alpha_{t-1}^k;\!\z_{t-1}^k),\!\tilde{\alpha}_{t-1}\!-\!\hat{\alpha}_t\!\rangle\\  
&\!+\!\frac{1}{2\eta} ((\bar{\alpha}_{t-1} \!-\! \alpha^*(\Tilde{\v}))^2 \!- \!(\bar{\alpha}_{t-1} \!-\!  \bar{\alpha}_t)^2 \!-\! (\bar{\alpha}_t \!-\!  \alpha^*(\Tilde{\v}))^2)
\!\\ 
& +\! \frac{\mu_{\alpha}}{3} (\bar{\alpha}_{t} \!-\! \alpha^*(\Tilde{\v}))^2 
 + \frac{1}{2\eta} ((\alpha^*(\Tilde{\v}) - \tilde{\alpha}_t)^2 -  (\alpha^*(\Tilde{\v})-\Tilde{\alpha}_{t+1})^2).
\end{split}
\end{equation*}
\end{small} 
\end{lemma}
\vspace*{-0.1in}
 The first two terms in the upper bounds of $A_1, A_2$ are the differences between individual solutions and their averages, the third term is the variance of stochastic gradient, and the expectation of the fourth term will diminish. 
The  lemma below will bound the difference between the averaged solution and the individual solutions. 
\begin{lemma}
\label{stich_lemma}
If $K$ machines communicate every $I$ iterations, and update with step size $\eta$, then\\
\vspace{-0.2in} 
\begin{small} 
\begin{align*} 
&\frac{1}{K}\sum\limits_{k=1}^{K} \E[\|\bar{\v}_t - \v_t^k\|^2] \leq 4\eta^2I^2 B_{\v}^2\mathbb I_{I>1}\\
&\frac{1}{K}\sum\limits_{k=1}^{K} \E[\|\bar{\alpha}_t - \alpha_t^k\|^2] \leq 4\eta^2 I^2 B_{\alpha}^2\mathbb I_{I>1}.
\end{align*}
\end{small}
\end{lemma}
\vspace{-0.3cm}
Combining the results in Lemma~\ref{lem:objgap}, Lemma~\ref{lem:var1}, Lemma~\ref{lem:var2} and Lemma~\ref{stich_lemma}, we can prove the key Lemma~\ref{lemma_one_stage}. 

\vspace{-0.4cm}
\section{Experiments}
In this section, we conduct some experiments to verify our theory. In our experiments, one ``machine'' corresponds to one GPU. We use a cluster of 4 computing nodes with each computer node having 4 GPUs, which gives a total of $16$ ``machines''. We would like to emphasize that even though 4 GPUs sit on one computing node, they only access to different parts of the data. For the experiment with $K=1$ GPU, We run one computing node by using one GPU.  For experiments with $K=4$ GPUs, we run one computing node by using all four GPUs, and for those experiments with $K=16$ GPUs, we use four computing nodes by using all GPUs. We notice that the communication costs among GPUs on one computing node may be less than that among GPUs on different computing nodes. Hence, it should be kept in mind that when comparing with $K=4$ GPUs on different computing nodes, the margin of using $K=16$ GPUs over using $K=4$ GPUs should be larger than what we will see in our experimental results.  All algorithms are implemented by PyTorch \cite{paszke2019pytorch}.


{\bf Data}. We do experiments on 3 datasets: Cifar10, Cifar100 and ImageNet.
For Cifar10, we split the original training data into two classes, i.e., the positive class contains 5 original classes and the negative class is composed of the other 5 classes. The Cifar100 dataset is split in a similar way, i.e., the positive class contains 50 original classes and the negative class is composed of the other 50 classes. Testing data for Cifar10 and Cifar100 is the same as the original dataset.
For the ImageNet dataset, we sample $1\%$ of the original training data as testing data and use the remaining data as the training data.
The training data is split in a similar way to Cifar10 and Cifar100, i.e., the positive class contains 500 original classes and the negative class is composed of the other 500 classes. For each dataset, we create two versions of training data with different positive ratios. By keeping all data in the positive and negative classes, we have $p=50\%$ for all three datasetes. To create imbalanced data, we drop some proportion of the negative data for each dataset and keep all the positive examples.  In particular,  by keeping all the positive data and $40\%$ of the negative data we construct three datasets with positive ratio $p=71\%$. Training data is shuffled and evenly divided to each GPU, i.e., each GPU has access to $1/K$ of the training data, where $K$ is the number of GPUs. For all data, we use ResNet50 as our neural network \cite{he2016deep} and initialize the model as the pretrained model from PyTorch. Due to the limit of space, we only report the results on datasets with $p=71\%$ positive ratio, and other results are included in the supplement.  

{\bf Baselines and Parameter Setting.} For baselines, we compare with the single-machine algorithm PPD-SG as proposed in~\cite{liu2019stochastic}, which is represented by $K=1$ in our results, and the naive parallel version of PPD-SG, which is denoted by $K=$ X, $I=1$ in our results. 
For all algorithms, we set $T_s = T_0 3^k$, $\eta_s = \eta_0/3^k$. $T_0$ and $\eta_0$ are tuned for PPD-SG and set to the same for all other algorithms for fair comparison. $T_0$ is tuned in $[2000, 5000, 10000]$, and $\eta_0$ is tuned in $[0.1, 0.01, 0.001]$.
We fix the batch size for each GPU as 32. For simplicity, in our experiments we use a fixed value of $I$ in order to see its tradeoff with the number of machines $K$. 

\begin{figure*}[ht]
      \centering
     \subfigure[Fix $I$, vary $K$]
       {\includegraphics[scale=0.25]{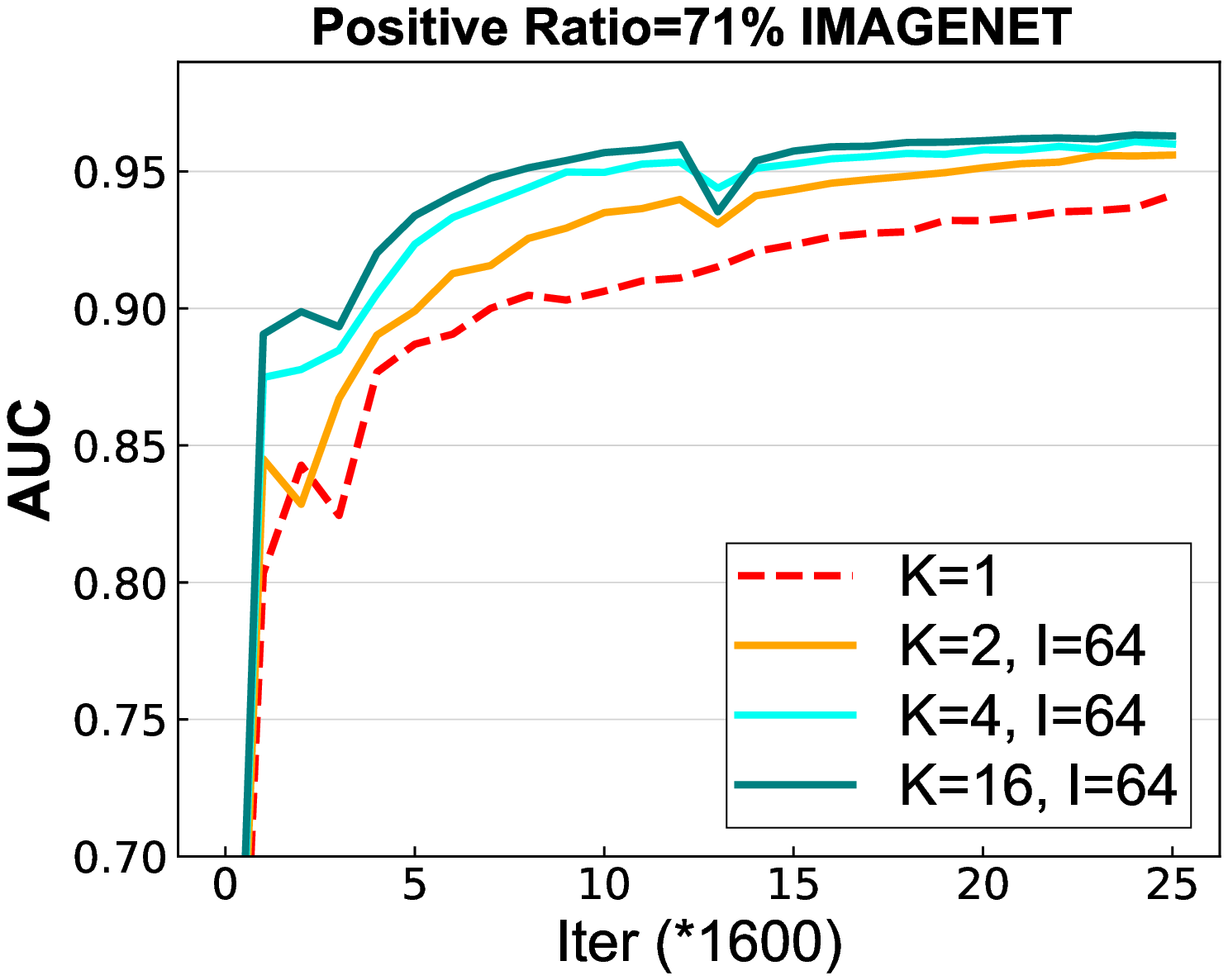} 
    \includegraphics[scale=0.25]{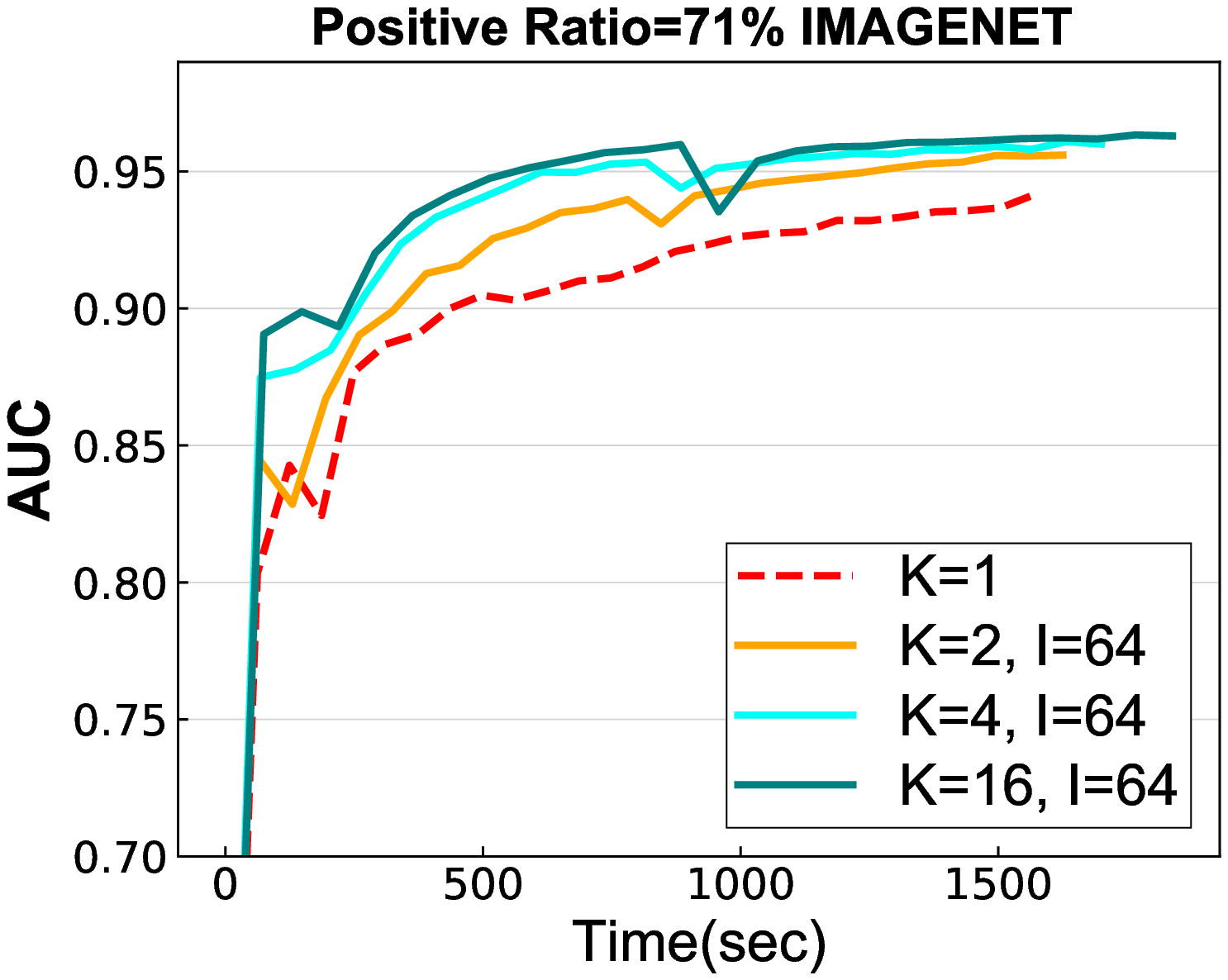}}
    \subfigure[Fix $K$, vary $I$]
    {\includegraphics[scale=0.25]{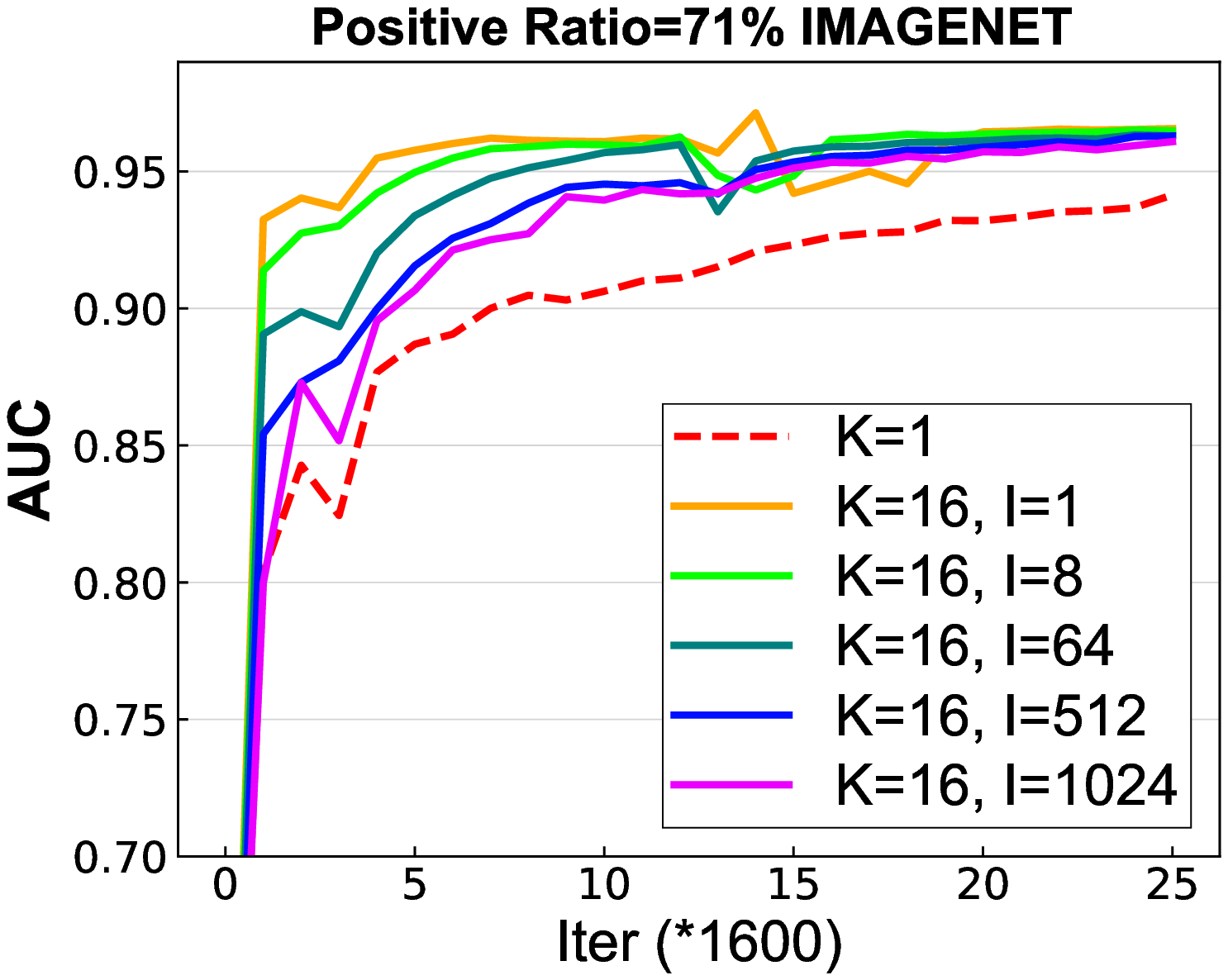}
    \includegraphics[scale=0.25]{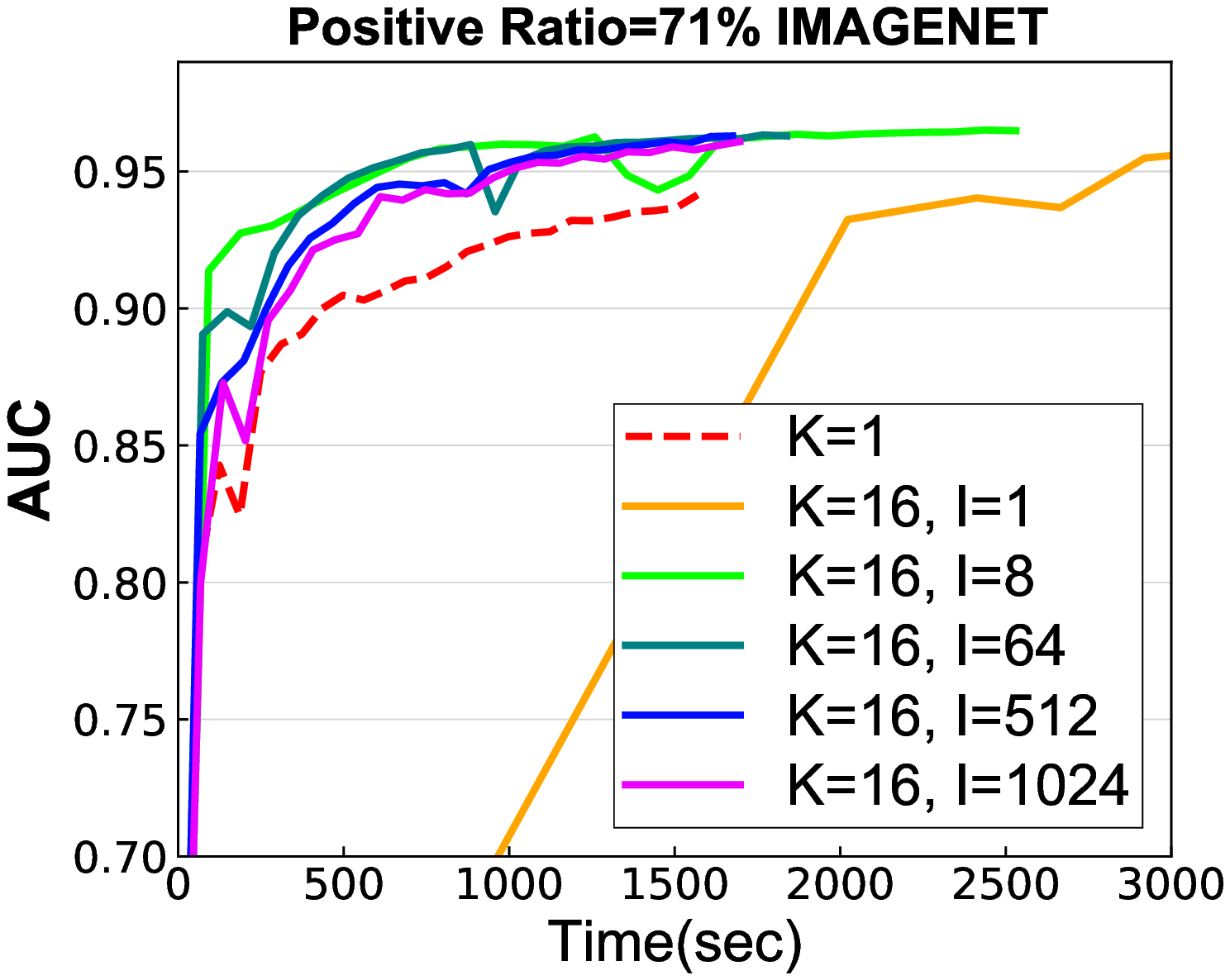}}
    \vspace{-0.4cm}
    \caption{ImageNet, positive ratio = 71\%.}
    \label{fig:imagenet_71}
\vspace{0.1cm}
     \centering
    \subfigure[Fix $I$, vary $K$]
    {\includegraphics[scale=0.25]{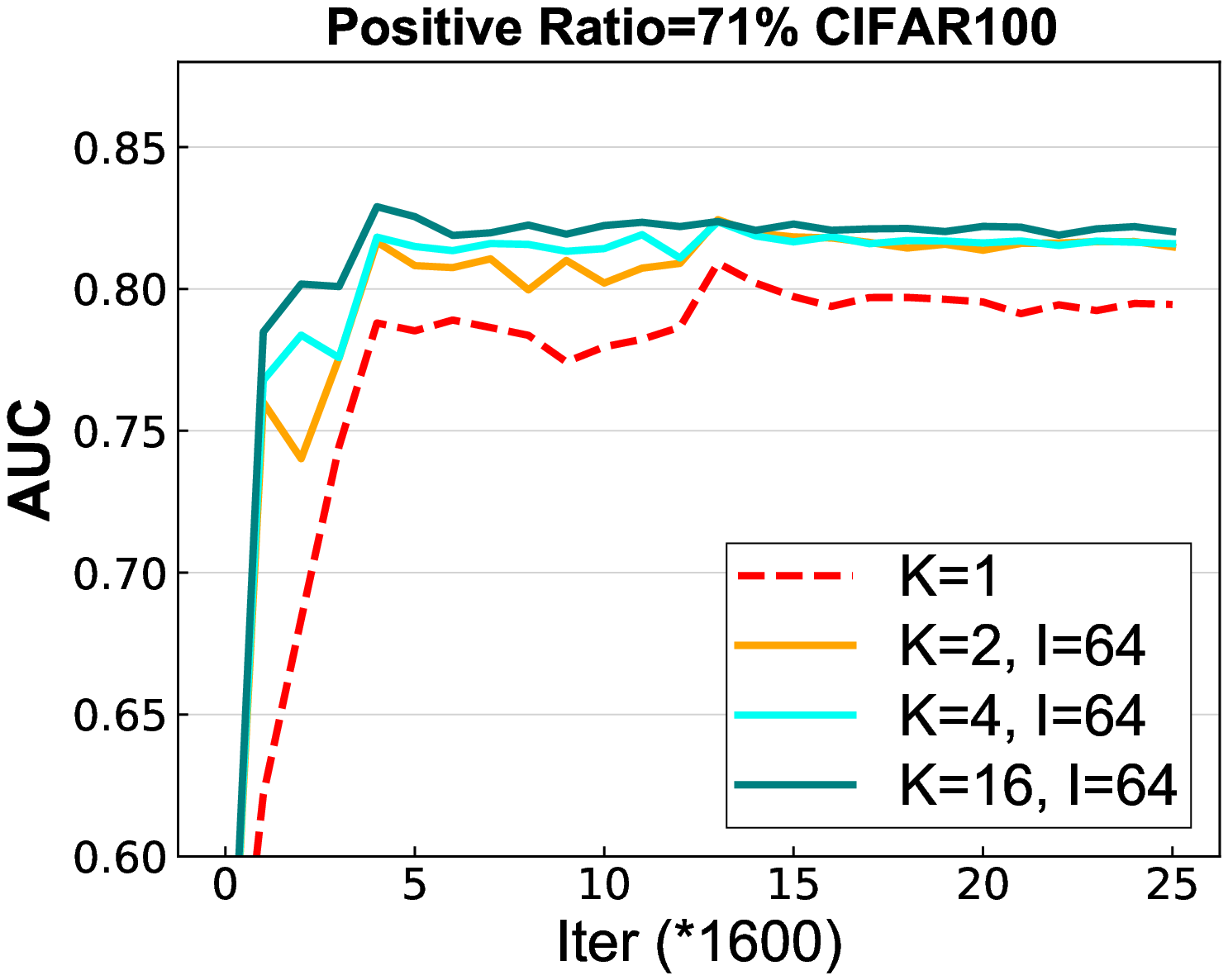}
    \includegraphics[scale=0.25]{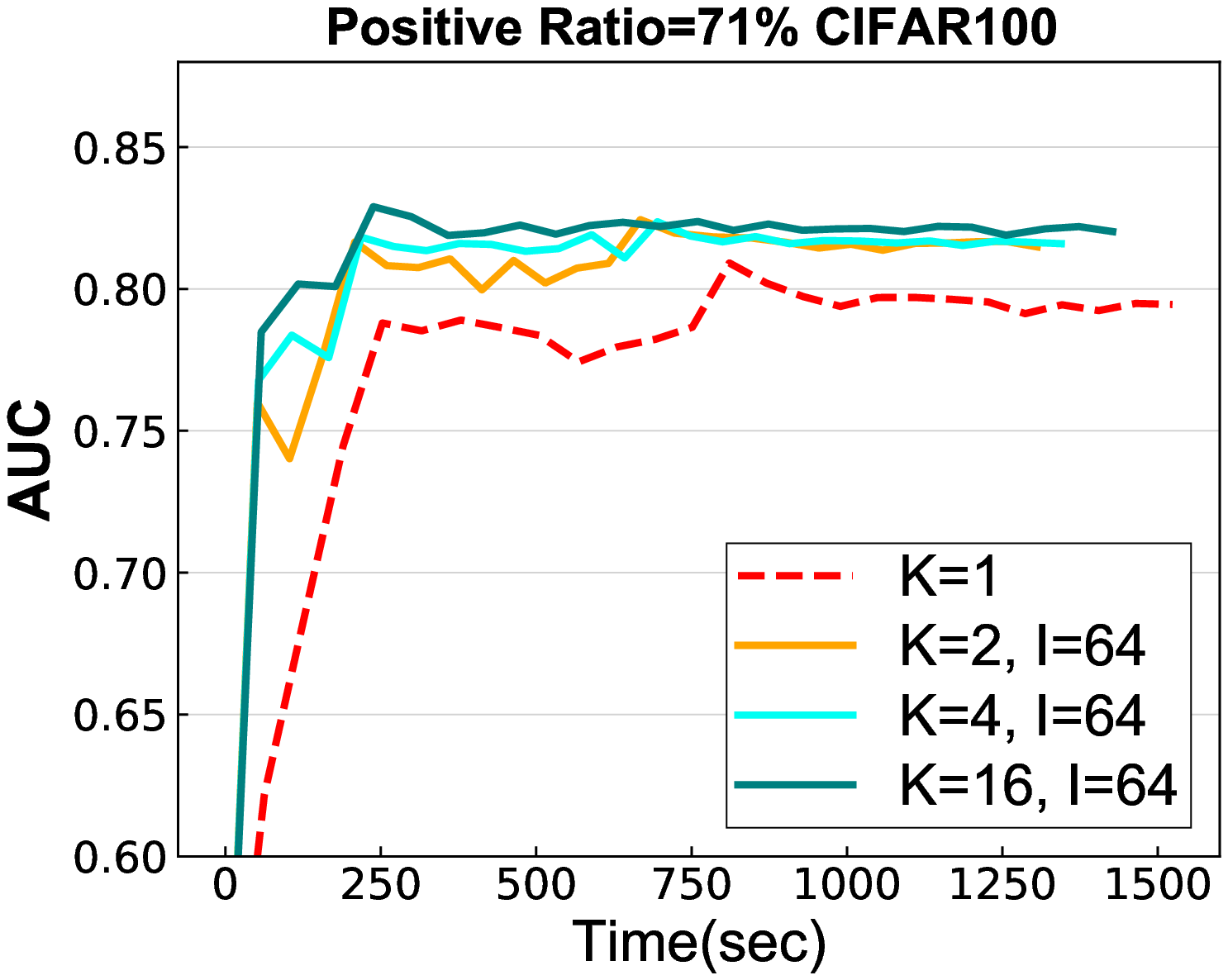}}
    \subfigure[Fix $K$, vary $I$]
    {\includegraphics[scale=0.25]{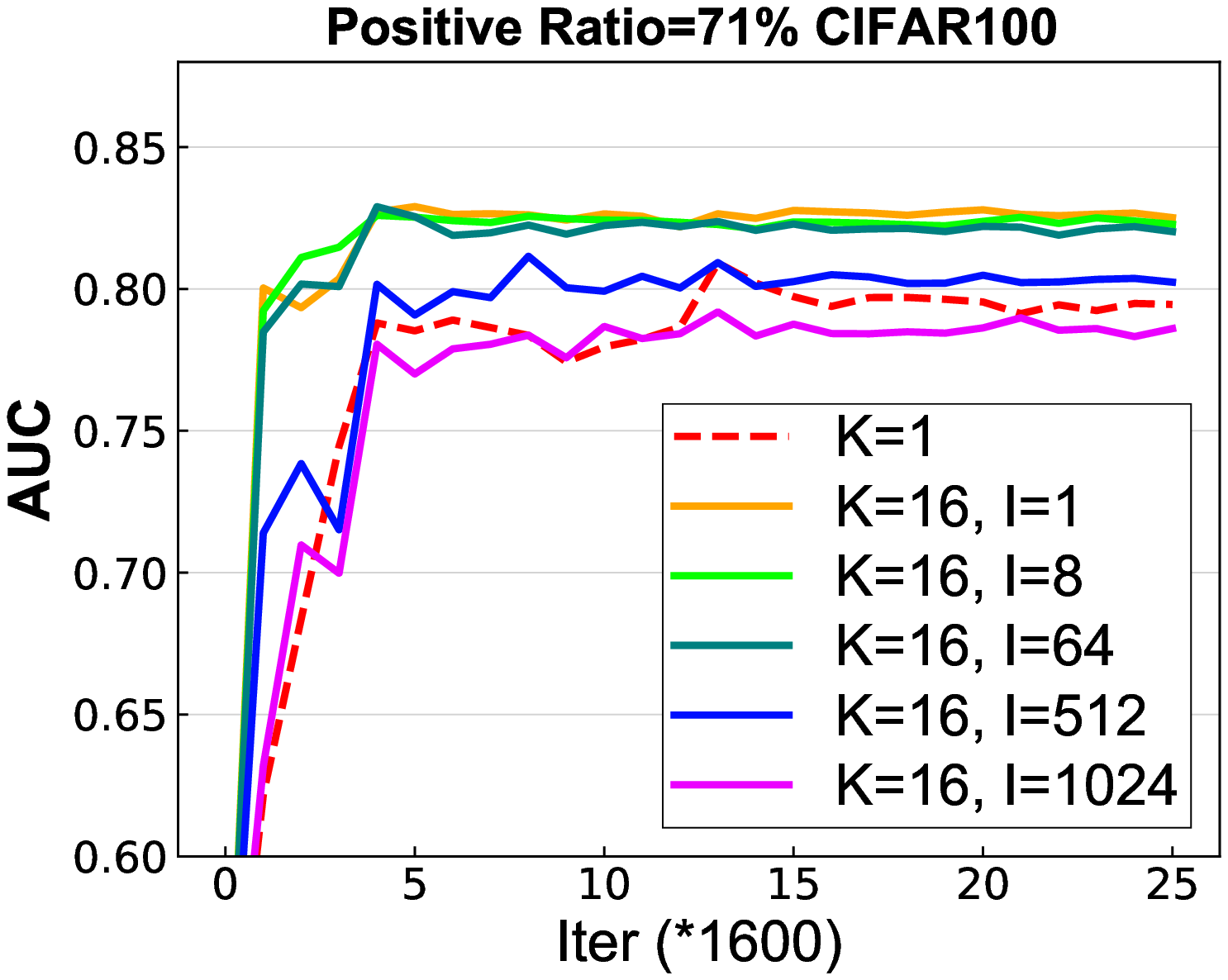}
    \includegraphics[scale=0.25]{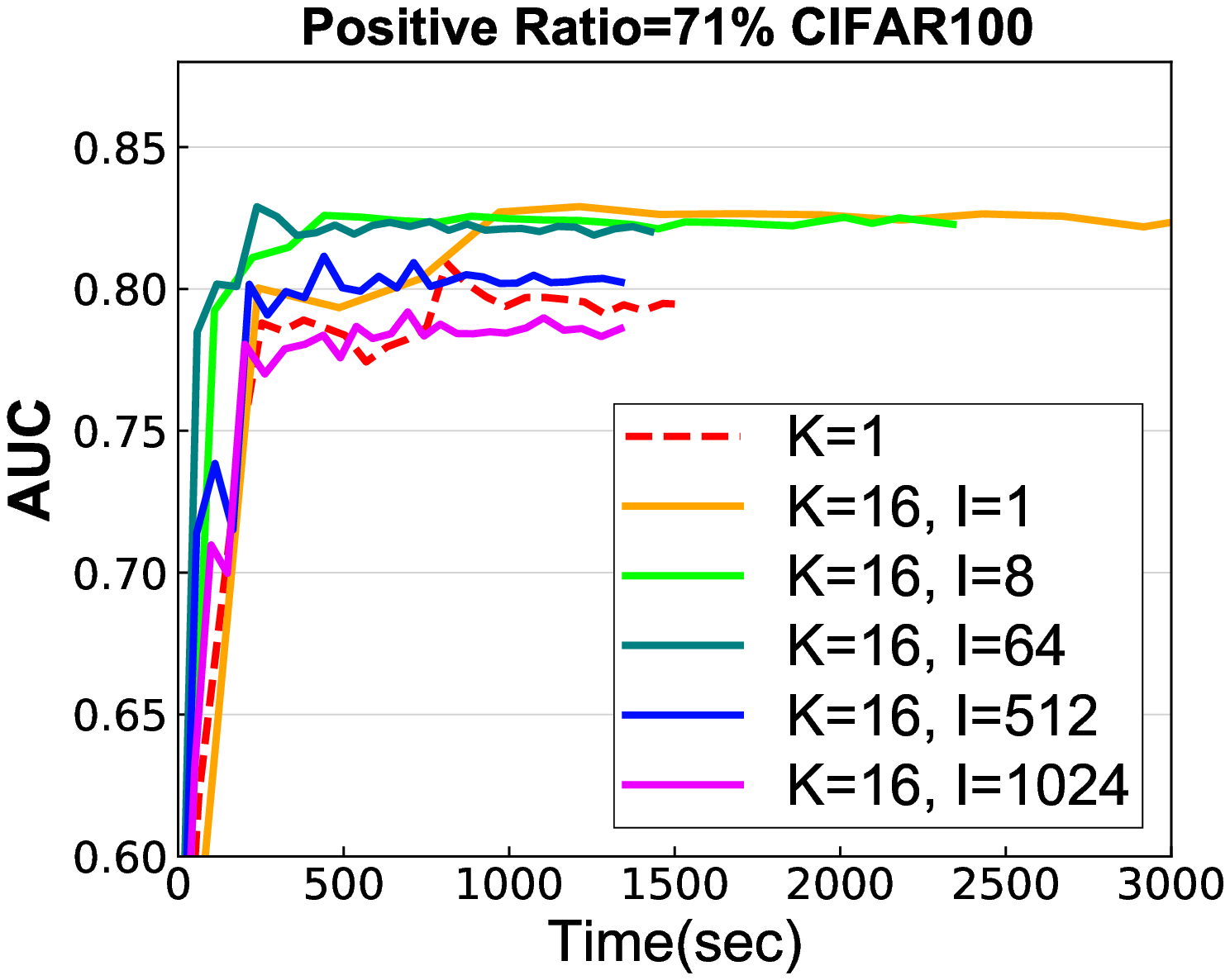}}
    \vspace{-0.4cm}
    \caption{Cifar100, positive ratio = 71\%.}
    \label{fig:cifar100_71}
  \vspace{-0.1cm}
    \subfigure[Fix $I$, vary $K$] {\includegraphics[scale=0.25]{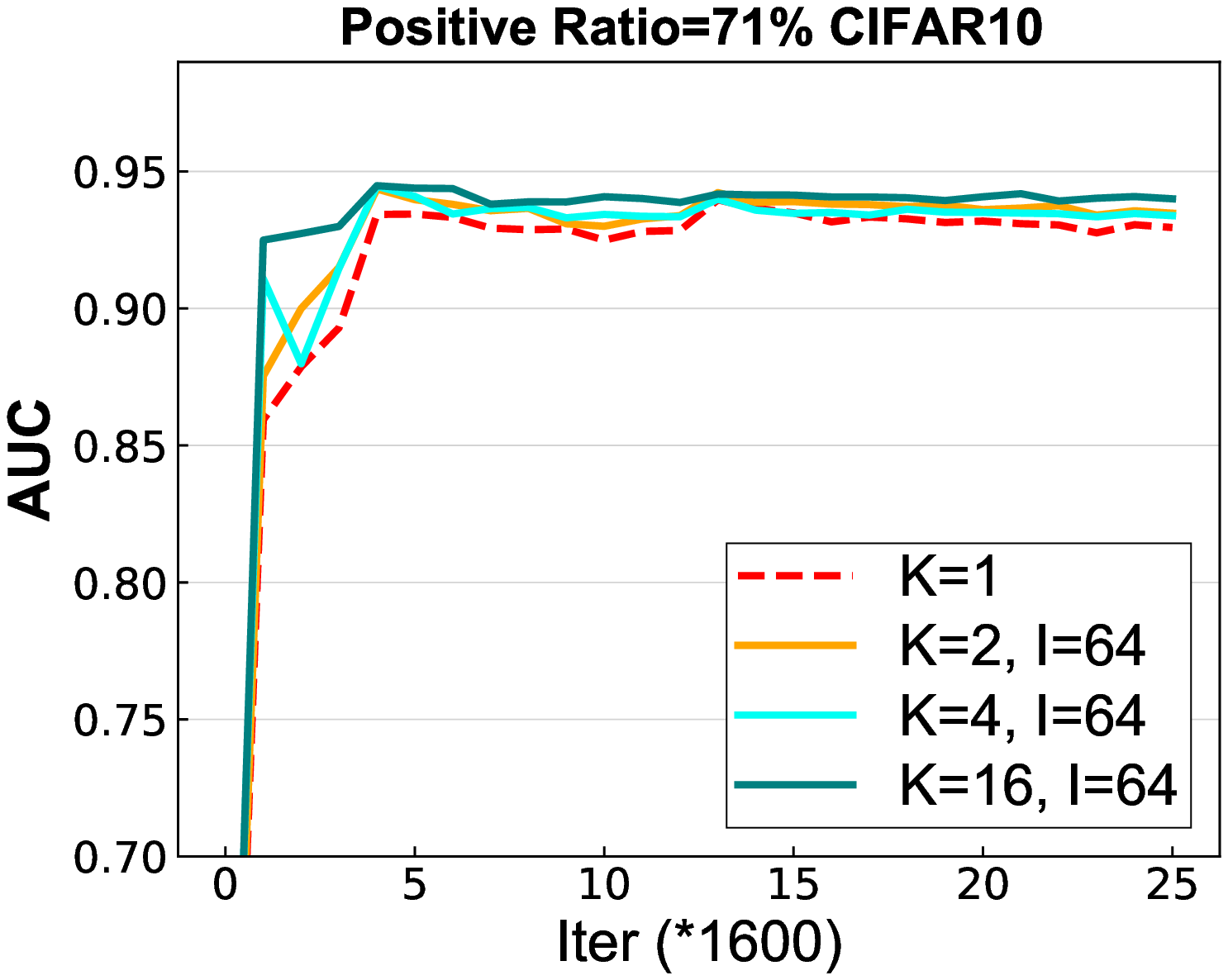}
    \includegraphics[scale=0.25]{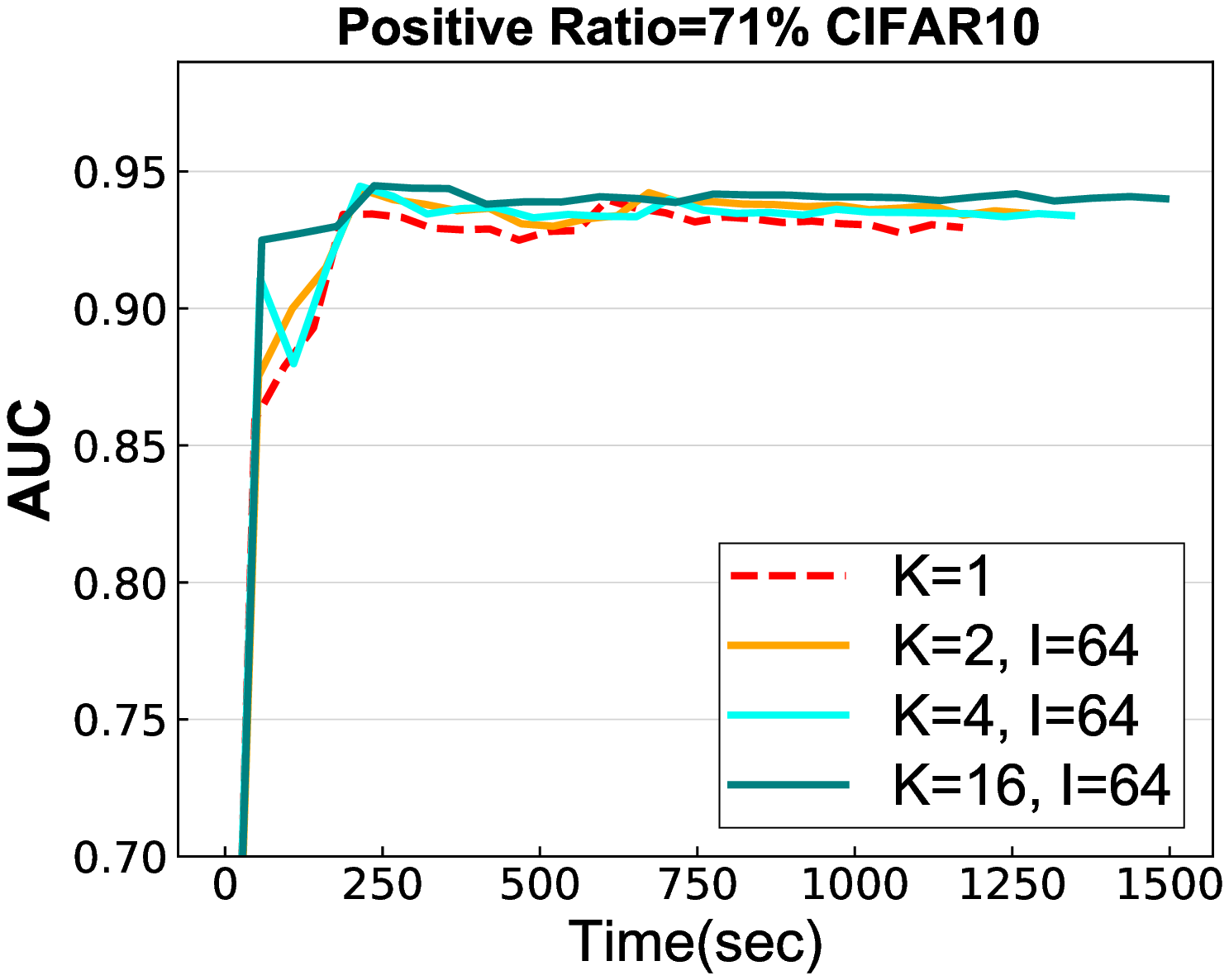}}
    \subfigure[Fix $K$, vary $I$]
    {\includegraphics[scale=0.25]{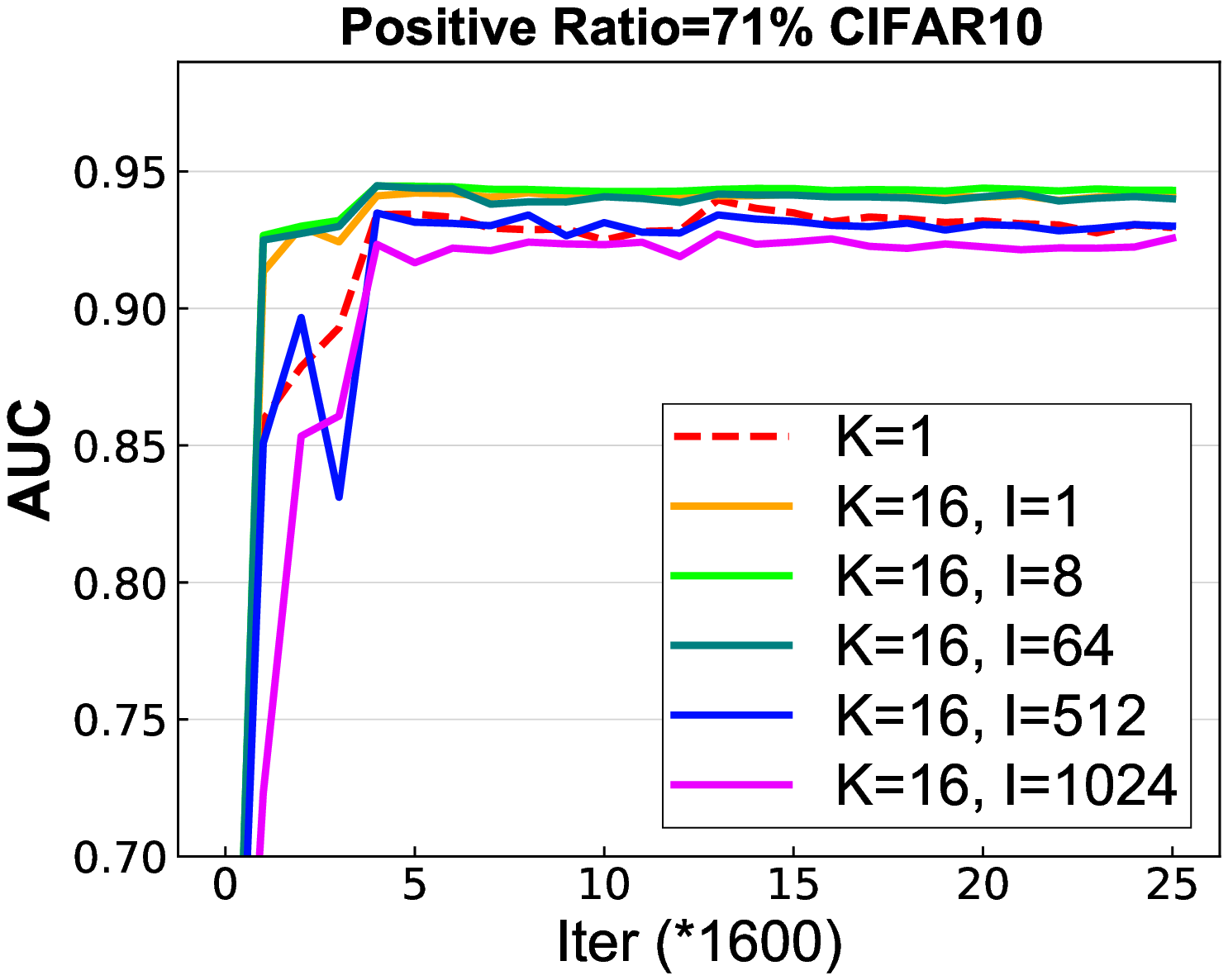}
    \includegraphics[scale=0.25]{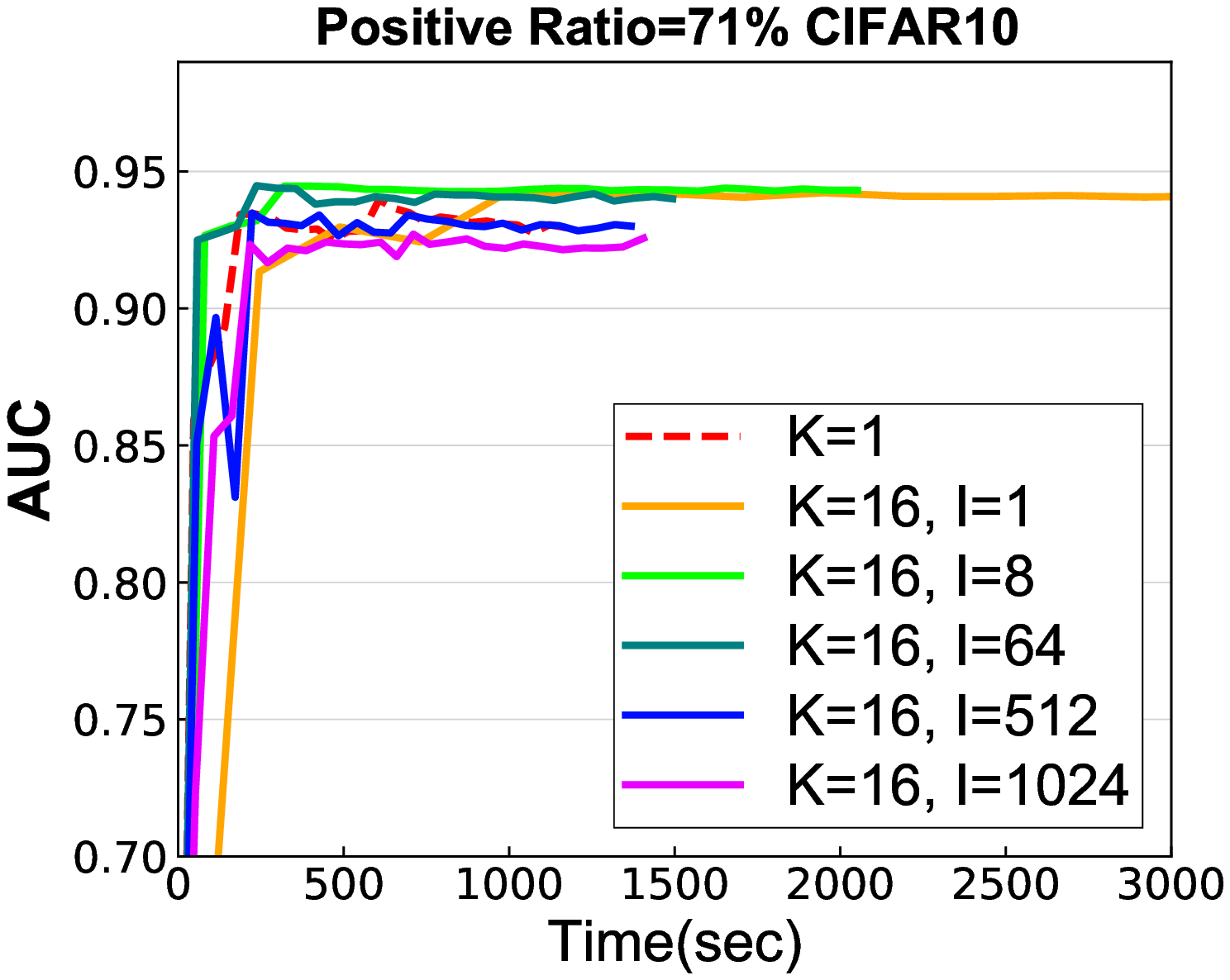}}
    \vspace{-0.4cm}
    \caption{Cifar10, positive ratio = 71\%.}
    \label{fig:cifar10_71} 
\end{figure*}

\begin{figure}[htpb]  
    \includegraphics[scale=0.25]{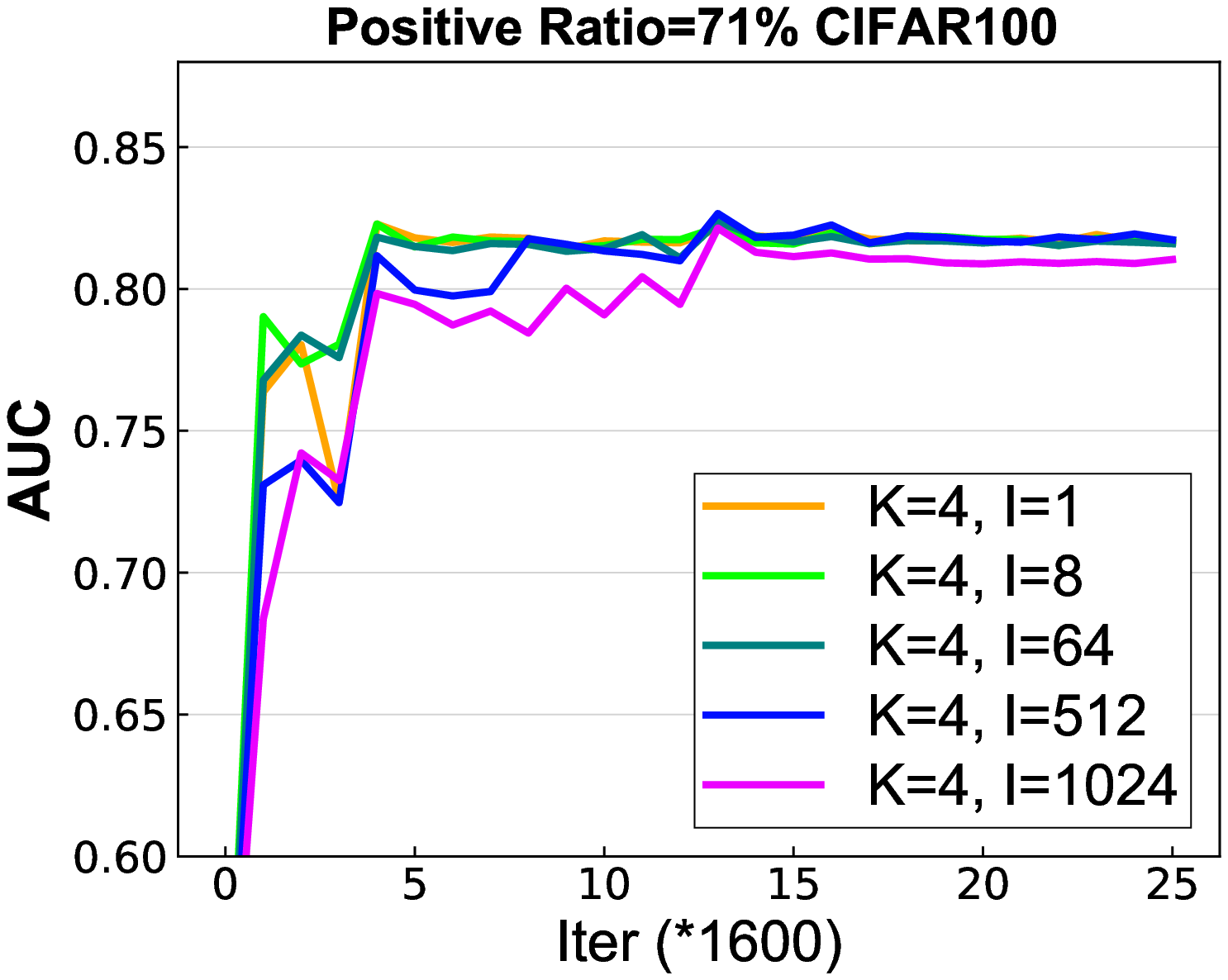}
    \includegraphics[scale=0.25]{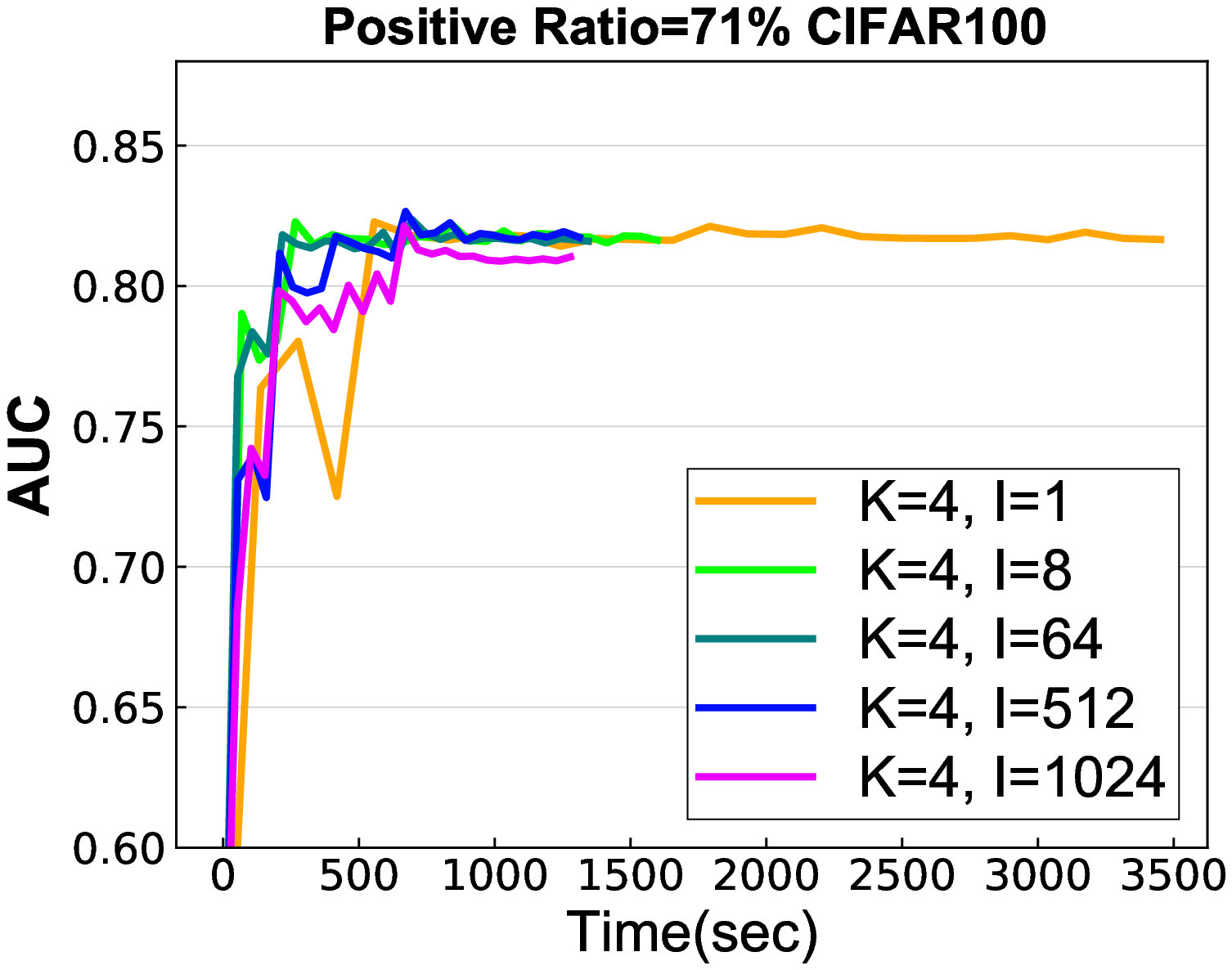}
    \vspace{-0.4cm}
    \caption{Cifar100, positive ratio = 71\%,  $K$=4.}
    \label{fig:ablation_c100_K_4_P_71} 
    \vspace*{0.1in} 
    \includegraphics[scale=0.25]{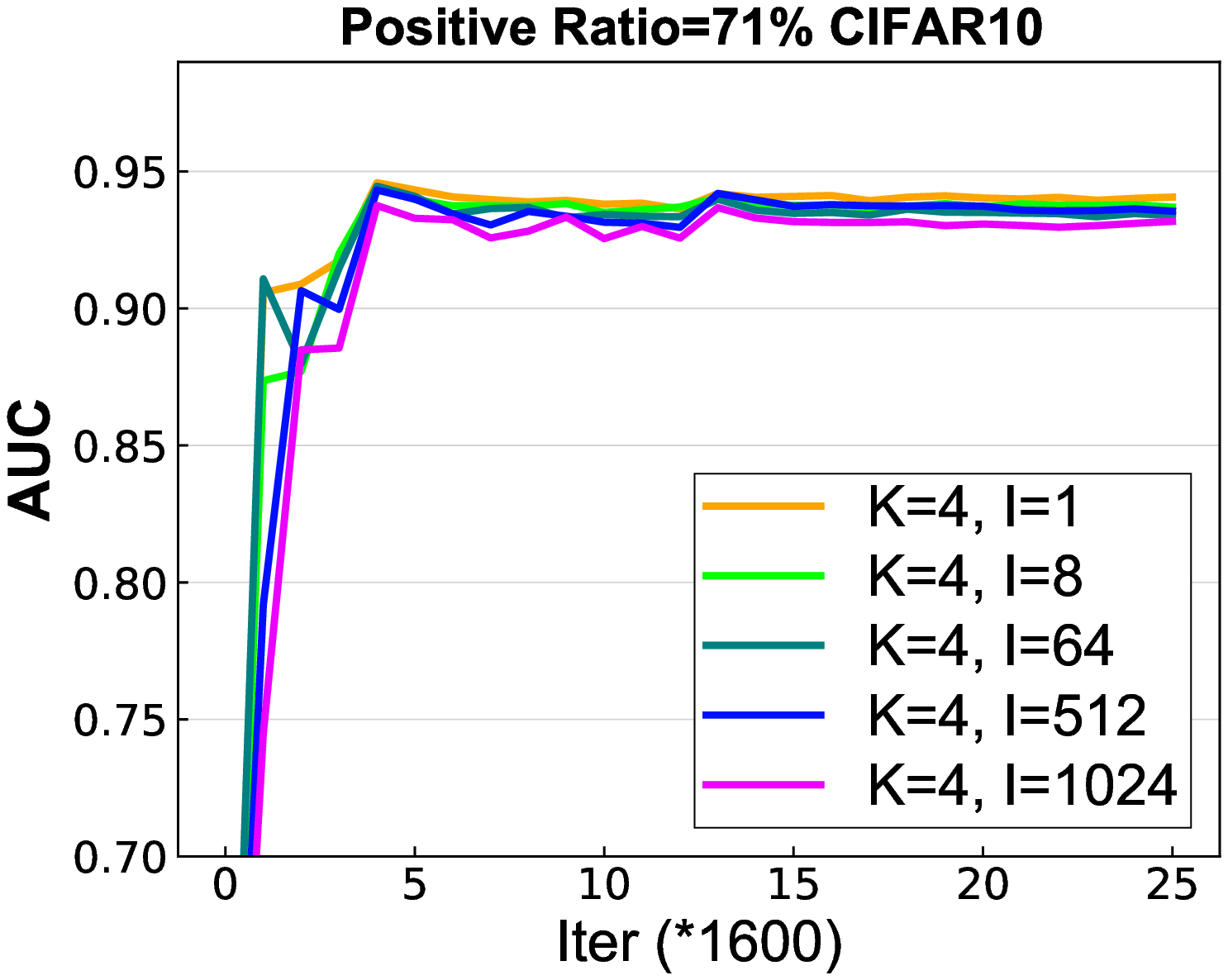} 
    \includegraphics[scale=0.25]{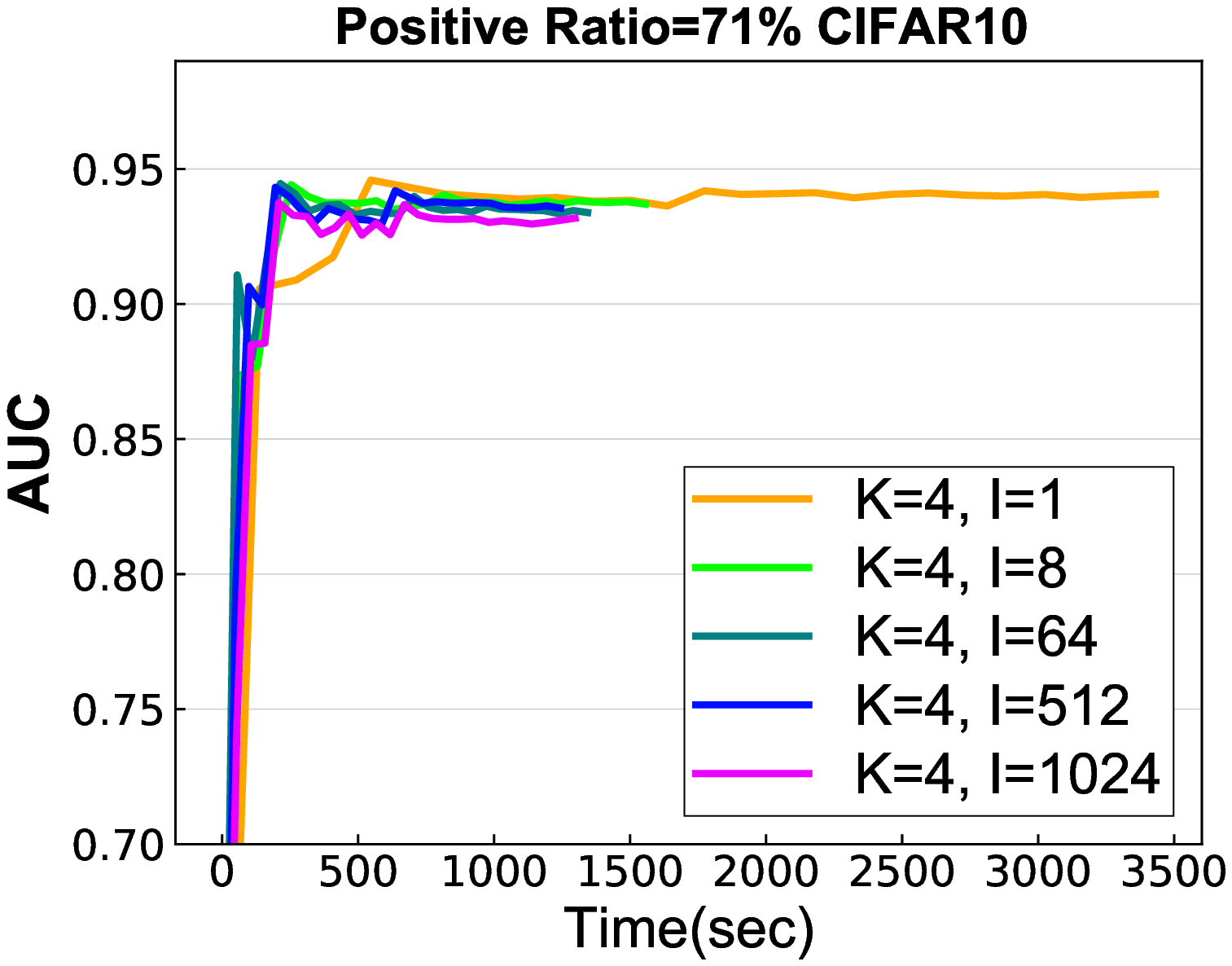} 
    \vspace{-0.4cm}  
    \caption{Cifar10, positive ratio = 71\%, $K$=4.}
    \label{fig:ablation_c10_K_4_P_71}
\end{figure}
{\bf Results.}  We plot the curve of testing AUC versus the number of iterations and versus running time. We notice that evaluating the training objective function value on all examples is very expensive, so we use the testing AUC as our evaluation metric. It may cause some gap between our results and the theory; however, the trend should be enough for our purpose to verify that our distributed algorithms can enjoy faster convergence in both the number of iterations and running time.  We have the following observations.
\vspace{-0.2cm}
\begin{itemize}
\item {\bf Varying $K$.} By varying $K$ and fixing the value of $I$, we aim to verify the parallel speedup. The results are shown in  Figure \ref{fig:imagenet_71}(a), Figure \ref{fig:cifar100_71}(a) and Figure \ref{fig:cifar10_71}(a). They show that when $K$ becomes larger, then our algorithm requires a less number of iterations to converge to the target AUC, which is consistent with the parallel speedup result as indicated by  Theorem~\ref{theorem_all_stage}. In addition,  CoDA with $K=16$ machines is also the most time-efficient algorithm among all settings. 
\vspace{-0.2cm}
\item {\bf Varying $I$.} By varying $I$ and fixing the value of $K$, we aim to verify that skipping communications  up to a certain number of iterations of CoDA does not hurt the iteration complexity but can dramatically reduce the total communication costs.  In particular, we fix $K=16$ and vary $I$ in the range $\{1, 8, 64, 512, 1024\}$. The results are shown in  Figures \ref{fig:imagenet_71}(b), Figures \ref{fig:cifar100_71}(b) and Figures \ref{fig:cifar10_71}(b). They exhibit that even when $I$ becomes moderately large, our algorithm is still able to deliver comparable performance in terms of the number of iterations compared with the case when $I=1$. The largest value of $I$ that does not cause a dramatic performance drop compared with $I=1$ is $I=1024$, $I=64$, $I=64$ on ImageNet, CIFAR100 and CIFAR10, respectively. However, up to these thresholds the running time of CoDA can be dramatically reduced than the naive parallel version with $I=1$.
\vspace{-0.2cm}

\item {\bf Trade-off between $I$ and $K$.} Finally, we verify the trade-off between $I$ and $K$ as indicated in Theorem~\ref{thm:main}. To this end, we conduct experiments by fixing $K=4$ GPUs and varying the value $I$, and comparing the limits of $I$ for $K=4$ and $K=16$. The results of using $K=4$ on CIFAR100 and CIFAR10 are reported in Figure \ref{fig:ablation_c100_K_4_P_71} and Figure \ref{fig:ablation_c10_K_4_P_71}.  We can observe that when $K=4$ the upper limit of $I$ that does not cause a dramatic performance drop compared with $I=1$ is $I=512$ for the two datasets, which is larger than the upper limit of $I=64$ for $K=16$. This is consistent with  our Theorem~\ref{theorem_all_stage}. 
\end{itemize}
\vspace{-0.5cm}
\section{Conclusion}
\vspace{-0.2cm}
In this paper, we have designed a communication-efficient distributed stochastic deep AUC maximization algorithm, in which each machine is able to do multiple iterations of local updates before communicating with the central node. We have proven the linear speedup property and shown that the communication complexity can be dramatically reduced for multiple machines up to a large threshold number. Our empirical studies verify the theory and also demonstrate the effectiveness of the proposed distributed algorithm on benchmark datasets.  

\section*{Acknowledgements}
This work is partially supported by National Science Foundation CAREER Award 1844403 and National Science Foundation Award 1933212.

\bibliography{paper}
\bibliographystyle{icml2020}

\onecolumn
\appendix

\section{Proof of Theorem \ref{theorem_all_stage}}
\textit{Proof. }
Define $\phi_s(\v) = \phi(\v) + \frac{1}{2\gamma}\|\v-\v_{s-1}\|^2$.
We can see that $\phi_s(\v)$ is convex and smooth since $\gamma \leq 1/L_{\v}$.
The smooth coefficient of $\phi_s$ is $\hat{L}_{\v} = L_{\v} + 1/\gamma$.
According to Theorem 2.1.5 of \cite{DBLP:books/sp/Nesterov04}, we have
\begin{equation}
\begin{split}
\|\nabla \phi_s(\v_s)\|^2 \leq 2\hat{L}_{\v} (\phi_{s}(\v_s) - \phi_{s}(\v_{\phi_s}^*)).
\end{split}
\label{by_nestrov}
\end{equation}
\vspace{-0.1in} 
Applying Lemma \ref{lemma_one_stage}, we have
\begin{equation*}
\begin{split}
E_{s-1}[\phi_s (\v_s) - \phi_s (\v_{\phi_s}^*)]
\leq \frac{2}{\eta_s T_s} \|\v_{s-1} - \v_{\phi_s}^*\|^2 + \frac{1}{\eta_s T_s} (\alpha_{s-1} -\alpha^*(\v_s))^2 + H \eta_s^2 I_s^2 B^2 \I_{I_s>1} + \frac{\eta_s (2\sigma_{\v}^2 + 3\sigma_{\alpha}^2)}{2K}. 
\end{split}
\end{equation*}
\vspace{-0.1in} 
Denote $\x_{1:m_s}^k = (\x_1^k, ..., \x_{m_s}^k)$, $y_{1:m_s}^k = (y_1^k, ..., y_{m_s}^k)$, and $\Tilde{f}_k (\x_{1:m_s}^k, y^k_{1:m_s}) =   \frac{\sum\limits_{i=1}^{m_s}h(\w_s;\x_i^k) \I_{y_i^k=y}  }{\sum\limits_{i=1}^{m_s}\I_{y^k_i=y}} - E_{\x^k}[h(\w_{s};\x^k)|y]$. 
If $\sum_{i=1}^{m_s} \I_{y_i^k=y}>0$, then  $\frac{\sum\limits_{i=1}^{m_s}h(\w_s;\x_i^k) \I_{y_i^k=y}   }{\sum\limits_{i=1}^{m_s}\I_{y^k_i=y}}$ is an unbiased estimation of $E_{\x^k}[h(\w_{s};\x^k)|y]$. 
Noting $0\leq h(\w; \x) \leq 1$, we have $Var(h(\w; \x^k)|y) \leq \tilde{\sigma}^2 \leq 1$. 
Then we know that
\begin{equation}
\begin{split} 
E_{\x_{1:m_s}^k}[(\Tilde{f}(\x_{1:m_s}^k, y_{1:m_s}^k))^2 | y_{1:m_s}^k]
&\leq \frac{\tilde{\sigma}^2}{\sum_{i=1}^{m_s} \I_{y_i^k=y}} \I_{(\sum_{i=1}^{m_s} \I_{y_i^k=y}>0)} +  1\cdot \I_{(\sum_{i=1}^{m_s} \I_{y_i^k=y}=0)}  \\
&\leq \frac{\I_{(\sum_{i=1}^{m_s} \I_{y_i^k=y}>0)}}{\sum_{i=1}^{m_s} \I_{y_i^k=y}} + \I_{(\sum_{i=1}^{m_s} \I_{y_i^k=y}=0)} .
\end{split} 
\end{equation} 
Hence, 
\begin{small}
\begin{equation} 
\begin{split} 
& E_{s-1}[ \Tilde{f}_k (\x_{1:m_s}^k, y_{1:m_s}^k) ]
= E_{y_{1:m_s}^k} \left[ E_{\x_{1:m_s}^k} [ (\Tilde{f}_k (\x_{1:m_s}^k, y_{1:m_s}^k))^2 | y_{1:m_s}^k]  \right] \\
&\leq E_{y_{1:m_s}^k} \left[\frac{\I_{(\sum_{i=1}^{m_s} \I_{y_i^k=y} > 0)} }{\sum_{i=1}^{m_s} \I_{y_i^k=y}} + \I_{\sum_{i=1}^{m_s} \I_{y_i^k=y} = 0} \right] \leq \frac{1}{m_s \text{Pr}(y_i^k = y)} + (1-\text{Pr}(y_i^k=y))^{m_s}.
\end{split}  
\end{equation} 
\end{small}

Denote 
\begin{small}
\begin{equation}
\begin{split}
\alpha^*(\v_s) &= \arg\max\limits_{\alpha} f(\v_s,\alpha) 
= \frac{1}{K}\sum\limits_{k=1}^{K}  E\left[ \frac{h(\w_s; \x^k)\I_{y^k=-1}}{1-p} -  \frac{h(\w_s; \x^k)\I_{y^k=1}}{p} \right]\\  
& =  \frac{1}{K} \sum\limits_{k=1}^{K}  
\left[E\left[h(\w_s; \x^k)|y^k=-1\right] - E\left[h(\w_s; \x^k)|y^k=1\right] \right].  
\end{split} 
\end{equation} 
\end{small}
Therefore,
\begin{small}
\begin{equation}
\begin{split}
E_{s-1} [(\alpha_{s-1}-\alpha^*(\v_{s-1}))^2] &= 
E_{s-1} \left[\frac{1}{K}   \sum\limits_{k=1}^{K}\frac{\sum\limits_{i=1}^{m_{s-1}}  h(\w_{s-1}; \x_i^k)  \I_{y_i^{k}=-1}}{\sum\limits_{i=1}^{m_{s-1}} \I_{y_i^{k}=-1}} - E\left[\frac{1}{K} \sum\limits_{k=1}^{K} h(\w_{s-1}; \x_i^k)|y=-1\right] \right.\\ 
&~~~~~~~~~~~~~~~~~  
\left. + E\left[\frac{1}{K}\sum\limits_{k=1}^{K} h(\w_{s-1}; \x_i^k)|y=1\right] 
-  \frac{1}{K}\sum\limits_{k=1}^K \frac{  \sum\limits_{i=1}^{m_{s-1}}h(\w_{s-1};  \x_i^k)\I_{y_i^k=1}}{\sum\limits_{i=1}^{m_{s-1}} \I_{y_i^k=1}} \right]^2 \\
& \leq \frac{2}{K m_{s-1} \text{Pr}(y_i^k=-1)} 
+ \frac{2(1-\text{Pr}(y_i^k=-1))^{m_{s-1}}}{K} + (1-\text{Pr}(y_i^k=-1))^{2m_{s-1}} \\
&~~~ + \frac{2}{K m_{s-1} \text{Pr}(y_i^k=1)}  
+ \frac{2(1-\text{Pr}(y_i=1))^{m_{s-1}} }{K} 
+ (1- \text{Pr}(y_i^k=1))^{2m_{s-1}}\\  
& \leq \frac{2}{K m_{s-1} p(1-p)} + \frac{3 p^{m_{s-1}}}{K} +  \frac{3(1-p)^{m_{s-1}}}{K}  
\leq 2\left(\frac{1}{K m_{s-1} p(1-p)} + \frac{3\Tilde{p}^{m_{s-1}}}{K} \right) \\ 
&\leq 2 \left(\frac{1}{K m_{s-1} p (1-p)} + \frac{C}{K m_{s-1}}\right) \leq \frac{2(1 + C)}{K m_{s-1} p(1-p)}, 
\end{split}    
\end{equation}  
\end{small} 
where $C = \frac{3\tilde{p}^{\frac{1}{\ln(1/\Tilde{p})}}}{2\ln(1/\Tilde{p})}$ and $\Tilde{p} = \max(p, 1-p)$.


Since $h(\w; \x)$ is $G_h$-Lipschitz, $E[h(\w, \x)|y=-1] - E[h(\w, \x)|y=1]$ is $2G_h$-Lipschitz.
It follows that
\begin{equation}
\begin{split}
&E_{s-1}[(\alpha_{s-1} - \alpha^*(\v_s))^2] = E_{s-1}[(\alpha_{s-1}-\alpha^*(\v_{s-1}) + \alpha^*(\v_{s-1}) -  \alpha^*(\v_{s}))^2] \\
&\leq E_{s-1}[2(\alpha_{s-1} - \alpha^*(\v_{s-1}))^2 + 2(\alpha^*(\v_{s-1}) - \alpha^*(\v_{s}))^2] \\
& = E_{s-1}[2(\alpha_{s-1} - \alpha^*(\v_s))^2] \\
&+ 2\left\|\frac{1}{K}\sum\limits_{k=1}^{K}  \bigg[E_{s-1}[h(\w_{s-1}; \x^k)|y^k=-1] -  E_{s-1}[h(\w_{s-1}; \x^k)|y^k=1]] - [E_{s-1}[h(\w_s; \x)|y^k=-1] - E_{s-1}[h(\w_s; \x^k)|y^k=1]\bigg] \right\|^2 \\ 
&\leq \frac{2(1+C)}{m_{s-1} K 4p^2(1-p)^2} + 8G_{h}^2 E_{s-1}[\|\v_{s-1} - \v_s \|^2]. 
\end{split} 
\end{equation}

Since $m_{s-1} \geq \frac{1+C}{\eta_s^2 T_s\sigma_{\alpha}^2 p^2(1-p)^2}$, then we have 
\begin{equation}
\begin{split}
E[\phi_s (\v_s) - \phi_s (\v_{\phi_s}^*)] &\leq \frac{2\|\v_{s-1} - \v_{\phi_s}^*\|^2 + 8 G_{h}^2E[\|\v_{s-1} -\v_{s} \|]}{\eta_s T_s} +  \frac{\eta_s\sigma_{\alpha}^2}{2 K}  + H \eta_s^2 I_s^2 B^2 \I_{I_s>1} + \frac{\eta_s (2\sigma_{\v}^2 + 3\sigma_{\alpha}^2)}{2 K} \\
&\leq \frac{2\|\v_{s-1} - \v_{\phi_s}^*\|^2 + 8 G_{h}^2 E[\|\v_{s-1} - \v_s\|^2]}{\eta_s T_s} + H \eta_s^2 I_s^2 B^2 \I_{I_s>1} + \frac{2\eta_s (\sigma_{\v}^2 + \sigma_{\alpha}^2)}{K}. \\
\end{split}  
\label{local_lemma_ved} 
\end{equation}  



We define $I'_s = 1/\sqrt{K\eta_s} = \frac{1}{K\sqrt{\eta_0}}\exp(\frac{c(s-1)}{2})$.
Applying this and (\ref{local_lemma_ved}) to (\ref{by_nestrov}), we get
\begin{equation}
\begin{split}
E[\|\nabla \phi_s (\v_{s})\|^2] &\leq 2\hat{L}_{\v} \bigg[ \frac{2\|\v_{s-1} - \v_{\phi_s}^*\|^2 + 8 G_{h}^2E[\|\v_{s-1} - \v_s\|^2]}{\eta_s T_s} + H\eta_s^2 {I'}_s^2 B^2 +  \frac{2\eta_s(\sigma_{\v}^2 + \sigma_{\alpha}^2)}{K}\bigg] \\
&\leq 2\hat{L}_{\v} \bigg[ \frac{2\|\v_{s-1} - \v_{\phi_s}^*\|^2 + 8 G_{h}^2E[\|\v_{s-1} - \v_s\|^2]}{\eta_s T_s} + H\eta_s^2 {I'}^2_s B^2 + \frac{2\eta_s(\sigma_{\v}^2 +  \sigma_{\alpha}^2)}{K} \bigg].
\end{split}
\label{random9}
\end{equation}
Taking $\gamma = \frac{1}{2L_{\v}}$, then $\hat{L}_{\v} = 3L_{\v}$.
Note that $\phi_{s}(\v)$ is $(\gamma^{-1} - L_{\v})$-strongly convex, we have
\begin{equation}
\begin{split}
\phi_{s}(\v_{s-1}) \geq \phi_s(\v_{\phi_s}^*) + \frac{L_{\v}}{2} \|\v_{s-1} - \v_{\phi_s}^*\|^2.
\end{split}
\label{random:strong_v}
\end{equation}

Plugging (\ref{random:strong_v}) into (\ref{local_lemma_ved}), we get
\begin{equation}
\begin{split}
&E_{s-1}[\phi(\v_s) + L_{\v}\|\v_s - \v_{s-1}\|^2]\\
&\leq \phi_s(\v_{\phi_s}^*) + \frac{2\|\v_{s-1} - \v_{\phi_s}^*\|^2 + 8 G_{h}^2E_{s-1}[\|\v_{s-1} - \v_s\|^2]}{\eta_s T_s} + H \eta_s^2 {I'}_s^2 B^2 + \frac{2\eta_s(\sigma_{\v}^2 + \sigma_{\alpha}^2)}{K}\\
&\leq \phi_s(\v_{s-1}) - \frac{L_{\v}}{2}\|\v_{s-1}-\v_{\phi_s}^*\|^2 +\\
&~~~~~~~~~~~ \frac{2\|\v_{s-1} - \v_{\phi_s}^*\|^2 + 8 G_{h}^2E_{s-1}[\|\v_{s-1} - \v_s\|^2]}{\eta_s T_s} + H\eta_s^2 {I'}_s^2 B^2 + \frac{2\eta_s(\sigma_{\v}^2 + \sigma_{\alpha}^2)}{K} .\\
\end{split}
\end{equation}

Noting $\eta_s T_s L_{\v} = \max(8, 8 G_{h}^2)$ and $\phi_{s}(\v_{s-1}) = \phi(\v_{s-1})$, we rearrange terms and get
\begin{equation}
\begin{split}
\frac{2\|\v_{s-1} - \v_{\phi_s}^*\|^2 + 8 G_{h}^2E_{s-1}[\|\v_{s-1} - \v_s\|^2]}{\eta_s T_s} \leq \phi(\v_{s-1}) - E_{s-1}[\phi(\v_{s})] + H\eta_s^2 {I'}_s^2 B^2 + \frac{2\eta_s(\sigma_{\v}^2 + \sigma_{\alpha}^2)}{K}.
\end{split}
\label{random_11}
\end{equation}

Combining (\ref{random9}) and (\ref{random_11}), we get
\begin{equation}
\begin{split}
E_{s-1}\|\nabla \phi_s(\v_s)\|^2 \leq 2\hat{L}_{\v}\bigg[ \phi(\v_{s-1}) - E_{s-1}[\phi(\v_s)] + 2H\eta_s^2 {I'}_s^2 B^2 + \frac{4\eta_s(\sigma_{\v}^2 + \sigma_{\alpha}^2)}{K} \bigg]\\
=6L_{\v}\bigg[ \phi(\v_{s-1}) -  E_{s-1}[\phi(\v_s)] + 2H\eta_s^2 {I'}_s^2 B^2 + \frac{4\eta_s(\sigma_{\v}^2 +  \sigma_{\alpha}^2)}{K} \bigg].\\
\end{split}
\end{equation}

Taking expectation on both sides over all randomness until $\v_{s-1}$ is generated and by tower property, we have
\begin{equation}
\begin{split}
E\|\nabla \phi_s(\v_s)\|^2 \leq 6L_{\v}\bigg( E[\phi(\v_{s-1}) - \phi(\v_{\phi}^*)] - E[\phi(\v_{s}) - \phi(\v_{\phi}^*)] + 2H\eta_s^2 {I'}_s^2 B^2 + \frac{4\eta_s(\sigma_{\v}^2 + \sigma_{\alpha}^2)}{K} \bigg)
\end{split}
\label{random13}
\end{equation}

Since $\phi(\v)$ is $L_{\v}$-smooth and hence is $L_{\v}$-weakly convex, we have
\begin{equation}
\begin{split}
&\phi(\v_{s-1}) \geq \phi(\v_{s}) + \langle \nabla\phi(\v_s), \v_{s-1}-\v_s \rangle - \frac{L_{\v}}{2}\|\v_{s-1}-\v_{s}\|^2\\
&=\phi(\v_s) + \langle \nabla \phi(\v_s) + 2L_{\v}(\v_s - \v_{s-1}), \v_{s-1} - \v_s\rangle + \frac{3}{2}L_{\v}\|\v_{s-1} - \v_{s}\|^2\\
& = \phi(\v_s) + \langle \nabla \phi_s(\v_s), \v_{s-1}-\v_{s}\rangle +\frac{3}{2}L_{\v}\|\v_{s-1}-\v_s\|^2\\
& = \phi(\v_s) - \frac{1}{2L_{\v}}\langle \nabla\phi_s(\v_s),\nabla\phi_s(\v_s) -\nabla\phi(\v_s)\rangle + \frac{3}{8L_{\v}}\|\nabla \phi_s(\v_s) - \nabla \phi(\v_s)\|^2 \\
& = \phi(\v_s) - \frac{1}{8L_{\v}}\|\nabla \phi_s(\v_s\|^2 - \frac{1}{4L_{\v}}\langle \nabla\phi_s(\v_s), \nabla\phi(\v_s)\rangle + \frac{3}{8L_{\v}}\|\nabla \phi (\v_s)\|^2
\end{split}
\end{equation}
Rearranging terms, it yields
\begin{equation}
\begin{split}
&\phi(\v_s) - \phi(\v_{s-1}) \leq \frac{1}{8L_{\v}}\|\nabla \phi_s(\v_s)\|^2 + \frac{1}{4L_{\v}}\langle \nabla\phi_s(\v_s), \nabla\phi(\v_s)\rangle - \frac{3}{8L_{\v}}\|\nabla \phi (\V_s)\|^2\\
&\leq \frac{1}{8L_{\v}}\|\nabla \phi_s(\v_s)\|^2 + \frac{1}{8L_{\v}} (\|\nabla \phi_s(\v_s)\|^2 + \|\nabla\phi(\v_s)\|^2) - \frac{3}{8L_{\v}}\|\nabla \phi (\V_s)\|^2\\
&=\frac{1}{4L_{\v}}\|\nabla \phi_s(\v_s)\|^2 - \frac{1}{4L_{\v}}\|\nabla \phi(\v_s)\|^2\\
&\leq \frac{1}{4L_{\v}}\|\nabla \phi_s(\v_s)\|^2 - \frac{\mu}{2L_{\v}}(\phi(\v_s) - \phi(\v_{\phi}^*))
\end{split}
\label{random15}
\end{equation}

Define $\Delta_s = \phi(\v_s) - \phi(\v_{\phi}^*)$.
Combining (\ref{random13}) and (\ref{random15}), we get
\begin{equation}
\begin{split}
E[\Delta_s - \Delta_{s-1}] \leq \frac{3}{2}E(\Delta_{s-1} - \Delta_s) + 3H\eta_s^2 {I'}_s^2 B^2 + \frac{6\eta_s(\sigma_{\v}^2+\sigma_{\alpha}^2)}{K} - \frac{\mu}{2L_{\v}}E[\Delta_s]
\end{split}
\end{equation}

Therefore,
\begin{equation}
\begin{split}
\bigg( \frac{5}{2} + \frac{\mu}{2L_{\v}}\bigg) E[\Delta_{s}] \leq \frac{5}{2}E[\Delta_{s-1}] + 3 H \eta_s^2 {I'}_s^2 B^2 + \frac{6\eta_s (\sigma_{\v}^2+\sigma_{\alpha}^2)}{K}
\end{split}
\end{equation}

Using $c = \frac{\mu/L_{\v}}{5+\mu/L_{\v}}$ as defined in the theorem,
\begin{equation}
\begin{split}
&E[\Delta_S] \leq \frac{5L_{\v}}{5L_{\v}+\mu} E[\Delta_{S-1}] + 
\frac{2L_{\v}}{5L_{\v}+\mu}\bigg[ 3H\eta_S^2 {I'}_S^2 B^2 + \frac{6\eta_S(\sigma_{\v}^2+\sigma_{\alpha}^2)}{K}\bigg]\\
&=(1-c)\bigg[E[\Delta_{S-1}] + \frac{2}{5}\bigg(3H\eta_S^2 {I'}_S^2 B^2 + \frac{6\eta_S(\sigma_{\v}^2+\sigma_{\alpha}^2)}{K}\bigg)\bigg]\\
&\leq (1-c)^S E[\Delta_0] + \frac{6H B^2}{5} \sum\limits_{j=1}^{S} \eta_j^2 {I'}_j^2 (1-c)^{S+1-j} +  \frac{12(\sigma_{\v}^2 + \sigma_{\alpha}^2)}{5K} \sum\limits_{j=1}^{S} \eta_j (1-c)^{S+1-j} \\
& = (1-c)^S E[\Delta_0] + \frac{6H B^2}{5} \sum\limits_{j=1}^{S} \eta_j^2 {I'}_j^2 (1-c)^{S+1-j} +  \frac{12(\sigma_{\v}^2 + \sigma_{\alpha}^2)}{5K} \sum\limits_{j=1}^{S} \eta_j (1-c)^{S+1-j}
\end{split}
\end{equation}

We then have
\begin{equation}
\begin{split}
&E[\Delta_S] \leq (1 - c )^S E[\Delta_0] + \left(\frac{6H B^2}{5K} + \frac{12(\sigma_{\v}^2 + \sigma_{\alpha}^2)}{5K}\right)\sum\limits_{j=1}^{S} \eta_j (1-c)^{S+1-j} \\
&\leq \exp(-c S)\Delta_0 +  \left(\frac{6H B^2}{5K} + \frac{12(\sigma_{\v}^2 + \sigma_{\alpha}^2)}{5K}\right) \sum\limits_{j=1}^{S}  \eta_j \exp(-c(S+1-j))\\
&=\exp(-c S)\Delta_0 +  \left(\frac{6H B^2}{5} + \frac{12(\sigma_{\v}^2 + \sigma_{\alpha}^2)}{5}\right) \eta_0 S\exp (-c S).\\
\end{split}
\end{equation}

To achieve $E[\Delta_S] \leq \epsilon$, it suffices to make
\begin{equation}
\begin{split}
\exp(-c S) \Delta_0 \leq \epsilon/2
\end{split}
\end{equation}
and 
\begin{equation}
\begin{split}
\left(\frac{6H B^2}{5} + \frac{12(\sigma_{\v}^2 + \sigma_{\alpha}^2)}{5}\right) \eta_0 S\exp (-c S)  \leq \epsilon/2.
\end{split}
\end{equation}
So, it suffices to make
\begin{equation}
\begin{split}
S\geq c^{-1} \max\bigg\{\log \left(\frac{2\Delta_0}{\epsilon}\right), \log S + \log\bigg[ \frac{2\eta_0}{\epsilon} \frac{6HB^2 + 12(\sigma_{\v}^2+\sigma_{\alpha}^2)}{5}\bigg]\bigg\}.
\end{split}
\end{equation}

Taking summation of iteration over $s=1, ..., S$, we have the total iteration complexity as
\begin{equation}
\begin{split}
&T = 
\sum\limits_{s=1}^{S} T_s \leq \frac{\max\{8, 8G_{h}^2\}}{L_{\v}\eta_0 K} \frac{\exp(cS) - 1}{\exp(c) - 1} \leq \frac{\max\{8, 8G_{h}^2\}}{L_{\v}\eta_0 K}\frac{5L_{\v} + \mu}{\mu} \exp(cS)  \\
&=\tilde{O}\left(\max\bigg(\frac{\Delta_0}{\mu \epsilon \eta_0 K }, \frac{S (6H B^2 + 12(\sigma_{\v}^2 + \sigma_{\alpha}^2))}{\mu \epsilon K} \bigg)\right)
=\tilde{O}\bigg(\max\left(\frac{\Delta_0}{\mu \epsilon \eta_0 K}, \frac{L_{\v
}}{\mu^2 K\epsilon}\right)\bigg).
\end{split}
\end{equation}

To analyze the total communication complexity, we will analyze two cases: (1) $\frac{1}{K\sqrt{\eta_0}} > 1$ ; (2) $\frac{1}{K\sqrt{\eta_0}} \leq 1$.

(1) If $\frac{1}{K\sqrt{\eta_0}} > 1$, then $I_s = \max(1, \frac{1}{K\sqrt{\eta_0}} \exp(\frac{c(s-1)}{2})) = \frac{1}{K\sqrt{\eta_0}} \exp(\frac{c(s-1)}{2}) $ for any $s \geq 1$.

The total number of communications:
\begin{equation}
\begin{split}
&\sum\limits_{s=1}^{S} \frac{T_s}{I_s} = \sum\limits_{s=1}^{S} \frac{\max(8, 8 G_h^2)}{L_{\v} {\eta^{1/2}_0}} \exp\bigg(\frac{c(s-1)}{2}\bigg) = \frac{\max(8, 8 G_h^2)}{L_{\v}{\eta^{1/2}_0}} \frac{\exp(c S/2) -  1}{\exp(c/2) - 1}\\
&=\tilde{O}\bigg(\max\bigg(\frac{(2\Delta_0/\epsilon)^{1/2}}{\mu {\eta^{1/2}_0}},\frac{(S(6H B^2+12(\sigma_{\v}^2+\sigma_{\alpha}^2))^{1/2}}{ \mu \epsilon^{1/2}}\bigg)\bigg)
=\tilde{O} \left(\frac{\Delta_0^{1/2}}{\mu (\eta_0 \epsilon)^{1/2}}, \frac{L_{\v}^{1/2}}{\mu^{3/2}\epsilon^{1/2}}\right).
\end{split}
\end{equation}

(2) If $\frac{1}{K\sqrt{\eta_0}} \leq 1$, then
$I_s = 1$ for $s \leq \left\lceil {2c^{-1}} \log(K\sqrt{\eta_0}) + 1 \right\rceil := S_1$
and $I_s = \frac{1}{K\sqrt{\eta_0}}\exp\left(\frac{s-1}{2}\right)$ for $s > \frac{2(5+\mu/L_{\v})}{\mu/L_{\v}} \log(K\sqrt{\eta_0}) + 1$.

Obviously, $S_1 \leq \frac{2(5+\mu/L_{\v})}{\mu/L_{\v}} \log(K\sqrt{\eta_0}) + 2$.
The number of iterations from $s=1$ to $S_1$ is
\begin{equation}
\begin{split}
& \sum\limits_{s=1}^{S_1} T_s = \sum\limits_{s=1}^{S_1} \frac{\max\{8, 8 G_h^2\}}{\eta_0 L_{\v} K} \exp(c(s-1))\\
& = \frac{\max\{8, 8 G_h^2\}}{\eta_0 L_{\v} K}  \frac{\exp(c S_1)-1}{\exp(c) - 1}\\
& \leq c^{-1} \frac{\max\{8, 8 G_h^2\}}{\eta_0 L_{\v} K} \exp\left( 2\log(K\sqrt{\eta_0}) +  2c \right) \\
& = c^{-1} \frac{\max\{8, 8 G_h^2\}}{\eta_0 L_{\v} K}  K^2 \eta_0 \exp\left(\frac{2\mu/L_{\v}}{5+\mu/L_{\v}}\right) \\
& \leq c^{-1} \max\{8, 8 G_h^2\} K  \exp\left(2\right). \\
\end{split} 
\end{equation} 

Thus, the total number of communications is 
\begin{equation}
\begin{split}
&\sum\limits_{s=1}^{S_1} T_s + \sum\limits_{s=S_1+1}^{S} \frac{T_s}{I_s}\\
& = c^{-1} \max\{8, 8G_h^2\} K  \exp\left(2\right) + \sum\limits_{s=S_1+1}^{S} \frac{\max(8, 8G_h^2)}{L_{\v} \eta_0^{1/2}} \exp\left( \frac{s-1}{2} \frac{\mu/L_{\v}}{5+\mu/L_{\v}} \right)\\
&\leq c^{-1} \max\{8, 8G_h^2\} K  \exp\left(2\right) + \sum\limits_{s=1}^{S} \frac{\max(8, 8G_h^2)}{L_{\v} \eta_0^{1/2}} \exp\left( \frac{s-1}{2} \frac{\mu/L_{\v}}{5+\mu/L_{\v}} \right)\\
& \leq c^{-1} \max\{8, 8G_h^2\} K  \exp\left(2\right) + \frac{\max(8, 8G_h^2)}{L_{\v} \eta_0^{1/2}} \frac{\exp(\frac{S}{2}\frac{\mu/L_{\v}}{5+\mu/L_{\v}}) - 1} {\exp(\frac{\mu/L_{\v}}{2(5+\mu/L_{\v}})) - 1}\\
& \in O\left(\max \left(\frac{K}{\mu} + \frac{\Delta_0}{\mu \eta_0^{1/2}\epsilon^{1/2}}, \frac{K}{\mu} +  \frac{L_{\v}^{1/2}}{\mu^{3/2} \epsilon^{1/2}}\right)\right).
\end{split} 
\end{equation}


\section{Proof of Lemma \ref{properties}}
To prove Lemma \ref{properties}, we need the following Lemma \ref{lem:alpha_bounded} and Lemma \ref{lem:ab_bounded} to show that the trajectories of $\alpha$, $a$ and $b$ are constrained in closed sets in Algorithm \ref{inner_loop}.

\begin{lemma}
\label{lem:alpha_bounded}
Suppose Assumption (\ref{assumption_1}) holds and $\eta \leq \frac{1}{2p(1-p)}$. Running Algorithm \ref{inner_loop} with the input given by Algorithm \ref{outer_loop}, we have $|\alpha_t^k| \leq \frac{\max\{p, (1-p)\}}{p(1-p)}$ for any iteration $t$ and any machine $k$.
\label{alpha_bounded}
\end{lemma}
\textit{Proof.} 
Firstly, we need to show that the input for any call of Algorithm (\ref{inner_loop}) satisfies $|\alpha_0| \leq \frac{\max\{p, (1-p)\}}{p(1-p)}$.
If Algorithm \ref{inner_loop} is called by Algorithm \ref{outer_loop} for the first time, we know $|\alpha_0| = 0 \leq \frac{\max\{p, (1-p)\}}{p(1-p)}$.
Otherwise, by the update of $alpha_s$ in Algorithm (\ref{outer_loop}) (lines 4-7), we know that the input for Algorithm (\ref{inner_loop}) satisfies $|\alpha_0| \leq 2 \leq \frac{\max\{p, (1-p)\}}{p(1-p)}$  since  $h(\w; \x^k) \in [0, 1]$ by Assumption \ref{assumption_1}($iv$).

Next, we will show by induction that $|\alpha_t^k| \leq \frac{\max\{p, (1-p)\}}{p(1-p)}$ for any iteration $t$ and any machine $k$ in Algorithm \ref{inner_loop}.
Obviously, $|a_0^k| \leq 2 \leq \frac{\max\{p, (1-p)\}}{p(1-p)}$ for any k.

Assume $|a_{t}^k| \leq \frac{\max\{p, (1-p)\}}{p(1-p)}$ for any $k$.

(1) If $t+1~ \text{mod}~ I \neq 0$, then we have
\begin{equation} 
\begin{split}
|\alpha_{t+1}^k| &= \bigg|\alpha_{t}^k + \eta (2(ph(\w_{t}^k; \x)\I_{[y=-1]} - (1-p)h(\w_{t}^k; \x)\I_{[y=1]}) - 2p(1-p)\alpha_{t})\bigg|\\
&\leq \bigg| (1-2\eta p(1-p))\alpha_{t}^k \bigg| + \bigg|2\eta(ph(\w_{t}^k; \x)\I_{[y=-1]} - (1-p)h(\w_{t}^k;\x)\I_{[y=1]}) \bigg|\\
&\leq (1-2\eta p(1-p)) \frac{\max\{p, (1-p)\}}{p(1-p)} + 2\eta \max \{p, (1-p)\}\\
& = (1-2\eta p(1-p) +2\eta p(1-p)) \frac{\max\{p, (1-p)\}}{p(1-p)}\\
&= \frac{\max\{p, (1-p)\}}{p(1-p)}.
\end{split}
\end{equation}

(2) If $t+1~ \text{mod}~ I = 0$, then by same analysis as above, we know that $|\alpha_{t+1}^k| \leq \frac{\max\{p, (1-p)\}}{p(1-p)}$ before being averaged across machines.
Therefore, after being averaged across machines, it still holds that $|\alpha_{t+1}^k| \leq \frac{\max\{p, (1-p)\}}{p(1-p)}$.

Therefore, $|\alpha_t^k| \leq \frac{\max\{p, (1-p)\}}{p(1-p)}$ holds for any iteration $t$ and any machine $k$ at any call of Algorithm (\ref{inner_loop}). $\Box$

\begin{lemma}
\label{lem:ab_bounded}
Suppose Assumption (\ref{assumption_1}) (1) holds and $\eta \leq \min(\frac{1}{2(1-p)}, \frac{1}{2p})$. Running Algorithm \ref{inner_loop} with the input given by Algorithm (\ref{outer_loop}), we have that $|a_t^k| \leq 1$ and $|b_t^k|\leq 1$ for any iteration $t$ and any machine $k$.
\end{lemma}
\textit{Proof.}
At the first call of Algorithm (\ref{inner_loop}), the input satisfies $|a_0|\leq 1$ and $|b_0| \leq 1$.
Thus $|a_0^k| \leq 1$ and $|b_0^k| \leq 1$ for any machine $k$.

Assume $|a_{t}^k| \leq 1$ and $|b_{t}^k| \leq 1$. Then:

(1) $t+1~\text{mod}~I\neq0$, then we have
\begin{equation}
\begin{split}
|a_t^k| &= \bigg|\frac{\gamma}{\eta + \gamma}a_{t-1}^k + \frac{\eta}{\eta+\gamma}a_0 - \frac{\eta\gamma}{\eta+\gamma}\nabla_{a} F_k(\v_{t-1}^k, \alpha_{t-1}^k, \z_{t-1}^k)\bigg| \\
&= \bigg| \frac{\gamma}{\eta + \gamma}a_{t-1}^k + \frac{\eta}{\eta+\gamma}a_0 + \frac{\eta\gamma}{\eta+\gamma} (2(1-p)(h(\w_{t-1}^k; \x^k_{t-1})-a_{t-1}^k))\I_{y^k=1} \bigg|\\
&=\bigg|\frac{\eta}{\eta+\gamma}a_0 + \frac{\gamma}{\eta+\gamma}a_{t-1}^{k}(1 - 2\eta(1-p))\I_{y^k=1} + \frac{\eta\gamma}{\eta+\gamma}2(1-p)h(\w_{t-1}^k; \x_{t-1}^k)\I_{y^k=1}\bigg|\\
&\leq \bigg|\frac{\eta}{\eta+\gamma}a_0\bigg| + \bigg|\frac{\gamma}{\eta+\gamma}a_{t-1}^{k}(1 - 2\eta(1-p))\I_{y^k=1} \bigg| + \bigg| \frac{\eta\gamma}{\eta+\gamma}2(1-p)h(\w_{t-1}^k; \x_{t-1}^k)\I_{y^k=1}\bigg| \\
&\leq \frac{\eta}{\eta+\gamma} + \frac{\gamma}{\eta+\gamma}(1-2\eta(1-p)) + \frac{\eta\gamma}{\eta+\gamma}2(1-p) \\
& = 1.
\end{split}
\end{equation}

(2) If $t+1~\text{mod}~I=0$, then by the same analysis as above, we have that $|a_{t+1}^k| \leq 1$ before being averaged across machines.
Therefore, after being averaged across machines, it still holds that $|a_{t+1}^k| \leq 1$.

Thus, we can see that $|a_t^k| \leq 1$ holds for any iteration $t$ and any machine $k$ in this call of Algorithm \ref{inner_loop}.
Therefore, the output of the stage also has $|\tilde{a}| \leq 1$.

Then we know that in the next call of Algorithm (\ref{inner_loop}), the input satisfies $|a_0| \leq 1$, by the same proof, we can see that $|a_t^k| \leq 1$ holds for any iteration $t$ and any machine $k$ in any call of Algorithm (\ref{inner_loop}).
With the same techniques, we can prove that $|b_t^k|$ holds for any iteration $t$ and any machine $k$ in any call of Algorithm (\ref{inner_loop}).
$\Box$

With the above lemmas, we are ready to prove Lemma \ref{properties} and derive the claimed constants.

By definition of $F(\v, \alpha; \z)$ and noting that $\v = (\w, a, b)$, we have
\begin{equation}
\begin{split}
&\nabla_{\v} F_k(\v, \alpha; \z) = [\nabla_{\w} F_k(\v, \alpha; \z)^T, \nabla_{a} F_k(\v, \alpha; \z), \nabla_{b} F_k(\v, \alpha; \z)]^T.
\end{split}
\label{decompose_nabla_v}
\end{equation}

Addressing each of the three terms on RHS, it follows that
\begin{equation}
\begin{split}
&\nabla_{\w} F_k(\v, \alpha; \z) = \bigg[2(1-p)(h(\w;\x^k)-a) - 2(1+\alpha)(1-p)\bigg] \nabla h(\w; \x^k)\I_{[y^k=1]}  \\
&~~~~~~~~~~~~~~~~~~~~~~~~~~~~~
+ \bigg[2p(h(\w; \x^k)-b) + 2(1+\alpha)p\bigg] \nabla h(\w; \x^k)\I_{[y^k=-1]}, \\
&\nabla_{a} F_k(\v, \alpha; \z) = -2(1-p)(h(\w;x^k)-a)\I_{[y^k=1]},\\
&\nabla_{b} F_k(\v, \alpha; \z) = -2p(h(\w; \x^k)-b).
\end{split}
\label{sub_nabla_v}
\end{equation}

Since $|h(\w; \x^k)| \in [0, 1]$, $\|\nabla h(\w; \x^k)\| \leq G_h$, $|\alpha| \leq \frac{\max\{p, 1-p\}}{p(1-p)}$, $|a| \leq 1$ and $b\leq 1$, we have
\begin{equation}
\begin{split}
\|\nabla_{\w} F_k(\v, \alpha; \z) \| &\leq \|2(1-p)(h(\w; \x^k)-a) - 2(1+\alpha)(1-p)\|G_h + \|2p(h(\w; \x^k)-b) + 2(1+\alpha)p\|G_h\\
&\leq |6+2\alpha|(1-p)G_h + |6+2\alpha| p G_h\\
&\leq \left(6 + 2\frac{\max\{p, 1-p\}}{p(1-p)}\right) G_h,
\end{split}
\end{equation}

\begin{equation}
\begin{split}
\|\nabla_{a} F_k(\v, \alpha; \z)\| \leq 4(1-p),
\end{split}
\end{equation}

\begin{equation}
\begin{split}
\|\nabla_{b} F_k(\v, \alpha; \z)\| \leq 4p.
\end{split}
\end{equation}

Thus,
\begin{equation}
\begin{split}
\|\nabla_{\v} F_k(\v, \alpha; \z)\|^2 &= \|\nabla_{\w} F_k(\v, \alpha; \z)\|^2 + \|\nabla_{a} F_k(\v, \alpha; \z)\|^2 + \|\nabla_{b} F_k(\v, \alpha; \z)\|^2 \\
&\leq \left(6+\frac{2\max\{p, 1-p\}}{p(1-p)}\right)^2 G_h^2 +16(1-p)^2 + 16p^2.
\end{split}
\end{equation}

\begin{equation}
\begin{split}
\|\nabla_{\alpha} F_k(\v, \alpha; \z)\|^2 &= \|2p h(\w; \x^k)\I_{y^k=-1} - 2(1-p)h(\w; \x^k)\I_{y^k=1} - 2p(1-p)\alpha\|^2 \\
&\leq (2p + 2(1-p) + 4\max\{p, 1-p\})^2 = (2+4\max\{p, 1-p\})^2.
\end{split} 
\end{equation} 

Thus, $B_{\v}^2 = \left(6+\frac{2\max\{p, 1-p\}}{p(1-p)}\right)^2 G_h^2 +16(1-p)^2 + 16p^2$ and $B_{\alpha}^2 = (2+4\max\{p, 1-p\})^2$.

It follow that 
\begin{equation}
\begin{split}
|\nabla_{\v} f_k(\v, \alpha)| = |E[\nabla_{\alpha} F_k (\v, \alpha; \z^k)]| \leq B_{\v}.
\end{split}
\end{equation}

Therefore,
\begin{equation}
\begin{split}
E[\|\nabla_{\v} f_k(\v, \alpha) - \nabla_{\v} F_k(\v, \alpha; \z^k)\|^2] \leq [2|\nabla_{\v} f_k(\v, \alpha)|^2 + 2|E[\nabla_{\v} F_k(\v, \alpha; \z^k)]|^2] \leq 4B_{\v}^2.
\end{split}
\end{equation}

Similarly,
\begin{equation}
\begin{split}
|\nabla_{\alpha} f_k(\w, a, b, \alpha)| = |E[\nabla_{\alpha} F_k (\w, a, b, \alpha; \z^k)]| \leq B_{\alpha}.
\end{split}
\end{equation}

Therefore,
\begin{equation}
\begin{split}
E[\|\nabla_{\alpha} f_k(\v, \alpha) - \nabla_{\alpha} F_k(\v, \alpha; \z^k)\|^2] \leq 2|\nabla_{\alpha} f_k(\v, \alpha)|^2 + 2E[F_k(\v, \alpha; \z^k)]|^2 \leq 4B_{\alpha}^2.
\end{split}
\end{equation}

Thus, $\sigma_{\v}^2 = 4B_{\v}^2$ and $\sigma_{\alpha}^2 = 4B_{\alpha}^2$.

Now, it remains to derive the constant $L_2$ such that $\|\nabla_{\v}F_k(\v_1, \alpha; \z) - \nabla_{\v} F_k(\v_2, \alpha; \z)\| \leq L_2 \|\v_1 - \v_2\|$.

By (\ref{sub_nabla_v}), we get
\begin{small}
\begin{equation}
\begin{split}
&\|\nabla_{\w} F_k(\v_1,\alpha;\z) - \nabla_{\w}F_k(\v_2, \alpha; \z)\|\\
&= \bigg\|\bigg[2(1-p)(h(\w_1;\x^k)-a_1) - 2(1+\alpha)(1-p)\bigg] \nabla h(\w_1; \x^k)\I_{[y^k=1]} 
+ \bigg[2p(h(\w_1; \x^k)-b_1) + 2(1+\alpha)p\bigg] \nabla h(\w_1; x^k)\I_{[y^k=-1]} \\
&~~~~~
- \bigg[2(1-p)(h(\w_2;\x^k)-a_2) - 2(1+\alpha)(1-p)\bigg] \nabla h(\w_2; \x^k)\I_{[y^k=1]}
- \bigg[2p(h(\w_2; \x^k)-b_2) + 2(1+\alpha)p\bigg] \nabla h(\w_2; \x^k)\I_{[y^k=-1]}
\bigg] \bigg\| \\
&= \Bigg\| 2(1-p)\bigg[h(\w_1; \x^k)\nabla h(\w_1;\x^k)-h(\w_2;\x^k)\nabla h(\w_2;\x^k)\bigg] \I_{[y^k=1]}
+ 2p \bigg[ h(\w_1; \x^k)\nabla h(\w_1; \x^k)-h(\w_2; \x^k)\nabla h(\w_2;\x^k) \bigg]\I_{[y^k=-1]}\\
& - (2(1+\alpha))(1-p)(\nabla h(\w_1; \x^k) - \nabla h(\w_2;\x^k))\I_{[y^k=1]}
+(2(1+\alpha)p)(\nabla h(\w_1; \x^k) - \nabla h(\w_2; \x^k))\I_{[y^k=-1]}\\
&- 2(1-p)(a_1 \nabla h(\w_1; \x^k) - a_2 \nabla h(\w_2; \x^k))\I_{y^k=1}
-2p(b_1 \nabla h(\w_1; \x^k) - b_2 \nabla h(\w_2;\x^k))\I_{[y^k=-1]}
\Bigg \| \\
&\leq  2(1-p)\|h(\w_1; \x^k)\nabla h(\w_1;\x^k) - h(\w_2; \x^k)\nabla h(\w_2;\x^k)\| + 2p \|h(\w_1; \x^k)\nabla h(\w_1;\x^k) - h(\w_2; \x^k)\nabla h(\w_2;\x^k)\| \\
&~~~~~+ \|2(1+\alpha)(1-p)\|\|\nabla h(\w_1; \x^k) - \nabla h(\w_2; \x^k)\| + \|2(1+\alpha)p\| \|\nabla h(\w_1; \x^k) - \nabla h(\w_2; \x^k)\|\\
&~~~~~+2(1-p)\|a_1 \nabla h(\w_1; \x^k) - a_2 \nabla h(\w_2; \x^k)\| + 2p\|b_1 \nabla h(\w_1; \x^k) - b_2 \nabla h(\w_2; \x^k)\|.
\end{split}
\label{H_grad}
\end{equation}
\end{small}

Denoting $\Gamma_1(\w; \x^k) = h(\w; \x^k) \nabla h(\w; \x^k)$,
\begin{equation}
\begin{split}
\|\nabla \Gamma_1(\w; \x^k)\| &= \|\nabla h(\w; \x^k) \nabla h(\w; \x^k)^T + h(\w; \x^k)\nabla^2 h(\w; \x^k)\| \\
& \leq \|\nabla h(\w; \x^k)\nabla h(\w; \x^k)^T\| + \|h(\w; \x^k) \nabla^2 h(\w; \x^k)\| \\
& \leq G_h^2 + L_h.
\end{split}
\end{equation}
Thus, $\|\Gamma_1(\w_1; \x^k) - \Gamma_1(\w_2; \x^k)\| =\|h(\w_1;\x^k) h'(\w_1; \x^k) - h(\w_2;\x^k) h'(\w_2; \x^k)\| \leq (G_h^2 + L_h)\|\w_1 - \w_2\|$.
Define $\Gamma_2(\w, \alpha; \x^k) = a \nabla h(\w; \x^k)$.
By Lemma \ref{lem:ab_bounded} and Assumption \ref{assumption_1}, we have
\begin{equation}
\begin{split}
\nabla_{\w, a} \Gamma_2 (\w, a; \x^k) \leq \|\nabla_{\w} \Gamma_2(\w, a; \z^k)\| + \|\nabla_{a} \Gamma_2(\w, a; \z^k)\| = \|a\nabla^2 h(\w; \x^k)\| + \|\nabla h(\w; \x^k)\| \leq L_h + G_h.
\end{split}
\end{equation}
Therefore, 
\begin{equation}
\begin{split}
    \|\! \Gamma_2 (\!\w_1, \! a_1;\! \x^k)\! -\! \Gamma_2 (\!\w_2,\! a_2;\! \x^k\!)\! \|\! =\! \| a_1\! \nabla\! h(\w_1;\! \x^k)\! -\! a_2\! \nabla\! h(\w_2;\! \x^k\!)\!\|\! \leq\! (L_h\! +\! G_h\!)\!\sqrt{\!\|\w_1\!-\!\w_2\|^2\! +\! \|a_1\! -\! a_2\|^2}.
\end{split}
\label{local_ah_lip}
\end{equation}

Similarly, we can prove that 
\begin{equation}
\begin{split}
\| b_1 \nabla h(\w_1; \x^k) - b_2 \nabla h(\w_2; \x^k)\| \leq (L_h+G_h)\sqrt{\|\w_1-\w_2\|^2+\|b_1-b_2\|^2}.
\end{split}
\label{local_bh_lip}
\end{equation}

Then plugging (\ref{local_ah_lip}), (\ref{local_bh_lip}) and Assumption \ref{assumption_1} into (\ref{H_grad}), we have
\begin{equation}
\begin{split}
&\|\nabla_{\w} F_k(\v_1,\alpha;\z) - \nabla_{\w}F_k(\v_2, \alpha; \z)\| \\
&\leq 2(G_h^2+L_h)\|\w_1-\w_2\| + 2|1+\alpha|G_h\|\w_1 - \w_2\| \\
&~~~~~+(L_h + G_h)\sqrt{\|\w_1 - \w_2\|^2 + \|a_1-a_2\|^2} + (L_h + G_h)\sqrt{\|\w_1-\w_2\|^2 + \|b_1-b_2\|^2}\\
&\leq (2(G_h^2+L_h) + |2(1+\alpha)|G_h + 2L_h + 2G_h)\sqrt{\|\w_1 - \w_2\|^2 + \|a_1-a_2\|^2 + \|b_1-b_2\|^2}\\
&\leq \left( 2 G_h^2 + 4L_h + \left(4 + \frac{2\max\{p, 1-p\}}{p(1-p)}\right) G_h\right) \|\v_1-\v_2\|.\\
\end{split}
\end{equation}

By (\ref{sub_nabla_v}), we also have
\begin{equation}
\begin{split}
&\|\nabla_{a} F_k(\v_1, \alpha; \z) - \nabla_{a} F_k(\v_2, \alpha; \z)\|^2 \leq 4(1-p)^2(\|h(\w_1; \x^k) - h(\w_2; \x^k)\|^2+\|a_1 - a_2\|^2)\\
&\leq 4(1-p)^2(G_h^2 \|\w_1-\w_2\|^2 + \|a_1 - a_2\|^2 + \|b_1 - b_2\|^2)
\leq 4(1-p)^2(G_h^2 + 1)\|\v_1 - \v_2\|^2,
\end{split}
\end{equation}
and 
\begin{equation}
\begin{split}
&\|\nabla_{b} F_k(\v_1, \alpha; \z) - \nabla_{b} F_k(\v_2, \alpha; \z)\|^2 \leq 4(1-p)^2(\|h(\w_1; \x^k) - h(\w_2; \x^k)\|^2+\|b_1 - b_2\|^2)\\
&\leq 4(1-p)^2(G_h^2 \|\w_1-\w_2\|^2 + \|a_1 - a_2\|^2 + \|b_1 - b_2\|^2)\leq 4(1-p)^2(G_h^2 + 1)\|\v_1 - \v_2\|^2.
\end{split}
\end{equation}

\begin{equation}
\begin{split}
\|\nabla_{\v} F_k (\v_1, \alpha; \z) - \nabla_{\v} F_k(\v_2, \alpha; \z)\|^2 &= \|\nabla_{\w}F_k(\v_1, \alpha; \z) - \nabla_{\w}F_k(\v_2, \alpha; \z)\|^2 \\
&~~~~~
+ \|\nabla_{a}F_k(\v_1, \alpha; \z) - \nabla_{b}F_k(\v_2, \alpha; \z)\|^2 + \|\nabla_{b} F_k(\v_1, \alpha; \z) - \nabla_{b}F_k(\v_1, \alpha; \z)\|^2\\
&\leq \left( G_h^2 + L_h + 4 + \frac{2\max\{p, 1-p\}}{p(1-p)} 8(1-p)^2 (G_h^2 + 1)\right) \|\v_1 - \v_2\|^2.
\end{split}
\end{equation}
Thus, we get $L_2 = \left( G_h^2 + L_h + 4 + \frac{2\max\{p, 1-p\}}{p(1-p)} 8(1-p)^2 (G_h^2 + 1)\right)^{1/2}$.

\section{Proof of Lemma \ref{lemma_one_stage}}
\noindent\textit{Proof. }
Plugging Lemma \ref{lem:var1}  and Lemma \ref{lem:var2}  into Lemma \ref{lem:objgap}, we get 

\begin{small}
\begin{equation}
\begin{split}
&\psi(\tilde{\v}) - \psi(\v_{\psi}^*) \\
&\leq \frac{1}{T}\sum\limits_{t=1}^{T} \Bigg[ \underbrace{ \left(\frac{L_{\v}+3G_{\alpha}^2/\mu_{\alpha}}{2} - \frac{1}{2\eta}\right) \|\bar{\v}_{t-1} - \bar{\v}_t\|^2 +  \left(\frac{L_{\alpha}+3G_{\v}^2/L_{\v}}{2} - \frac{1}{2\eta}\right)(\bar{\alpha}_t -  \bar{\alpha}_{t-1})^2}_{C_1}\\
&+\underbrace{\left(\frac{1}{2\eta} - \frac{\mu_{\alpha}}{3} \right) (\bar{\alpha}_{t-1} - \alpha^*(\tilde{\v}))^2 - \left(\frac{1}{2\eta} - \frac{\mu_{\alpha}}{3}\right)(\bar{\alpha}_t-\alpha^*(\tilde{\v}))^2}_{C_2}
+\underbrace{\left(\frac{2L_{\v}}{3} + \frac{1}{2\eta}\right) \|\bar{\v}_{t-1} - \v_{\psi}^*\|^2 - \left(\frac{1}{2\eta} + \frac{2L_{\v}}{3}\right)\|\bar{\v}_t - \v_
{\psi}^*\|^2}_{C_3}\\
& +\underbrace{\frac{1}{2\eta}((\alpha^*(\Tilde{\v}) - \Tilde{\alpha}_{t-1})^2 -  (\alpha^*(\Tilde{\v})-\Tilde{\alpha}_t)^2)}_{C_4} +\underbrace{\left(\frac{3G_{\v}^2}{2\mu_{\alpha}} + \frac{3L_{\v}}{2}\right)\frac{1}{K}\sum\limits_{k=1}^{K}\|\bar{\v}_{t-1}-\v_{t-1}^k\|^2 
+ \left(\frac{3 G_{\alpha}^2}{2L_{\v}} + \frac{3 L_{\alpha}^2}{2\mu_{\alpha}}\right)\frac{1}{K}\sum\limits_{k=1}^{K} (\bar{\alpha}_{t-1} - \alpha_{t-1}^k)^2}_{C_5}\\
&+\underbrace{\eta \left\|\frac{1}{K}\sum\limits_{k=1}^{K} [\nabla_{\v} f_k(\v_{t-1}^k, \alpha_{t-1}^k) - \nabla_{\v}F_k(\v_{t-1}^k, \alpha_{t-1}^k; \z_{t-1}^k)]\right\|^2}_{C_6} +  \underbrace{\frac{3\eta}{2} \left\|  \frac{1}{K}\sum\limits_{k=1}^{K} \nabla_{\alpha}f_k(\v_{t-1}^k, \alpha_{t-1}^k) - \nabla_{\alpha} F_k(\v_{t-1}^k, \alpha_{t-1}^k; \z_{t-1}^k) \right\|^2}_{C_7} \\
&\!+\!\underbrace{\left\langle\! \frac{1}{K}\sum\limits_{k=1}^{K}[\nabla_{\v}f_k(\v_{t-1}^k, \alpha_{t-1}^k)\! -\! \nabla_{\v}F_k(\v_{t-1}^k, \alpha_{t-1}^k; \z_{t-1}^k)],\! \hat{\v}_t\! -\! \v_{\psi}^* \right\rangle}_{C_8} 
+\underbrace{\left\langle\! \!\frac{1}{K}\sum\limits_{k=1}^{K}[\nabla_{\alpha} f_k(\v_{t-1}^k, \alpha_{t-1}^k)\! -\! F_k(\v_{t-1}^k, \alpha_{t-1}^k; \z_{t-1}^k)],  \tilde{\alpha}_{t-1}\! -\! \hat{\alpha}_t)  \right\rangle}_{C_9} \Bigg].\\ 
\end{split}
\label{before_summation}
\end{equation}
\end{small}

Since $\eta\leq \min(\frac{1}{L_{\v} + 3G_{\alpha}^2/\mu_{\alpha}}, \frac{1}{L_{\alpha}+3G_{\v}^2/L_{\v}})$,
thus in the RHS of (\ref{before_summation}), $C_1$ can be cancelled.
$C_2$, $C_3$ and $C_4$ will be handled by telescoping sum.
$C_5$ can be bounded by Lemma \ref{stich_lemma}.



Taking expectation over $C_6$,
\begin{small}
\begin{equation}
\begin{split}
&E\left[\eta \left\|\frac{1}{K}\sum\limits_{k=1}^{K}[\nabla_{\v} f_k(\v_{t-1}^k, \alpha_{t-1}^k) - \nabla_{\v}F_k(\v_{t-1}^k, \alpha_{t-1}^k; \z_{t-1}^k)]\right\|^2\right]\\
&=E\left[\frac{\eta}{K^2} \left\|\sum\limits_{k=1}^{K}[\nabla_{\v} f_k(\v_{t-1}^k, \alpha_{t-1}^k) - \nabla_{\v}F_k(\v_{t-1}^k, \alpha_{t-1}^k; \z_{t-1}^k)]\right\|^2\right]\\
&=E\left[\frac{\eta}{K^2}\left(\sum\limits_{k=1}^K \|\nabla_{\v} f_k(\v_{t-1}^k, \alpha_{t-1}^k) - \nabla_{\v}F_k(\v_{t-1}^k, \alpha_{t-1}^k; \z_{t-1}^k)\|^2\right.\right.\\
&~~~~~~\left.\left.+ 2\sum\limits_{i=1}^{K}\sum\limits_{j=i+1}^{K} \left\langle \nabla_{\v} f_i(\v_{t-1}^i, \alpha_{t-1}^i) 
- \nabla_{\v} F_i(\v_{t-1}^{i}, \alpha_{t-1}^{i}; \z_{t-1}^i), 
\nabla_{\v} f_j(\v_{t-1}^j, \alpha_{t-1}^j) 
- \nabla_{\v} F_j(\v_{t-1}^{j}, \alpha_{t-1}^{j}; \z_{t-1}^j)
\right\rangle \right)  \right]\\
&\leq \frac{\eta \sigma_{\v}^2}{K}.
\end{split}  
\label{local_variance_v}
\end{equation}
\end{small}
The last inequality holds because $\|\nabla_{\v}f_k(\v_{t-1}^k,\alpha_{t-1}^k) - \nabla_{\v}F_k (\v_{t-1}^k, \alpha_{t-1}^k;\z_{t-1}^k)\|^2 \leq \sigma_{\v}^2$
for any $k$ and $E\left[ \left\langle \nabla_{\v} f_i(\v_{t-1}^i, \alpha_{t-1}^i) 
- \nabla_{\v} F_i(\v_{t-1}^{i}, \alpha_{t-1}^{i}; \z_{t-1}^i), 
\nabla_{\v} f_j(\v_{t-1}^j, \alpha_{t-1}^j) 
- \nabla_{\v} F_j(\v_{t-1}^{j}, \alpha_{t-1}^{j}; \z_{t-1}^j)
\right\rangle \right] = 0$ for any $i \neq j$ as each machine draws data independently.
Similarly, we take expectation over $C_7$ and have
\begin{small}
\begin{equation}
\begin{split}
&E\left[\frac{3\eta}{2}  \left(\frac{1}{K}\sum\limits_{k=1}^{K}[\nabla_{\alpha} f_k(\v_{t-1}^k, \alpha_{t-1}^k) - \nabla_{\alpha}F_k(\v_{t-1}^k, \alpha_{t-1}^k; \z_{t-1}^k)]\right)^2\right]
\leq \frac{3\eta \sigma_{\alpha}^2}{2K}.
\end{split}  
\label{local_variance_alpha}
\end{equation}
\end{small}

Note $E\bigg[\left\langle \frac{1}{K}\sum\limits_{k=1}^{K}[\nabla_{\v}f_k(\v_{t-1}^k, \alpha_{t-1}^k) - \nabla_{\v}F_k(\v_{t-1}^k, \alpha_{t-1}^k; \z_{t-1}^k)], \hat{\v}_t - \v_{\psi}^* \right\rangle\bigg] = 0 $ and \\
$E\bigg[\left\langle -\frac{1}{K}\sum\limits_{k=1}^{K}[\nabla_{\alpha} f_k(\v_{t-1}^k, \alpha_{t-1}^k) - F_k(\v_{t-1}^k, \alpha_{t-1}^k; \z_{t-1}^k)], \tilde{\alpha}_{t-1} - \hat{\alpha}_t \right\rangle\bigg] = 0$.
Therefore, $C_8$ and $C_9$ will diminish after taking expectation.


As $\eta \leq \frac{1}{L_{\v} + 3G_{\alpha}^2/\mu_{\alpha}}$, we have $L_{\v} \leq \frac{1}{\eta}$.
Plugging (\ref{local_variance_v}) and  (\ref{local_variance_alpha})  into (\ref{before_summation}), and taking expectation, it yields
\begin{equation*}
\begin{split}
E[\psi(\tilde{\v}) - \psi(\v_{\psi}^*)]
&\leq E\bigg\{\frac{1}{T}\left(\frac{2L_{\v}}{3} + \frac{1}{2\eta}\right) \|\bar{\v}_0-\v_{\psi}^*\|^2 +  \frac{1}{T}\left(\frac{1}{2\eta} - \frac{\mu_{\alpha}}{3} \right)(\bar{\alpha}_0 - \alpha^*(\tilde{\v}))^2 
+ \frac{1}{2\eta T} (\tilde{\alpha}_0 - \alpha^*(\Tilde{\v}))^2 \\ 
&~~~~~+ \frac{1}{T}\sum\limits_{t=1}^{T}\left(\frac{3G_{\v}^2}{2 \mu_{\alpha}} + \frac{3L_{\v}}{2}\right)\frac{1}{K}\sum\limits_{k=1}^{K}\|\bar{\v}_{t-1} - \v_{t-1}^k\|^2 + \frac{1}{T}\sum\limits_{t=1}^{T}\left(\frac{3G_{\alpha}^2}{2L_{\v}} + \frac{3L_{\alpha}^2}{2\mu_{\alpha}}\right)\frac{1}{K}\sum\limits_{k=1}^{K}\|\bar{\alpha}_{t-1} - \alpha_{t-1}^k\|^2\\
&~~~~~+\frac{1}{T} \sum\limits_{t=1}^{T}\frac{\eta \sigma_{\v}^2}{K}
+\frac{1}{T} \sum\limits_{t=1}^{T}\frac{3\eta \sigma_{\alpha}^2}{2 K}\bigg\}\\ 
&\leq \frac{2}{\eta T} \|\v_0 - \v_{\psi}^*\|^2 + \frac{1}{\eta T} (\alpha_0 -\alpha^*(\tilde{\v}))^2 + \left(\frac{6G_{\v}^2}{\mu_{\alpha}} +6L_{\v} + \frac{6G_{\alpha}^2}{L_{\v}} + \frac{6L_{\alpha}^2}{\mu_{\alpha}}\right) \eta^2 I^2 B^2\I_{I>1}  + \frac{\eta(2\sigma_{\v}^2 + 3\sigma_{\alpha}^2)}{2K},
\end{split} 
\end{equation*} 
where we use Lemma \ref{stich_lemma}, $\v_0 = \bar{\v}_0$, $\alpha_0 = \bar{\alpha}_0$ and $B^2 = \max\{B^2_{\v}, B^2_{\alpha}\}$ in the last inequality.
$\Box$

\section{Proof of Lemma \ref{lem:objgap}}
\noindent \textit{Proof.}
Define $\alpha^*(\tilde{\v}) = \arg\max\limits_{\alpha} f(\tilde{\v}, \alpha)$ and $\tilde{\alpha} = \frac{1}{K}\sum\limits_{k=1}^{K} \frac{1}{T}\sum\limits_{t=1}^{T}\alpha_t^k$.
\begin{equation}
\begin{split}
\psi(\tilde{\v}) - \min\limits_{\v}\psi (\v) &= \max\limits_{\alpha}\left[f(\tilde{\v}, \alpha) + \frac{1}{2\gamma} \|\tilde{\v} - \v_{0}\|^2\right] - \min\limits_{\v} \max\limits_{\alpha}\left[f(\v, \alpha)+\frac{1}{2\gamma}\|\v-\v_{0}\|^2\right]\\
&= \left[f(\tilde{\v}, \alpha^*(\tilde{\v})) + \frac{1}{2\gamma} \|\tilde{\v} - \v_{0}\|^2 \right] - \max\limits_{\alpha}\left[f(\v_{\psi}^*, \alpha) + \frac{1}{2\gamma}\|\v_{\psi}^*-\v_{0}\|^2\right]\\
&\leq \left[f(\tilde{\v}, \alpha^*(\tilde{\v})) + \frac{1}{2\gamma} \|\tilde{\v} - \v_{0}\|^2 \right] - \left[f(\v_{\psi}^*, \tilde{\alpha}) + \frac{1}{2\gamma}\|\v_{\psi}^*-\v_{0}\|^2\right]\\
&\leq \frac{1}{T}\sum\limits_{t=1}^{T} \left[\left(f(\bar{\v}_t, \alpha^*(\tilde{\v})) +\frac{1}{2\gamma} \|\bar{\v}_t - \v_{0}\|^2\right) - \left(f(\v_{\psi}^*, \bar{\alpha}_t) + \frac{1}{2\gamma}\|\v_{\psi}^* - \v_{0}\|^2\right)\right],\\
\end{split}
\label{objgap_relax}
\end{equation}
where the last inequality uses Jensen's inequality and the fact that $f(\v, \alpha) + \frac{1}{2\gamma}\|\v-\v_{0}\|^2$ is convex w.r.t. $\v$ and concave w.r.t. $\alpha$.

By $L_{\v}$-weakly convexity of $f(\cdot)$ w.r.t. $\v$, we have
\begin{equation}
\begin{split}
 f(\bar{\v}_{t-1}, \bar{\alpha}_{t-1}) + \langle \nabla_{\v}f(\bar{\v}_{t-1}, \bar{\alpha}_{t-1}), \v_{\psi}^* - \bar{\v}_{t-1}\rangle - \frac{L_{\v}}{2}\|\bar{\v}_{t-1}-\v_{\psi}^*\|^2 \leq f(\v_{\psi}^*, \bar{\alpha}_{t-1}),
\end{split}
\label{weakly_convex_v}
\end{equation}
and by $L_{\v}$-smoothness of $f(\cdot)$ w.r.t. $\v$, we have
\begin{equation}
\begin{split}
f(\bar{\v}_t, \alpha^*(\Tilde{\v}))  &\leq f(\bar{\v}_{t-1}, \alpha^*(\Tilde{\v}))  
+ \langle \nabla_{\v}f(\bar{\v}_{t-1}, \alpha^*(\Tilde{\v})), \bar{\v}_t - \bar{\v}_{t-1}\rangle + \frac{L_{\v}}{2} \|\bar{\v}_t - \bar{\v}_{t-1}\|^2\\
&= f(\bar{\v}_{t-1}, \alpha^*(\Tilde{\v})) + \langle \nabla_{\v}f(\bar{\v}_{t-1}, \alpha^*(\Tilde{\v})), \bar{\v}_t - \bar{\v}_{t-1}\rangle  + \frac{L_{\v}}{2}\|\bar{\v}_t - \bar{\v}_{t-1}\|^2 \\
&~~~~ + \langle \nabla_{\v}f(\bar{\v}_{t-1}, \bar{\alpha}_{t-1}), \bar{\v}_t - \bar{\v}_{t-1}\rangle -  \langle \nabla_{\v}f(\bar{\v}_{t-1}, \bar{\alpha}_{t-1}), \bar{\v}_t - \bar{\v}_{t-1}\rangle \\
&=  f(\bar{\v}_{t-1}, \alpha^*(\Tilde{\v})) + \langle \nabla_{\v}f(\bar{\v}_{t-1}, \bar{\alpha}_{t-1}), \bar{\v}_t - \bar{\v}_{t-1}\rangle + \frac{L_{\v}}{2}\|\bar{\v}_t - \bar{\v}_{t-1}\|^2 \\
&~~~~ + \langle\nabla_{\v} f(\bar{\v}_{t-1}, \alpha^*(\Tilde{\v}))-\nabla_{\v}f(\bar{\v}_{t-1}, \bar{\alpha}_{t-1}) , \bar{\v}_t - \bar{\v}_{t-1} \rangle\\
&\overset{(a)}\leq f(\bar{\v}_{t-1}, \alpha^*(\Tilde{\v})) + \langle \nabla_{\v}f(\bar{\v}_{t-1}, \bar{\alpha}_{t-1}), \bar{\v}_t - \bar{\v}_{t-1}\rangle + \frac{L_{\v}}{2}\|\bar{\v}_t - \bar{\v}_{t-1}\|^2 \\
&~~~~ + G_{\alpha} |\bar{\alpha}_{t-1} - \alpha^*(\Tilde{\v})| \|\bar{\v}_t - \bar{\v}_{t-1}\| \\
&\overset{(b)}\leq f(\bar{\v}_{t-1}, \alpha^*(\Tilde{\v})) + \langle \nabla_{\v}f(\bar{\v}_{t-1}, \bar{\alpha}_{t-1}), \bar{\v}_t - \bar{\v}_{t-1}\rangle + \frac{L_{\v}}{2}\|\bar{\v}_t - \bar{\v}_{t-1}\|^2 \\
&~~~~ + \frac{\mu_{\alpha}}{6} |\bar{\alpha}_{t-1} - \alpha^*(\Tilde{\v})|^2 +  \frac{3G_{\alpha}^2}{2\mu_{\alpha}} \|\bar{\v}_t - \bar{\v}_{t-1}\|^2,\\ 
\end{split}
\label{smooth_v}
\end{equation}
where $(a)$ holds because we know that $\nabla_{\v} f(\cdot)$ is $G_{\alpha} = 2\max\{p, 1-p\}$-Lipshitz w.r.t. $\alpha$ by the definition of $f(\cdot)$, and $(b)$ holds by Young's inequality. 

By $\frac{1}{\gamma}$-strong convexity of $\frac{1}{2\gamma} \|\v - \v_{0}\|^2$ w.r.t. $\v$, we have
\begin{equation}
\begin{split}
\frac{1}{2\gamma}\|\bar{\v}_t - \v_{0}\|^2 + \frac{1}{\gamma}\langle \bar{\v}_t - \v_{0}, \v_{\psi}^* - \v_t \rangle + \frac{1}{2\gamma}\|\v_{\psi}^* - \v_t\|^2 \leq \frac{1}{2\gamma}\|\v_{\psi}^* - \v_{0}\|^2.
\end{split}
\label{convex_v}
\end{equation}

Adding (\ref{weakly_convex_v}), (\ref{smooth_v}), (\ref{convex_v}), and rearranging terms, we have
\begin{equation}
\begin{split}
 &f(\bar{\v}_{t-1}, \bar{\alpha}_{t-1}) 
+ 
f(\bar{\v}_t, \alpha^*(\Tilde{\v})) 
+
\frac{1}{2\gamma}\|\bar{\v}_t - \v_{0}\|^2 
-
\frac{1}{2\gamma}\|\v_{\psi}^* - \v_{0}\|^2\\
\leq
&f(\v_{\psi}^*, \bar{\alpha}_{t-1})
+
f(\bar{\v}_{t-1}, \alpha^*(\Tilde{\v})) + \langle \nabla_{\v}f(\bar{\v}_{t-1}, \bar{\alpha}_{t-1}), \bar{\v}_t - \bar{\v}_{\psi}^*\rangle 
+
 \frac{L_{\v} + 3G_{\alpha}^2/\mu_{\alpha}}{2}\eta^2 \|\bar{\v}_t-\bar{\v}_{t-1}\|^2 
+ \frac{L_{\v}}{2}\|\bar{\v}_{t-1}-\v_{\psi}^*\|^2 \\
&~~~ + \frac{\mu_{\alpha}}{6} (\bar{\alpha}_{t-1} - \alpha^*(\Tilde{\v}))
- \frac{1}{2\gamma}\|\v_{\psi}^* - \v_t\|^2 
+\frac{1}{\gamma}\langle \bar{\v}_t - \v_{0}, \v_t - \v_{\psi}^*\rangle.
\end{split}
\label{sum_v}
\end{equation}

By definition, we know that $f(\cdot)$ is $\mu_{\alpha}:=2p(1-p)$-strong concavity w.r.t. $\alpha$ ($-f(\cdot)$ is $\mu_{\alpha}$-strong convexity w.r.t. $\alpha$).
Thus, we have
\begin{equation}
\begin{split} 
-f(\bar{\v}_{t-1}, \bar{\alpha}_{t-1})  
- \nabla_{\alpha}f(\bar{\v}_{t-1}, \bar{\alpha}_{t-1})^T(\alpha^*(\Tilde{\v}) - \bar{\alpha}_{t-1}) + \frac{\mu_{\alpha}}{2}(\alpha^*(\Tilde{\v}) - \bar{\alpha}_{t-1})^2 \leq -f(\bar{\v}_{t-1}, \alpha^*(\Tilde{\v})) 
\end{split}
\label{concave_alpha}
\end{equation}

By definition, we know that $f(\cdot)$ is smooth in $\alpha$ (with coefficient $L_{\alpha}: = 2p(1-p)$), we get
\begin{equation}
\begin{split}
&- f(\v_{\psi}^*, \bar{\alpha}_t) \leq -f(\v_{\psi}^*, \bar{\alpha}_{t-1}) - \langle \nabla_{\alpha}f(\v_{\psi}^*, \bar{\alpha}_{t-1}), \bar{\alpha}_t - \bar{\alpha}_{t-1} \rangle + \frac{L_{\alpha}}{2}(\bar{\alpha}_t - \bar{\alpha}_{t-1})^2\\
&= -f(\v_{\psi}^*, \bar{\alpha}_{t-1}) - \langle \nabla_{\alpha} f(\v_{\psi}^*, \bar{\alpha}_{t-1}), \bar{\alpha}_t - \bar{\alpha}_{t-1}\rangle + \frac{L_{\alpha}}{2}(\bar{\alpha}_t - \bar{\alpha}_{t-1})^2 \\
&~~~~~ - \langle \nabla_{\alpha} f(\bar{\v}_{t-1}, \bar{\alpha}_{t-1}), \bar{\alpha}_t - \bar{\alpha}_{t-1}\rangle + \langle \nabla_{\alpha} f(\bar{\v}_{t-1}, \bar{\alpha}_{t-1}), \bar{\alpha}_t - \bar{\alpha}_{t-1}\rangle \\
&\overset{(a)}\leq -f(\v_{\psi}^*, \bar{\alpha}_{t-1}) - \langle \nabla_{\alpha} f(\bar{\v}_{t-1}, \bar{\alpha}_{t-1}), \bar{\alpha}_t - \bar{\alpha}_{t-1}\rangle + \frac{L_{\alpha}}{2} (\bar{\alpha}_t - \bar{\alpha}_{t-1})^2 + G_{\v}|\langle \v_{\psi}^* - \bar{\v}_{t-1} , \bar{\alpha}_t - \bar{\alpha}_{t-1}\rangle|\\
&\leq -f(\v_{\psi}^*, \bar{\alpha}_{t-1}) - \langle \nabla_{\alpha} f(\bar{\v}_{t-1}, \bar{\alpha}_{t-1}), \bar{\alpha}_t - \bar{\alpha}_{t-1}\rangle + \frac{L_{\alpha}}{2} (\bar{\alpha}_t - \bar{\alpha}_{t-1})^2 + \frac{L_{\v}}{6} \| \bar{\v}_{t-1} - \v_{\psi}^*\|^2 + \frac{3G_{\v}^2}{2L_{\v}}(\bar{\alpha}_t - \bar{\alpha}_{t-1})^2,\\
\end{split}
\label{smooth_alpha}
\end{equation}
where (a) holds because $\nabla_{\alpha} f(\cdot)$ is Lipshitz in $\alpha$ with coefficient $G_{\v} = 2\max\{p, 1-p\} G_h$ by definition of $f(\cdot)$.

Adding (\ref{concave_alpha}), (\ref{smooth_alpha}) and arranging terms, we have
\begin{equation}
\begin{split}
-f(\bar{\v}_{t-1}, \bar{\alpha}_{t-1}) 
- f(\v_{\psi}^*, \bar{\alpha}_t) 
&\leq-f(\bar{\v}_{t-1}, \alpha^*(\Tilde{\v}))  
-f(\v_{\psi}^*, \bar{\alpha}_{t-1}) - \langle \nabla_{\alpha} f(\bar{\v}_{t-1},  \bar{\alpha}_{t-1}), \bar{\alpha}_t -  \alpha^*(\Tilde{\v})\rangle + \frac{L_{\alpha}}{2} \|\bar{\alpha}_t - \bar{\alpha}_{t-1}\|^2 \\  
&~~~~~+ \frac{L_{\v}}{6} \| \bar{\v}_{t-1} - \v_{\psi}^*\|^2 + \frac{3G_{\v}^2}{2L_{\v}}(\bar{\alpha}_t - \bar{\alpha}_{t-1})^2
- \frac{\mu_{\alpha}}{2}(\alpha^*(\Tilde{\v}) - \bar{\alpha}_{t-1})^2.
\end{split} 
\label{sum_alpha}
\end{equation}

Adding (\ref{sum_v}) and (\ref{sum_alpha}), we get
\begin{equation}
\begin{split}
&\bigg[f(\bar{\v}_t, \alpha^*(\Tilde{\v})) + \frac{1}{2\gamma}\|\bar{\v}_t-\v_{0}\|^2 \bigg] - \bigg[f(\v_{\psi}^*, \bar{\alpha}_t) + \frac{1}{2\gamma}\|\v_{\psi}^* - \v_{0}\|^2\bigg]
\leq \\
&\langle \nabla_{\v}f(\bar{\v}_{t-1}, \bar{\alpha}_{t-1}), \bar{\v}_t - \v_{\psi}^*\rangle - \langle \nabla_{\alpha} f(\bar{\v}_{t-1}, \bar{\alpha}_{t-1}),  \bar{\alpha}_t - \alpha^*(\Tilde{\v})\rangle \\
&+\frac{L_{\v} + 3G_{\alpha}^2/\mu_{\alpha}}{2} \eta^2 \|\bar{\v}_t-\bar{\v}_{t-1}\|^2
+ \bigg(\frac{L_{\v}}{6} + \frac{L_{\v}}{2}\bigg) \| \bar{\v}_{t-1} - \v_{\psi}^*\|^2 
- \frac{1}{2\gamma}\|\v_{\psi}^* - \v_t\|^2  \\
&+ \frac{L_{\alpha} + 3G_{\v}^2/L_{\v}}{2} \eta^2 \|\bar{\alpha}_t - \bar{\alpha}_{t-1}\|^2 -  \frac{\mu_{\alpha}}{3}(\bar{\alpha}_{t-1} - \alpha^*(\Tilde{\v}))^2\\
& +\frac{1}{\gamma}\langle \bar{\v}_t - \v_{0}, \bar{\v}_t - \v_{\psi}^*\rangle. 
\end{split}
\label{sum_v_alpha}
\end{equation}

Applying $\gamma = \frac{1}{2L_{\v}}$ to (\ref{sum_v_alpha}) and then plugging it into (\ref{objgap_relax}), we get

\begin{equation*} 
\begin{split}
&\psi(\tilde{\v}) - \min\limits_{\v} \psi(\v) \leq \frac{1}{T} \sum\limits_{t=1}^{T}\bigg[ \langle \nabla_{\v}f(\bar{\v}_{t-1}, \bar{\alpha}_{t-1}), \bar{\v}_t - \v_{\psi}^*\rangle +2L_{\v} \langle \bar{\v}_t-\v_0, \bar{\v}_t-\v_{\psi}^*\rangle + \langle \nabla_{\alpha} f(\bar{\v}_{t-1}, \bar{\alpha}_{t-1}),  \alpha^*(\Tilde{\v}) - \bar{\alpha}_t \rangle  \\
&~~~~~~~~~~~~~~~~~~~~~~~~~~~~~~~~~~~
+\frac{L_{\v} + 3G_{\alpha}^2/\mu_{\alpha}}{2}\|\bar{\v}_t - \bar{\v}_{t-1}\|^2  + \frac{L_{\alpha} + 3G_{\v}^2/L_{\v}}{2} (\bar{\alpha}_t - \bar{\alpha}_{t-1})^2\\
&~~~~~~~~~~~~~~~~~~~~~~~~~~~~~~~~~~~
+ \frac{2L_{\v}}{3} \| \bar{\v}_{t-1} - \v_{\psi}^*\|^2 - L_{\v} \| \bar\v_t - \v_{\psi}^* \|^2   - \frac{\mu_{\alpha}}{3} (\bar{\alpha}_{t-1} - \alpha^*(\Tilde{\v}))^2\bigg]. ~\Box\\
\end{split}
\end{equation*}

\section{Proof of Lemma \ref{lem:var1}} 
\noindent \textit{Proof.}
According to the update rule of $\v$ and taking $\gamma = \frac{1}{2L_{\v}}$, we have
\begin{equation}
\begin{split}
2L_{\v}(\v_t^k - \v_{0}) = -\nabla_{\v} F_k(\v_{t-1}^k, \alpha_{t-1}^k; \z_{t-1}^k) - \frac{1}{\eta} (\v_t^k - \v_{t-1}^k).
\end{split}
\end{equation}
Taking average over $K$ machines, we have
\begin{equation}
\begin{split}
2L_{\v} (\bar{\v}_t - \v_{0}) = -\frac{1}{K}\sum\limits_{k=1}^{K}\nabla_{\v} F_k(\v_{t-1}^k, \alpha_{t-1}^k; \z_{t-1}^k) - \frac{1}{\eta} (\bar{\v}_t - \bar{\v}_{t-1}).
\end{split}
\end{equation}

It follows that
\begin{equation}
\begin{split}
&\langle \nabla_{\v}f(\bar{\v}_{t-1}, \bar{\alpha}_{t-1}), \bar{\v}_t - \v_{\psi}^*\rangle +2L_{\v}\langle \bar{\v}_t - \v_{0}, \bar{\v}_t - \v_{\psi}^*\rangle\\
&= \bigg\langle \frac{1}{K}\sum\limits_{k=1}^{K} \nabla_{\v}f_k(\bar{\v}_{t-1}, \bar{\alpha}_{t-1}), \bar{\v}_t - \v_{\psi}^*\bigg\rangle
- \bigg\langle \frac{1}{K}\sum\limits_{k=1}^{K} \nabla_{\v}F_k(\v_{t-1}, \alpha_{t-1}^k; \z_{t-1}^k), \bar{\v}_t - \v_{\psi}^*\bigg\rangle 
+ \frac{1}{\eta}\langle \bar{\v}_t - \bar{\v}_{t-1}, \bar{\v}_t - \v_{\psi}^*\rangle\\
&\leq \bigg\langle \frac{1}{K}\sum\limits_{k=1}^{K} [\nabla_{\v}f_k(\bar{\v}_{t-1}, \bar{\alpha}_{t-1}) - \nabla_{\v} f_k(\bar{\v}_{t-1}, \alpha_{t-1}^k)], \bar{\v}_t - \v_{\psi}^*\bigg\rangle ~~~~~~~~~~~\textcircled{\small{1}}\\
&~~~ +\bigg\langle \frac{1}{K}\sum\limits_{k=1}^{K} [\nabla_{\v}f_k(\bar{\v}_{t-1}, \alpha_{t-1}^k) - \nabla_{\v}f_k(\v_{t-1}^k, \alpha_{t-1}^k)], \bar{\v}_t - \v_{\psi}^*\bigg\rangle ~~~~~~~~~~~\textcircled{\small{2}}\\
&~~~ +\bigg\langle \frac{1}{K}\sum\limits_{k=1}^{K}[\nabla_{\v}f_k(\v_{t-1}^k, \alpha_{t-1}^k) - \nabla_{\v}F_k(\v_{t-1}, \alpha_{t-1}^k; \z_{t-1}^k)], \bar{\v}_t - \v_{\psi}^*\bigg\rangle~~~~~~~~~~~\textcircled{\small{3}} \\
&~~~ + \frac{1}{2\eta}(\|\bar{\v}_{t-1} - \v_{\psi}^*\|^2 - \|\bar{\v}_{t-1} - \bar{\v}_{t}\|^2 - \|\bar{\v}_t - \v_{\psi}^*\|^2).
\end{split}
\label{circledv}
\end{equation}

Then we will bound \textcircled{\small{1}}, \textcircled{\small{2}} and \textcircled{\small{3}} separately,
\begin{equation}
\begin{split}
\textcircled{\small{1}} &\overset{(a)}\leq \frac{3}{2L_{\v}} \left\| \frac{1}{K} \sum\limits_{k=1}^{K}[\nabla_{\v}f_k(\bar{\v}_{t-1}, \bar{\alpha}_{t-1}) - \nabla_{\v}f_k(\bar{\v}_{t-1}, \alpha_{t-1}^{k})] \right\|^2 + \frac{L_{\v}}{6}\|\bar{\v}_t - \v_{\psi}^*\|^2\\
&\overset{(b)}\leq \frac{3}{2 L_{\v}} \frac{1}{K}\sum\limits_{k=1}^{K} \|\nabla_{\v}f_k(\bar{\v}_{t-1}, \bar{\alpha}_{t-1}) - \nabla_{\v} f_k(\bar{\v}_{t-1}, \alpha_{t-1}^k)\|^2 + \frac{L_{\v}}{6}\|\bar{\v}_t - \v_{\psi}^*\|^2\\
&\overset{(c)}\leq \frac{3G_{\alpha}^2}{2 L_{\v}}\frac{1}{K}\sum\limits_{k=1}^{K}\|\bar{\alpha}_{t-1} - \alpha_{t-1}^k\|^2 + \frac{L_{\v}}{6}\|\bar{\v}_t - \v_{\psi}^*\|^2,\\
\end{split}
\label{circled1}
\end{equation}
where (a) follows from Young's inequality and (b) follows from Jensen's inequality.
(c) holds because $\nabla_{\v} f_k(\v, \alpha)$ is Lipschitz in $\alpha$ with coefficient $G_{\alpha} = 2\max(p, 1-p)$ for any $\v$ by definition of $f_k(\cdot)$. 
By similar techniques, we have
\begin{equation}
\begin{split}
\textcircled{\small{2}} &\leq \frac{3}{2L_{\v}} \frac{1}{K}\sum\limits_{k=1}^{K}\| \nabla_{\v} f_k(\bar{\v}_{t-1}, \alpha_{t-1}^{k}) - \nabla_{\v} f_k(\v_{t-1}^k, \alpha_{t-1}^{k})\|^2 + \frac{L_{\v}}{6}\|\bar{\v}_t - \v_{\psi}^*\|^2 \\
& \leq \frac{3 L_{\v}}{2}\frac{1}{K}\sum\limits_{k=1}^{K} \|\bar{\v}_{t-1} - \v_{t-1}^{k}\|^2 + \frac{L_{\v}}{6} \|\bar{\v}_t - \v_{\psi}^*\|^2.
\end{split}
\label{circled2}
\end{equation}

Let $\hat{\v}_t = \arg\min\limits_{\v} \left(\frac{1}{K}\sum\limits_{k=1}^{K} \nabla_{\v} f(\v^k_{t-1}, \alpha^k_{t-1})\right)^T \v + \frac{1}{2\eta} \|\v - \bar{\v}_{t-1}\|^2 + \frac{1}{2\gamma}\|\v - \v_{0}\|^2$.
Then we have 
\begin{equation}
\begin{split}
\bar{\v}_t - \hat{\v}_t = \frac{\eta \gamma}{\eta + \gamma}\bigg(\nabla_{\v}f(\v^{k}_{t-1}, \alpha^k_{t-1}) - \frac{1}{K}\sum\limits_{k=1}^{K}\nabla_{\v}f_k(\v^{k}_{t-1}, \alpha^k_{t-1}; \z_{t-1}^k)\bigg).
\end{split}
\end{equation}

Hence we get
\begin{equation}
\begin{split}
&\textcircled{\small{3}} = \left\langle \frac{1}{K}\sum\limits_{k=1}^{K}[\nabla_{\v}f_k(\v_{t-1}^k, \alpha_{t-1}^k) - \nabla_{\v} F_k(\v_{t-1}^{k}, \alpha_{t-1}^k; \z_{t-1}^k)], \bar{\v}_t - \hat{\v}_t \right\rangle \\
&~~~~+ \left\langle \frac{1}{K}\sum\limits_{k=1}^{K}[\nabla_{\v}f_k(\v_{t-1}^k, \alpha_{t-1}^k) - \nabla_{\v} F_k(\v_{t-1}^{k}, \alpha_{t-1}^k; \z_{t-1}^k)], \hat{\v}_t - \v_{\psi}^* \right\rangle\\
& = \frac{\eta\gamma}{\eta + \gamma} \left\|\frac{1}{K} \sum\limits_{k=1}^{K}[\nabla_{\v}f_k(\v_{t-1}^k, \alpha_{t-1}^k) - \nabla_{\v} F_k(\v_{t-1}^k, \alpha_{t-1}^k; \z_{t-1}^k)]  \right\|^2\\
&~~~~+ \left\langle \frac{1}{K}\sum\limits_{k=1}^{K}[\nabla_{\v}f_k(\v_{t-1}^k, \alpha_{t-1}^k) - \nabla_{\v} F_k(\v_{t-1}^{k}, \alpha_{t-1}^k; \z_{t-1}^k)], \hat{\v}_t - \v_{\psi}^* \right\rangle\\
&\leq \eta \left\|\frac{1}{K} \sum\limits_{k=1}^{K}[\nabla_{\v}f_k(\v_{t-1}^k, \alpha_{t-1}^k) - \nabla_{\v} F_k(\v_{t-1}^k, \alpha_{t-1}^k; \z_{t-1}^k)]  \right\|^2\\
&~~~~+ \left\langle \frac{1}{K}\sum\limits_{k=1}^{K}[\nabla_{\v}f_k(\v_{t-1}^k, \alpha_{t-1}^k) - \nabla_{\v} F_k(\v_{t-1}^{k}, \alpha_{t-1}^k; \z_{t-1}^k)], \hat{\v}_t - \v_{\psi}^* \right\rangle\\
\end{split}
\label{circled3}
\end{equation}

Plugging (\ref{circled1}), (\ref{circled2}) and (\ref{circled3}) into (\ref{circledv}), we get
\begin{equation}
\begin{split}
& \left\langle \nabla_{\v} f(\bar{\v}_{t-1}, \bar{\alpha}_{t-1}), \bar{\v}_t - \v_{\psi}^*\right\rangle + \frac{1}{\gamma}\langle \bar{\v}_t - \v_{0}, \bar{\v}_t - \v_{\psi}^*\rangle \\
& \leq \frac{3G_{\alpha}^2}{2L_{\v}}\frac{1}{K} \sum\limits_{k=1}^{K} \|\bar{\alpha}_{t-1}-\alpha_{t-1}^{k}\|^2 + \frac{L_{\v}}{6}\|\bar{\v}_t - \v_{\psi}^*\|^2 + \frac{3 L_{\v}}{2}\frac{1}{K}\sum\limits_{k=1}^{K}\|\bar{\v}_{t-1}-\v_{t-1}^k\|^2 + \frac{L_{\v}}{6}\|\bar{\v}_t - \v_{\psi}^*\|^2\\
&~~~~~+ \eta \left\|\frac{1}{K}\sum\limits_{k=1}^{K}[\nabla_{\v}f_k(\v_{t-1}^k, \alpha_{t-1}^k) - \nabla_{\v}F_k(\v_{t-1}^{k}, \alpha_{t-1}^k; \z_{t-1}^k)]\right\|^2 \\
&~~~~~+\left\langle \frac{1}{K}\sum\limits_{k=1}^{K}[\nabla_{\v}f_k(\v_{t-1}^k, \alpha_{t-1}^k)-\nabla_{\v}F_k(\v_{t-1}^{k}, \alpha_{t-1}^k;\z_{t-1}^k)], \hat{\v}_t - \v_{\psi}^*\right\rangle\\
&~~~~~+\frac{1}{2\eta} (\|\bar{\v}_{t-1}-\v_{\psi}^*\|^2 - \|\bar{\v}_{t-1} - \bar{\v}_t\|^2 - \|\bar{\v}_t - \v_{\psi}^*\|^2). \Box
\end{split}
\label{grad_v}
\end{equation}


\section{Proof of Lemma \ref{lem:var2}}
\noindent\textit{Proof.}
\begin{equation}
\begin{split}
\langle \nabla_{\alpha} f(\bar{\v}_{t-1}, \bar{\alpha}_{t-1}), \alpha^*(\Tilde{\v}) - \bar{\alpha}_t \rangle &= \bigg\langle \frac{1}{K} \sum\limits_{k=1}^{K}\nabla_{\alpha}  f_k(\bar{\v}_{t-1}, \bar{\alpha}_{t-1}), \alpha^*(\Tilde{\v}) - \bar{\alpha}_t \bigg\rangle\\
&= \bigg\langle \frac{1}{K} \sum\limits_{k=1}^{K}[\nabla_{\alpha} f_k(\bar{\v}_{t-1}, \bar{\alpha}_{t-1})-\nabla_{\alpha}f_k(\bar{\v}_{t-1}, \alpha_{t-1}^k)],  \alpha^*(\Tilde{\v}) - \bar{\alpha}_t \bigg\rangle ~~~~~~\textcircled{\small{4}}\\
&~~~~+ \bigg\langle \frac{1}{K}\sum\limits_{k=1}^{K}[\nabla_{\alpha}f_k(\bar{\v}_{t-1}, \alpha_{t-1}^k) - \nabla_{\alpha} f_k(\v_{t-1}^k, \alpha_{t-1}^k)], \alpha^*(\Tilde{\v}) - \bar{\alpha}_t \bigg\rangle ~~~~~~\textcircled{\small{5}}\\
&~~~~+ \bigg\langle \frac{1}{K}\sum\limits_{k=1}^{K}[\nabla_{\alpha} f_k(\v_{t-1}^k, \alpha_{t-1}^k) - \nabla_{\alpha}f_k (\v_{t-1}^k, \alpha_{t-1}^k; \z_{t-1}^k)], \alpha^*(\Tilde{\v}) - \bar{\alpha}_t \bigg\rangle ~~~~~~\textcircled{\small{6}}\\
&~~~~+ \bigg\langle \frac{1}{K}\sum\limits_{k=1}^{K}\nabla_{\alpha}F_k (\v_{t-1}^k, \alpha_{t-1}^k; \z_{t-1}^k), \alpha^*(\Tilde{\v}) - \bar{\alpha}_t \bigg\rangle ~~~~~~\textcircled{\small{7}}\\
\end{split}
\end{equation}

\begin{equation}
\begin{split} 
\textcircled{\small{4}} &\overset{(a)}\leq \frac{3}{2\mu_{\alpha}} \bigg(\frac{1}{K}\sum\limits_{k=1}^{K}[\nabla_{\alpha}f_k(\bar{\v}_{t-1}, \bar{\alpha}_{t-1}) - \nabla_{\alpha}f_k(\bar{\v}_{t-1}, \alpha_{t-1}^k)] \bigg)^2 + \frac{\mu_{\alpha}}{6}(\bar{\alpha}_t - \alpha^*(\Tilde{\v}))^2\\
&\overset{(b)}\leq \frac{3}{2 \mu_{\alpha}} \frac{1}{K}\sum\limits_{k=1}^{K}(\nabla_{\alpha}f_k(\bar{\v}_{t-1}, \bar{\alpha}_{t-1}) - \nabla_{\alpha}f_k(\bar{\v}_{t-1}, \alpha_{t-1}^k))^2 + \frac{\mu_{\alpha}}{6}(\bar{\alpha}_t - \alpha^*(\Tilde{\v}))^2\\
&\overset{(c)}\leq \frac{3 L_{\alpha}^2}{2\mu_{\alpha}} \frac{1}{K}\sum\limits_{k=1}^{K}(\bar{\alpha}_{t-1} - \alpha_{t-1}^k)^2 + \frac{\mu_{\alpha}}{6}(\bar{\alpha}_t - \alpha^*(\Tilde{\v}))^2,\\ 
\end{split}
\label{circled_4} 
\end{equation}
where (a) follows from Young's inequality, (b) follows from Jensen's inequality, and 
(c) holds because $f_k(\v, \alpha)$ is smooth in $\alpha$ with coefficient $L_{\alpha} = 2p(1-p)$ for any $\v$ by definition of $f_k(\cdot)$.

\begin{equation}
\begin{split}
\textcircled{\small{5}} &\overset{(a)}\leq \frac{3}{2\mu_{\alpha}}\bigg\|\frac{1}{K}\sum\limits_{k=1}^{K}[\nabla_{\alpha}f_k(\bar{\v}_{t-1},\alpha_{t-1}^k) - \nabla_{\alpha}f_k(\v_{t-1}^k,\alpha_{t-1}^k)] \bigg\|^2 + \frac{\mu_{\alpha}}{6}(\alpha^*(\Tilde{\v})-\bar{\alpha}_t)^2 \\
&\overset{(b)}\leq \frac{3}{2\mu_{\alpha}} \frac{1}{K}\sum\limits_{k=1}^{K}\bigg\|\nabla_{\alpha}f_k(\bar{\v}_{t-1},\alpha_{t-1}^k) - \nabla_{\alpha}f_k(\v_{t-1}^k,\alpha_{t-1}^k)\bigg\|^2 + \frac{\mu_{\alpha}}{6}(\alpha^*(\Tilde{\v})-\bar{\alpha}_t)^2 \\
& \overset{(c)}\leq \frac{3G_{\v}^2}{2\mu_{\alpha}}\frac{1}{K}\sum\limits_{k=1}^{K}\|\bar{\v}_{t-1}-\v_{t-1}^{k}\|^2 + \frac{\mu_{\alpha}}{6}(\alpha^*(\Tilde{\v})-\bar{\alpha}_t)^2,
\end{split}
\label{circled_5}
\end{equation}
where (a) follows from Young's inequality, (b) follows from Jensen's inequality.
(c) holds because $\nabla_{\alpha} f_k(\v, \alpha)$ is Lipschitz in $\v$ with coefficient $G_{\v} = 2\max(p, 1-p) G_h$ by definition of $f_k(\cdot)$.

Let $\hat{\alpha}_t = \bar{\alpha}_{t-1} + \frac{\eta}{K} \sum\limits_{k=1}^{K}\nabla_{\alpha} f_k(\v^k_{t-1}, \alpha^k_{t-1})$. Then we have
\begin{equation}
\begin{split}
\bar{\alpha}_t - \hat{\alpha}_t = \eta  \bigg(\frac{1}{K}\sum\limits_{k=1}^{K}\nabla_{\alpha} F_k(\v^k_{t-1}, \alpha^k_{t-1};\z_{t-1}^k) - \nabla_{\alpha} f_k(\v^k_{t-1}, \alpha^k_{t-1}) \bigg).
\end{split}
\end{equation}

And for the auxiliary sequence $\tilde{\alpha}_t$, we can verify that 
\begin{small} 
\begin{align} 
\begin{split} 
\tilde{\alpha}_{t} = \arg\min\limits_{\alpha}  \bigg(\frac{1}{K}\sum\limits_{k=1}^{K}(
 \nabla_{\alpha} F_k(\v_{t-1}^{k},\alpha_{t-1}^k; \z_{t-1}^k) 
-\nabla_{\alpha}  f_k(\v_{t-1}^{k},\alpha_{t-1}^k))\bigg)^T \alpha + \frac{1}{2\eta}(\alpha - \tilde{\alpha}_{t-1})^2 := \lambda_{t-1}(\alpha).
\end{split}
\end{align}
\end{small}

Since $\lambda_{t-1}(\alpha)$ is $\frac{1}{\eta}$-strongly convex, we have
\begin{small}
\begin{align} 
\begin{split}
& \frac{1}{2}(\alpha^*(\tilde{\v}) - \tilde{\alpha}_{t})^2 \leq \lambda_{t-1}(\alpha^*(\tilde{\v})) - \lambda_{t-1}(\tilde{\alpha}_{t}) \\
& = \bigg(\frac{1}{K}\sum\limits_{k=1}^{K}( 
 \nabla_{\alpha} F_k(\v_{t-1}^{k},\alpha_{t-1}^k; \z_{t-1}^k) 
-\nabla_{\alpha} f_k(\v_{t-1}^{k},\alpha_{t-1}^k))\bigg)^T \alpha^*(\tilde{\v}) + \frac{1}{2\eta}(\alpha^*(\tilde{\v}) - \tilde{\alpha}_{t-1})^2 \\ 
&~~~ -\bigg(\frac{1}{K}\sum\limits_{k=1}^{K}( 
 \nabla_{\alpha} F_k(\v_{t-1}^{k},\alpha_{t-1}^k; \z_{t-1}^k) 
-\nabla_{\alpha} f_k(\v_{t-1}^{k},\alpha_{t-1}^k))\bigg)^T \tilde{\alpha}_{t} 
- \frac{1}{2\eta}(\tilde{\alpha}_{t}  - \tilde{\alpha}_{t-1})^2 \\ 
& = \bigg(\frac{1}{K}\sum\limits_{k=1}^{K}(  
 \nabla_{\alpha} F_k(\v_{t-1}^{k},\alpha_{t-1}^k; \z_{t-1}^k) 
-\nabla_{\alpha} f_k(\v_{t-1}^{k},\alpha_{t-1}^k))\bigg)^T (\alpha^*(\tilde{\v}) - \tilde{\alpha}_{t-1}) 
+ \frac{1}{2\eta}(\alpha^*(\tilde{\v}) - \tilde{\alpha}_{t-1})^2 \\
&~~~ -\bigg(\frac{1}{K} \sum\limits_{k=1}^{K}(  
 \nabla_{\alpha} F_k(\v_{t-1}^{k},\alpha_{t-1}^k; \z_{t-1}^k) 
-\nabla_{\alpha} f_k(\v_{t-1}^{k},\alpha_{t-1}^k))\bigg)^T 
(\tilde{\alpha}_{t} - \tilde{\alpha}_{t-1})  
- \frac{1}{2\eta}(\tilde{\alpha}_{t}  - \tilde{\alpha}_{t-1})^2 \\
&\leq \bigg(\frac{1}{K}\sum\limits_{k=1}^{K}(
 \nabla_{\alpha} F_k(\v_{t-1}^{k},\alpha_{t-1}^k; \z_{t-1}^k) 
-\nabla_{\alpha} f_k(\v_{t-1}^{k},\alpha_{t-1}^k))\bigg)^T (\alpha^*(\tilde{\v}) - \tilde{\alpha}_{t-1})+ \frac{1}{2\eta}(\alpha^*(\tilde{\v}) - \tilde{\alpha}_{t-1})^2 \\ 
&~~~+ \frac{\eta}{2}\bigg(\frac{1}{K}\sum\limits_{k=1}^{K}(
 \nabla_{\alpha} F_k(\v_{t-1}^{k},\alpha_{t-1}^k; \z_{t-1}^k) 
-\nabla_{\alpha} f_k(\v_{t-1}^{k},\alpha_{t-1}^k))\bigg)^2.
\end{split} 
\label{local:before_circled6_pre}
\end{align} 
\end{small}

Hence we get
\begin{equation}
\begin{split}
\textcircled{\small{6}}
&=\bigg\langle \frac{1}{K}\sum\limits_{k=1}^{K}[\nabla_{\alpha} f_k(\v_{t-1}^k, \alpha_{t-1}^k) - \nabla_{\alpha}F_k (\v_{t-1}^k, \alpha_{t-1}^k; \z_{t-1}^k)], \hat{\alpha}_t -\bar{\alpha}_t \bigg\rangle \\
&~~~~+\bigg\langle \frac{1}{K}\sum\limits_{k=1}^{K}[\nabla_{\alpha} f_k(\v_{t-1}^k, \alpha_{t-1}^k) - \nabla_{\alpha}F_k (\v_{t-1}^k, \alpha_{t-1}^k; \z_{t-1}^k)], \alpha^*(\Tilde{\v}) - \hat{\alpha}_t \bigg\rangle \\
&=\eta \bigg(\frac{1}{K}\sum\limits_{k=1}^{K}[\nabla_{\alpha} f_k(\v_{t-1}^k, \alpha_{t-1}^k) - \nabla_{\alpha}F_k (\v_{t-1}^k, \alpha_{t-1}^k; \z_{t-1}^k)]\bigg)^2\\
&~~~~+ \bigg\langle \frac{1}{K}\sum\limits_{k=1}^{K}[\nabla_{\alpha} f_k(\v_{t-1}^k, \alpha_{t-1}^k) - \nabla_{\alpha}F_k (\v_{t-1}^k, \alpha_{t-1}^k; \z_{t-1}^k)], \alpha^*(\Tilde{\v}) - \hat{\alpha}_t\bigg\rangle.
\end{split}
\label{circled_6_pre}
\end{equation}

Combining (\ref{local:before_circled6_pre}) and (\ref{circled_6_pre}), we get
\begin{small}  
\begin{align}
\begin{split}
\textcircled{\small{6}} \leq& \frac{3\eta}{2} \bigg(\frac{1}{K}\sum\limits_{k=1}^{K}[\nabla_{\alpha} f_k(\v_{t-1}^k, \alpha_{t-1}^k) - \nabla_{\alpha}F_k (\v_{t-1}^k, \alpha_{t-1}^k; \z_{t-1}^k)]\bigg)^2\\
& + \bigg\langle \frac{1}{K}\sum\limits_{k=1}^{K}[\nabla_{\alpha} f_k(\v_{t-1}^k, \alpha_{t-1}^k) - \nabla_{\alpha}F_k (\v_{t-1}^k, \alpha_{t-1}^k; \z_{t-1}^k)], 
\tilde{\alpha}_{t-1} - \hat{\alpha}_t\bigg\rangle \\
& + \frac{1}{2\eta}(\alpha^*(\tilde{\v}) - \tilde{\alpha}_{t-1})^2 
- \frac{1}{2\eta}(\alpha^*(\tilde{\v}) - \tilde{\alpha}_{t})^2.
\end{split}
\label{circled_6}
\end{align}
\end{small}

\textcircled{\small{7}} can be bounded as
\begin{equation}
\begin{split}
\textcircled{\small{7}} &= \langle \bar{\alpha}_t - \bar{\alpha}_{t-1}, \alpha^*(\Tilde{\v}) - \bar{\alpha}_t\rangle  =\frac{1}{2\eta}((\bar{\alpha}_{t-1}-\alpha^*(\Tilde{\v}))^2 - (\bar{\alpha}_{t-1} - \bar{\alpha}_t)^2 - (\bar{\alpha}_t-\alpha^*(\Tilde{\v}))^2 ).
\end{split} 
\label{circled_7}
\end{equation}

Adding (\ref{circled_4}), (\ref{circled_5}), (\ref{circled_6}) and (\ref{circled_7}), we get 
\begin{equation*}
\begin{split}
\langle \nabla_{\alpha} f(\bar{\v}_{t-1}, \bar{\alpha}_{t-1}), \alpha^*(\Tilde{\v}) - \bar{\alpha}_t \rangle &\leq \frac{3G_{\v}^2}{2\mu_{\alpha}} \frac{1}{K}\sum\limits_{k=1}^{K}\|\bar{\v}_{t-1}-\v_{t-1}^k\|^2 + \frac{3L_{\alpha}^2}{2\mu_{\alpha}}\frac{1}{K}\sum\limits_{k=1}^{K} (\bar{\alpha}_{t-1} - \alpha_{t-1}^k)^2\\
&~~~+\frac{3\eta}{2} \left( \frac{1}{K} \sum\limits_{k=1}^{K}[ \nabla_{\alpha} f_k(\v_{t-1}^k, \alpha_{t-1}^k) -  \nabla_{\alpha} F_k (\v_{t-1}^k, \alpha_{t-1}^k; \z_{t-1})] \right)^2\\
&~~~+\!\frac{1}{K}\sum\limits_{k=1}^{K}\!\langle\! \nabla_{\alpha}\! f_k(\v_{t-1}^k,\!\alpha_{t-1}^k)\!- \! \nabla_{\alpha}\!F_k(\v_{t-1}^k,\!\alpha_{t-1}^k;\!\z_{t-1}^k),\tilde{\alpha}_{t-1}-\!\hat{\alpha}_t\!\rangle\\ 
&~~~+\frac{1}{2\eta} ((\bar{\alpha}_{t-1} - \alpha^*(\Tilde{\v}))^2 - (\bar{\alpha}_{t-1} - \bar{\alpha}_t)^2 - (\bar{\alpha}_t - \alpha^*(\Tilde{\v})))^2)+ \frac{\mu_{\alpha}}{3} (\bar{\alpha}_t - \alpha^*(\Tilde{\v})))^2 \\
&~~~ + \frac{1}{2\eta}( (\alpha^*(\Tilde{\v}) - \Tilde{\alpha}_{t-1}) - (\alpha^*(\Tilde{\v}) - \Tilde{\alpha}_t)).
~~\Box 
\end{split} 
\label{grad_alpha}
\end{equation*}

\section{Proof of Lemma \ref{stich_lemma}}
\noindent \textit{Proof.}
If $I=1$, $\|\v_t^k - \bar{\v}_t^k \| = 0$ and $|\alpha_t^k - \bar{\alpha}_t^k| = 0$ for any iteration $t$ and any machine $k$ since $\v$ and $\alpha$ are averaged across machines at each iteration.

We prove the case when $I > 1$ in the following.
For any iteration $t$, there must be an iteration with index $t_0$ before $t$ such  that $t~ \text{mod}~ I =0 $ and $t - t_0 \leq I$. Since $\v$ and $\alpha$ are averaged across machines at $t_0$, we have $\bar{\v}_{t_0} = \v_{t_0}^k$.

(1) For $\v$, according to the update rule,
\begin{equation}
\begin{split}
\v_{t}^k = -\frac{\eta\gamma}{\eta + \gamma} \nabla_{\v} F_k(\v_{t-1}^k, \alpha_{t - 1}^k; \z_{t-1}^k)+\frac{\gamma}{\eta+\gamma}\v_{t-1}^k + \frac{\eta}{\eta+\gamma}\v_0,
\end{split}
\end{equation}
and hence
\begin{equation}
\begin{split}
\bar{\v}_{t} = -\frac{\eta\gamma}{\eta + \gamma}\frac{1}{K}\sum\limits_{k=1}^{K} \nabla_{\v} F_k(\v_{t-1}^k, \alpha_{t - 1}^k; \z_{t-1}^k)+\frac{\gamma}{\eta+\gamma}\bar{\v}_{t-1} + \frac{\eta}{\eta+\gamma}\v_0.
\end{split}
\end{equation}

Thus,
\begin{equation}
\begin{split}
\|\bar{\v}_{t} - \v_{t}^k\| &\leq \frac{\eta\gamma}{\eta+\gamma}\left\|\nabla_{\v} F_k(\v_{t-1}^k, \alpha_{t}^k; \z_{t}^k) - \frac{1}{K}\sum\limits_{i=1}^{K}\nabla_{\v}F_i(\v_{t-1}^i, \alpha_{t-1}^i; \z_{t-1}^i)\right\| + \frac{\gamma}{\eta + \gamma} \|\bar{\v}_{t-1} - \v_{t-1}^k\|\\
& \leq 2B_{\v} \frac{\eta \gamma}{\eta+\gamma} + \frac{\gamma}{\eta+\gamma}\|\bar{\v}_{t - 1} - \v_{t-1}^k\|.
\end{split}
\label{local_v_induction}
\end{equation}

Since $\bar{\v}_{t_0} = \v_{t_0}^k$ (for any $k$), we can see $\|\bar{\v}_{t_0+1} - \v_{t_0+1}^k\| \leq 2\frac{\eta\gamma}{\gamma + \eta} B_{\v} \leq 2\eta B_{\v}$,
Assuming $\|\bar{\v}_{t-1} - \v_{t-1}^k\| \leq 2(t-1-t_0)\eta B_{\v}$, then $\|\bar{\v}_{t} - \v_{t}^k\| \leq 2(t-t_0) \eta  B_{\v}$ by (\ref{local_v_induction}).
Thus, by induction, we know that for any $t$,
$\|\bar{\v}_{t} - \v_{t}^k\| \leq 2(t-t_0)\eta  B_{\v} \leq 2\eta I B_{\v}$.
Hence proved.

(ii) 
\begin{equation}
\alpha_t^k = \alpha_{t-1}^k + \eta \nabla_{\alpha} F_{k}(\v_{t-1}^k, \alpha_{t-1}^k; \z_{t-1}^k), 
\end{equation}
and 
\begin{equation}
\bar{\alpha}_t = \bar{\alpha}_{t-1} + \eta \frac{1}{K}\sum\limits_{k=1}^{K} \nabla_{\alpha} F_{k}(\v_{t-1}^k, \alpha_{t-1}^k; \z_{t-1}^k).
\end{equation}

Thus, 
\begin{equation}
\begin{split}
|\bar{\alpha}_t - \alpha_t^k| &\leq |\bar{\alpha}_{t-1}-\alpha_{t-1}^k| + \eta \left|\nabla_{\alpha}F_k(\v_{t-1}^k, \alpha_{t-1}^k; \z_{t-1}^k) - \frac{1}{K}\sum\limits_{i=1}^{K} \nabla_{\alpha} F_i(\v_{t-1}^i, \alpha_{t-1}^i; \z_{t-1}^i) \right|\\
&\leq |\bar{\alpha}_{t-1}-\alpha_{t-1}^k | + 2\eta B_{\alpha}.
\end{split}
\end{equation}

Since $\bar{\alpha}_{t_0} = \alpha_{t_0}^k$ (for any $k$), we can see that $\|\bar{\alpha}_{t_0+1} - \alpha_{t_0+1}^k\| \leq 2\eta B_{\alpha}$.
Assuming $|\bar{\alpha}_{t-1}-\alpha_{t-1}^k| \leq 2(t-1-t_0)\eta B_{\alpha}$, then $|\bar{\alpha}_t-\alpha_t^k|\leq 2(t-t_0)\eta B_{\alpha}$.
Thus, by induction, we know that for any $t$, $\|\bar{\alpha}_t-\alpha_t^k\|\leq 2(t-t_0)\eta B_{\alpha} \leq 2\eta I B_{\alpha}$.
Hence proved.
$\Box$

\section{More Experiments}
\begin{figure*}[htbp]
    \centering
    \subfigure[ImageNet]{
    \includegraphics[scale=0.25]{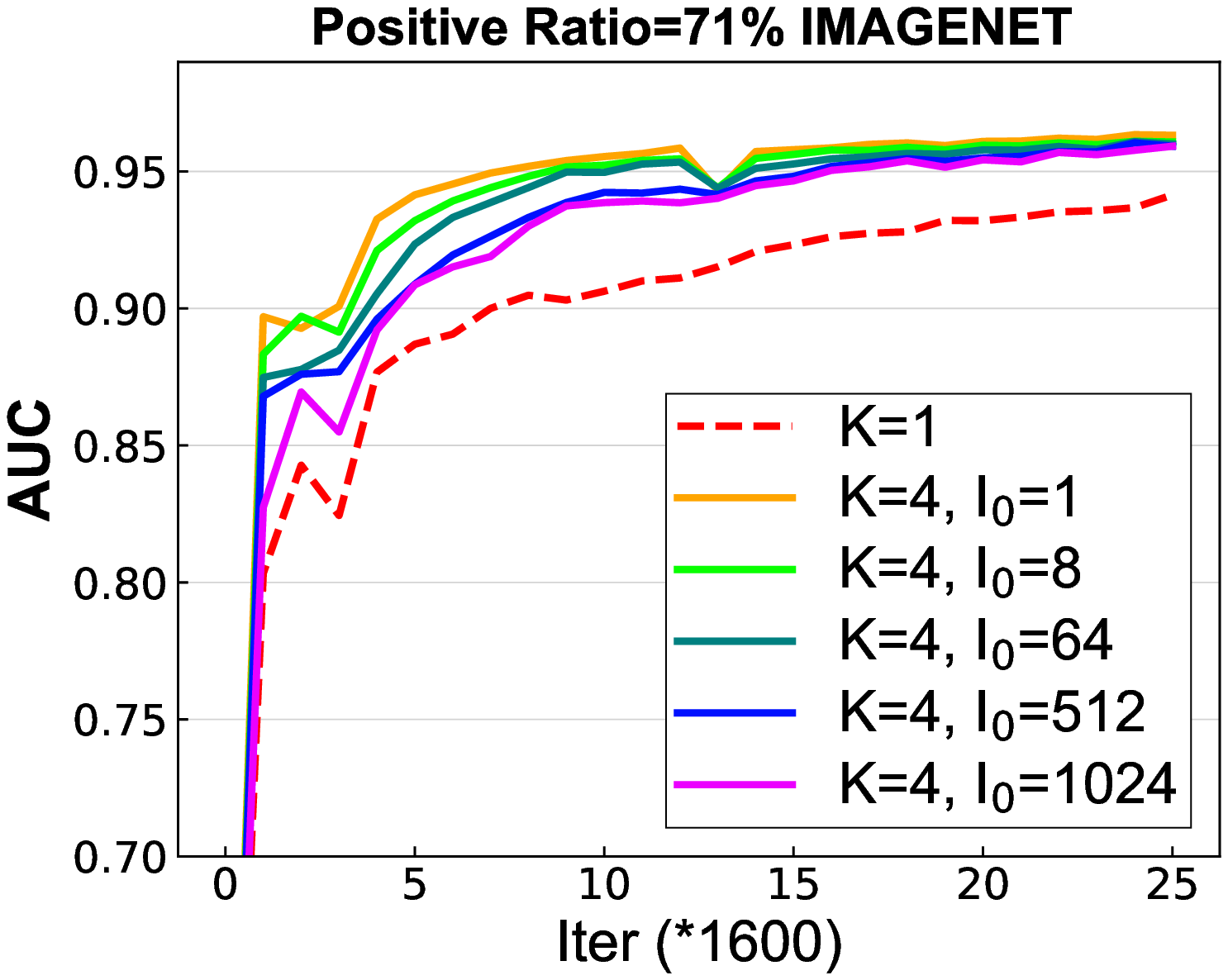}
    \includegraphics[scale=0.25]{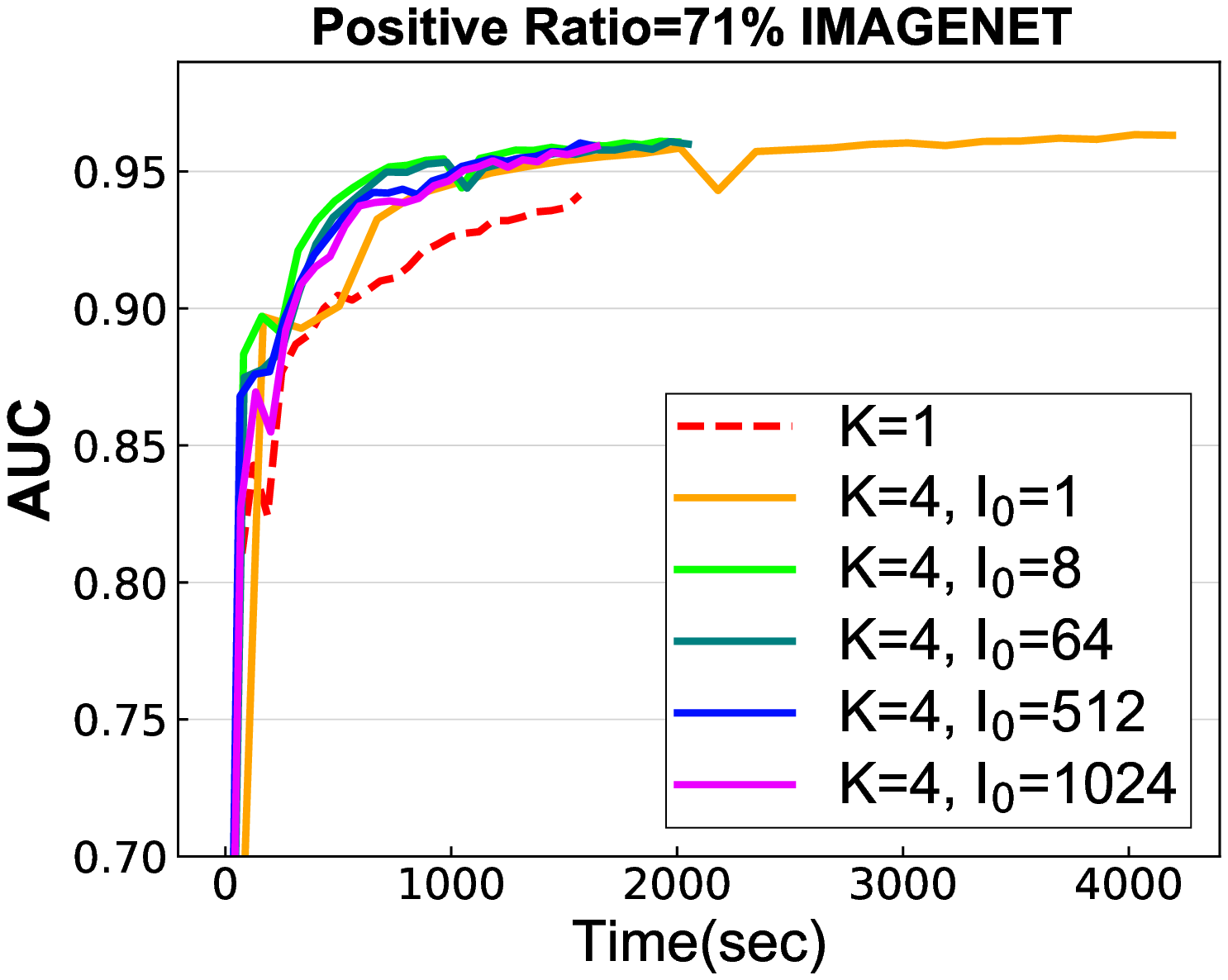}
    \vspace{-0.4cm}
    }
    \vspace{0.4cm}
     \subfigure[Cifar100]{
    \includegraphics[scale=0.25]{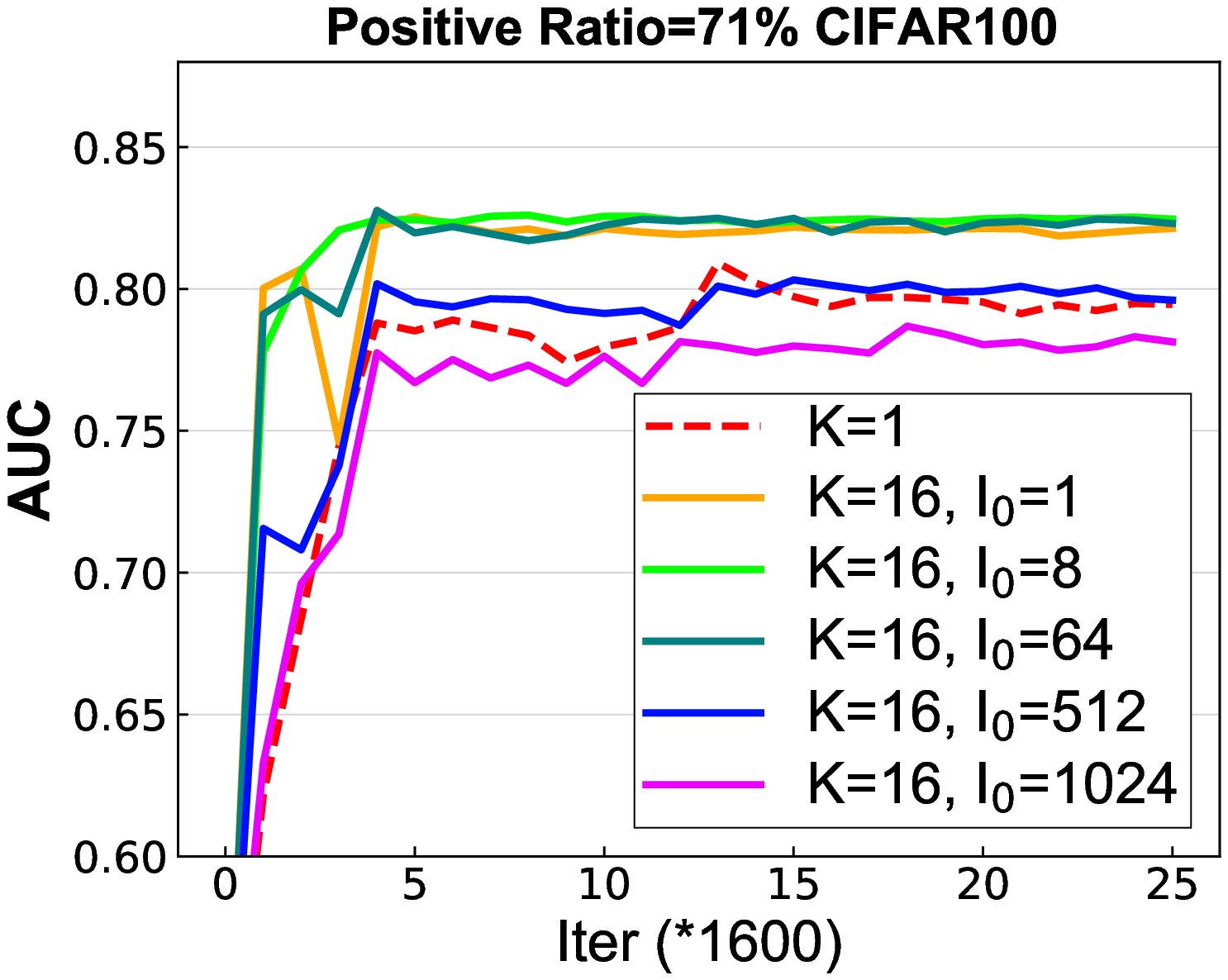}
    \includegraphics[scale=0.25]{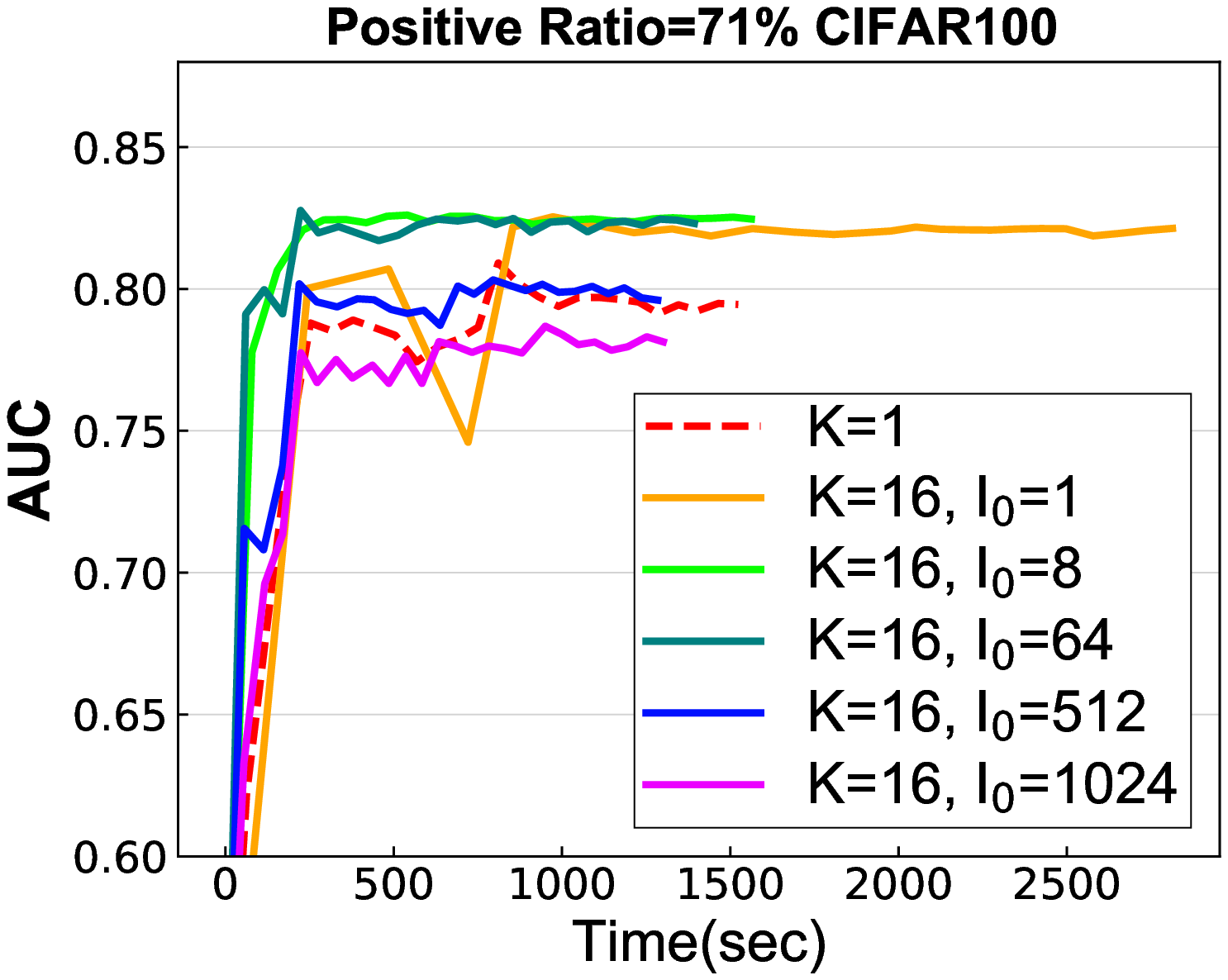}
    \vspace{-0.4cm}
    } 
    \subfigure[Cifar10]{
    \includegraphics[scale=0.25]{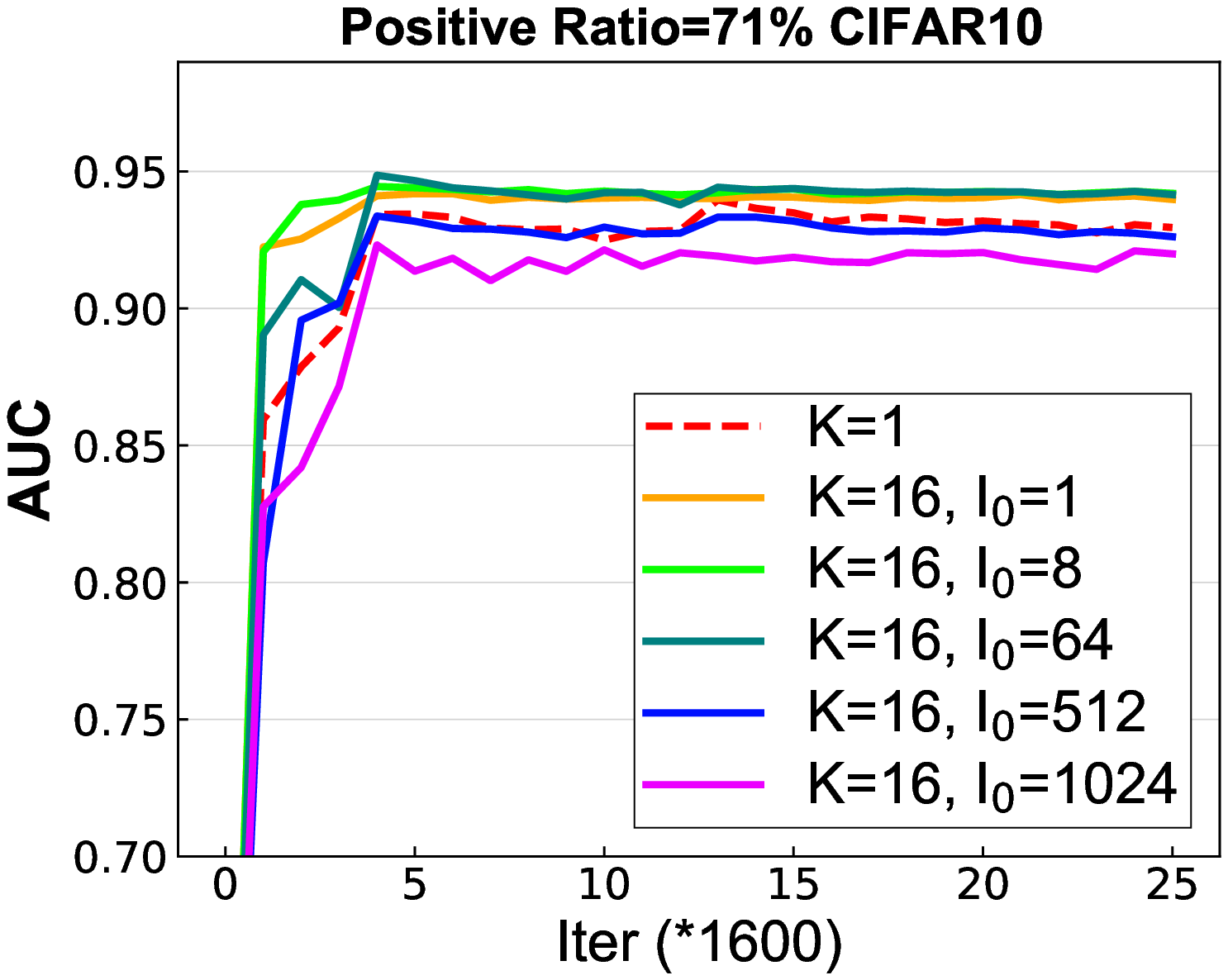}
    \includegraphics[scale=0.25]{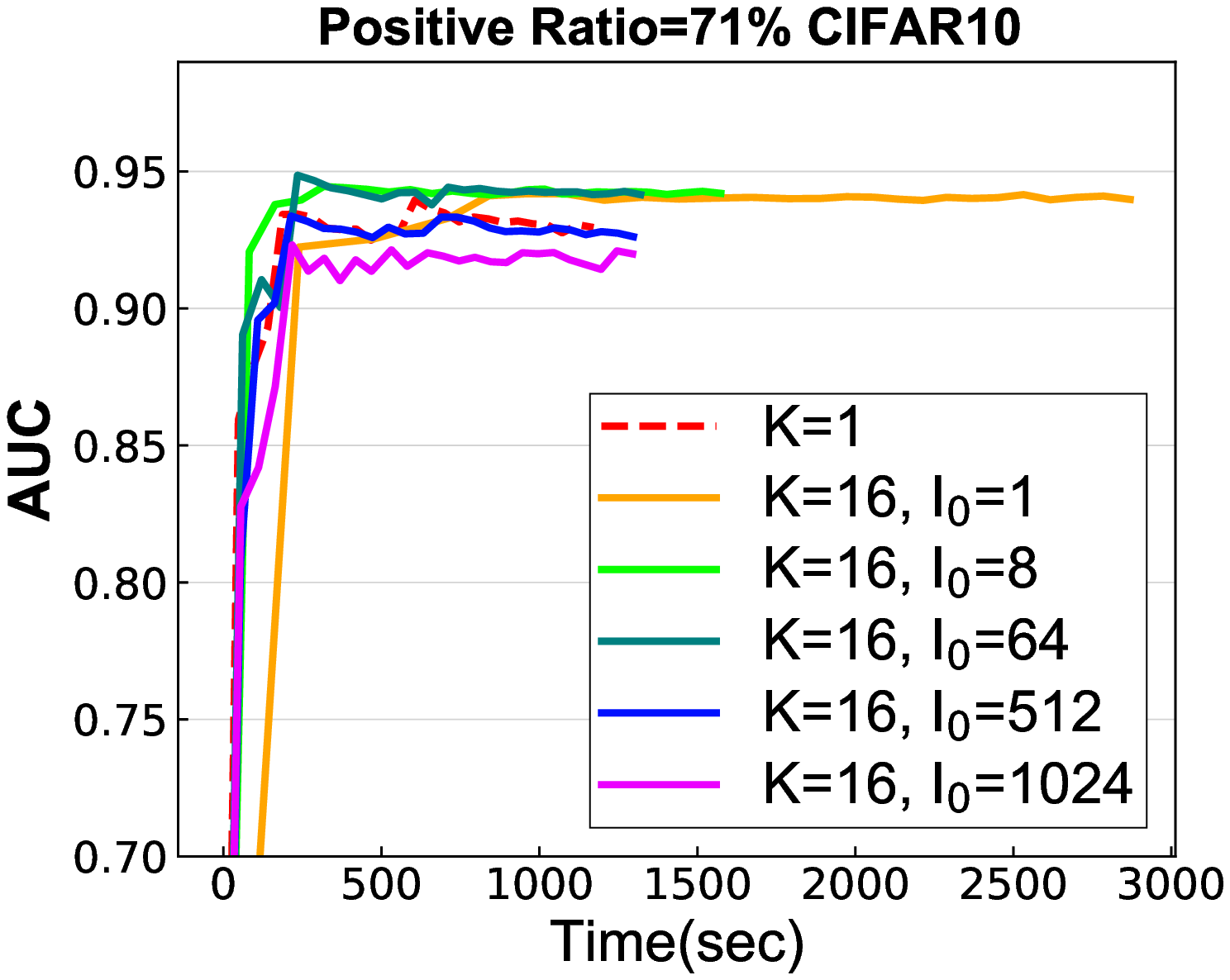}
    \vspace{-0.4cm}
    }
    \caption{$I_s = I_0 3^{(s-1)}$, positive ratio = 71\%.}
    \label{fig:thm_I}
\end{figure*}

In this section, we include more experimental results. 
Most of the settings are the same as in the Experiments section in the main paper, except that in Figure \ref{fig:thm_I}, we set $I = I_0 *3^{(s-1)}$, other than set $I$ to be a constant.
This means that a later stage will communicate less frequently since the step size is decreased after each stage (see the first remark of Theorem \ref{thm:main}).

\begin{figure*}[ht]
        \centering
    \subfigure[Fix $I$, vary $K$]
        {\includegraphics[scale=0.25]{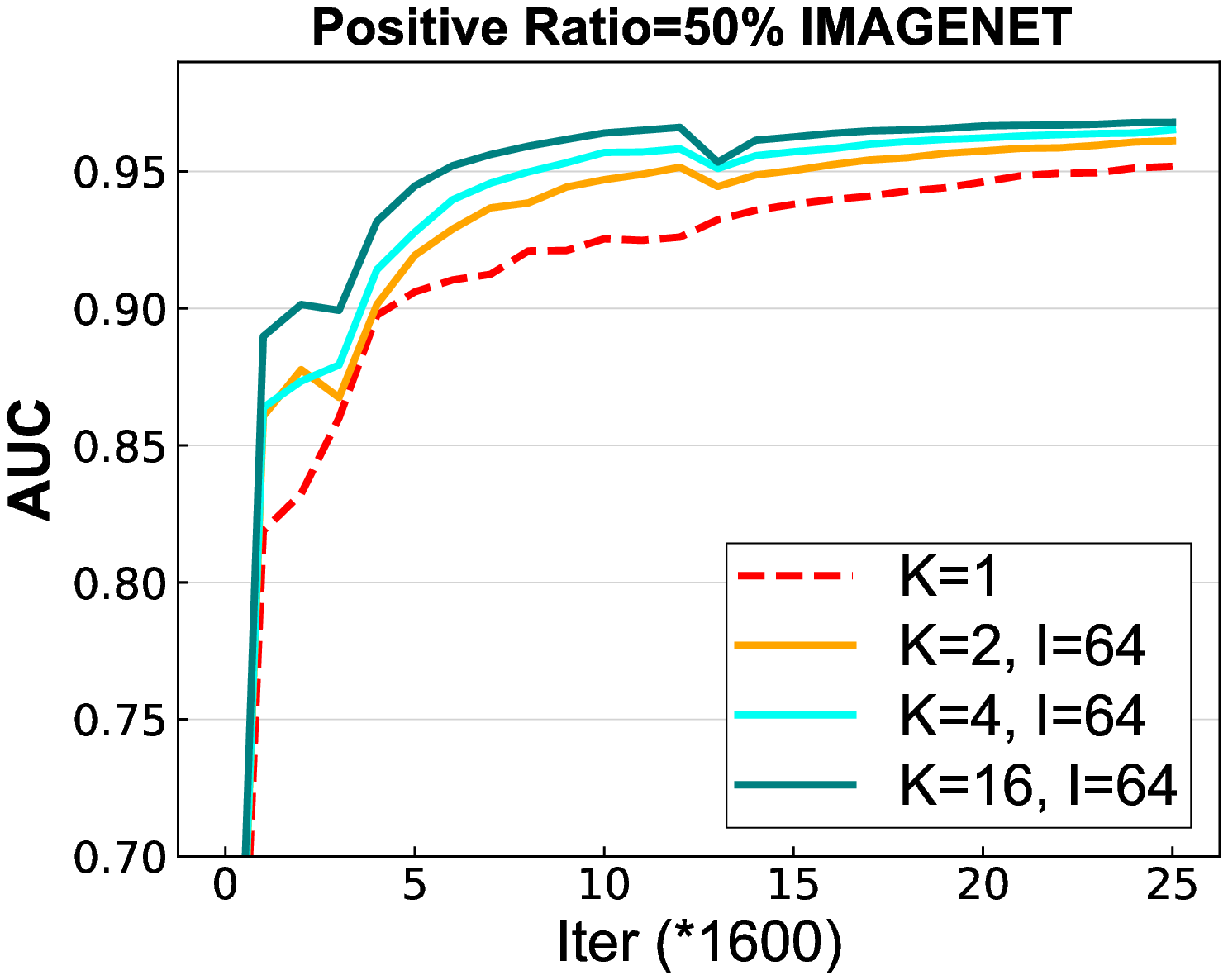}
    \includegraphics[scale=0.25]{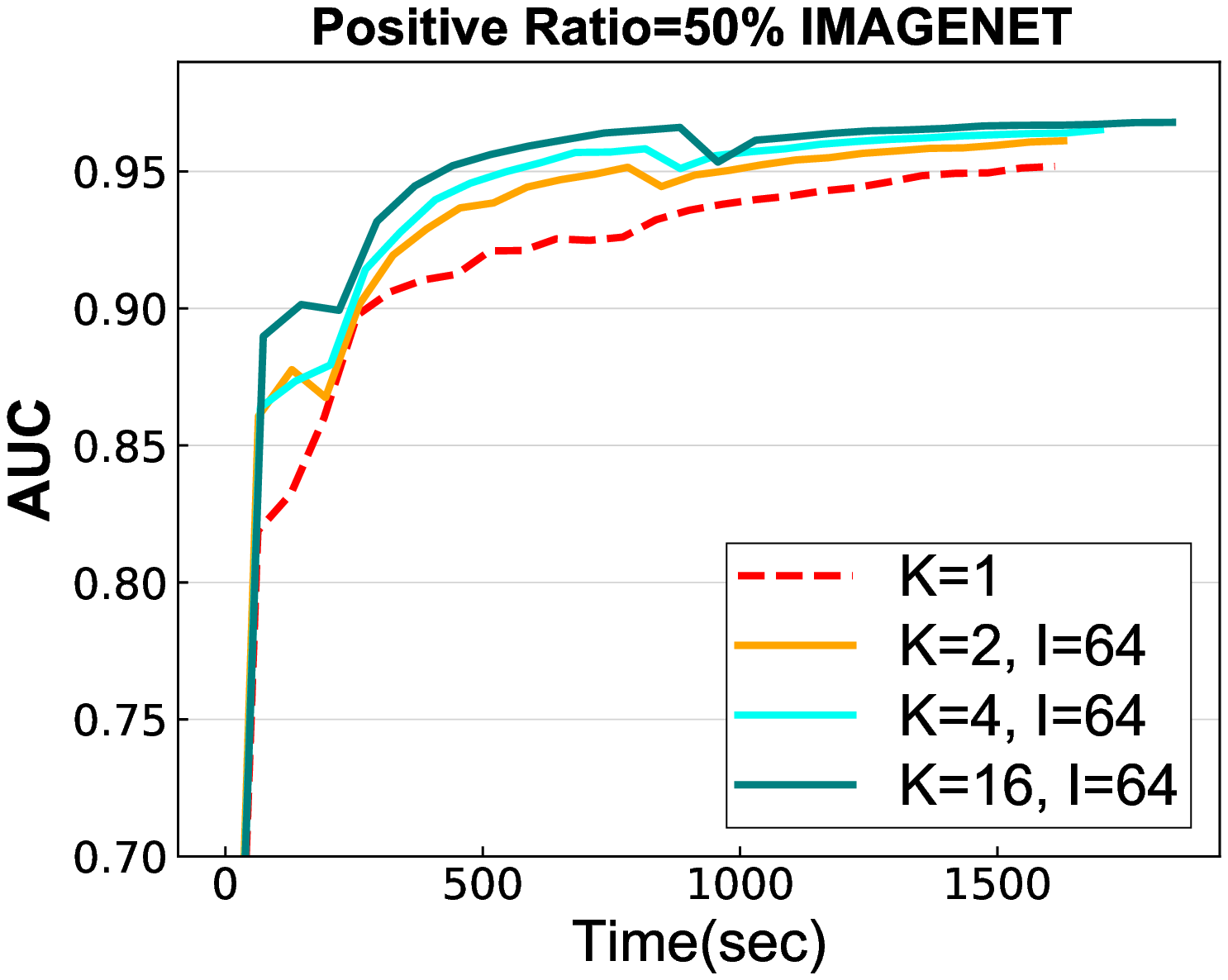}}
    \subfigure[Fix $K$, vary $I$]
    {\includegraphics[scale=0.25]{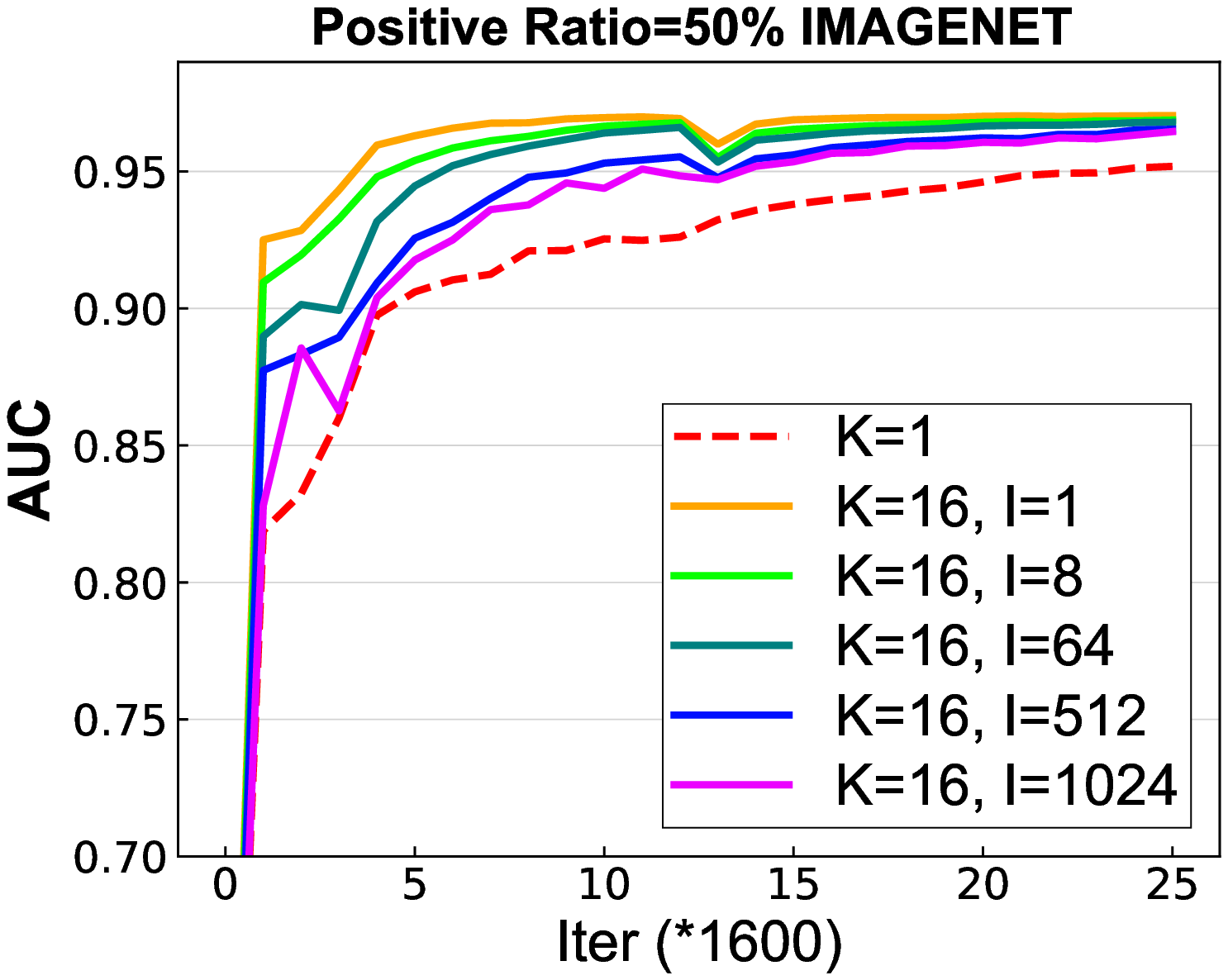}
    \includegraphics[scale=0.25]{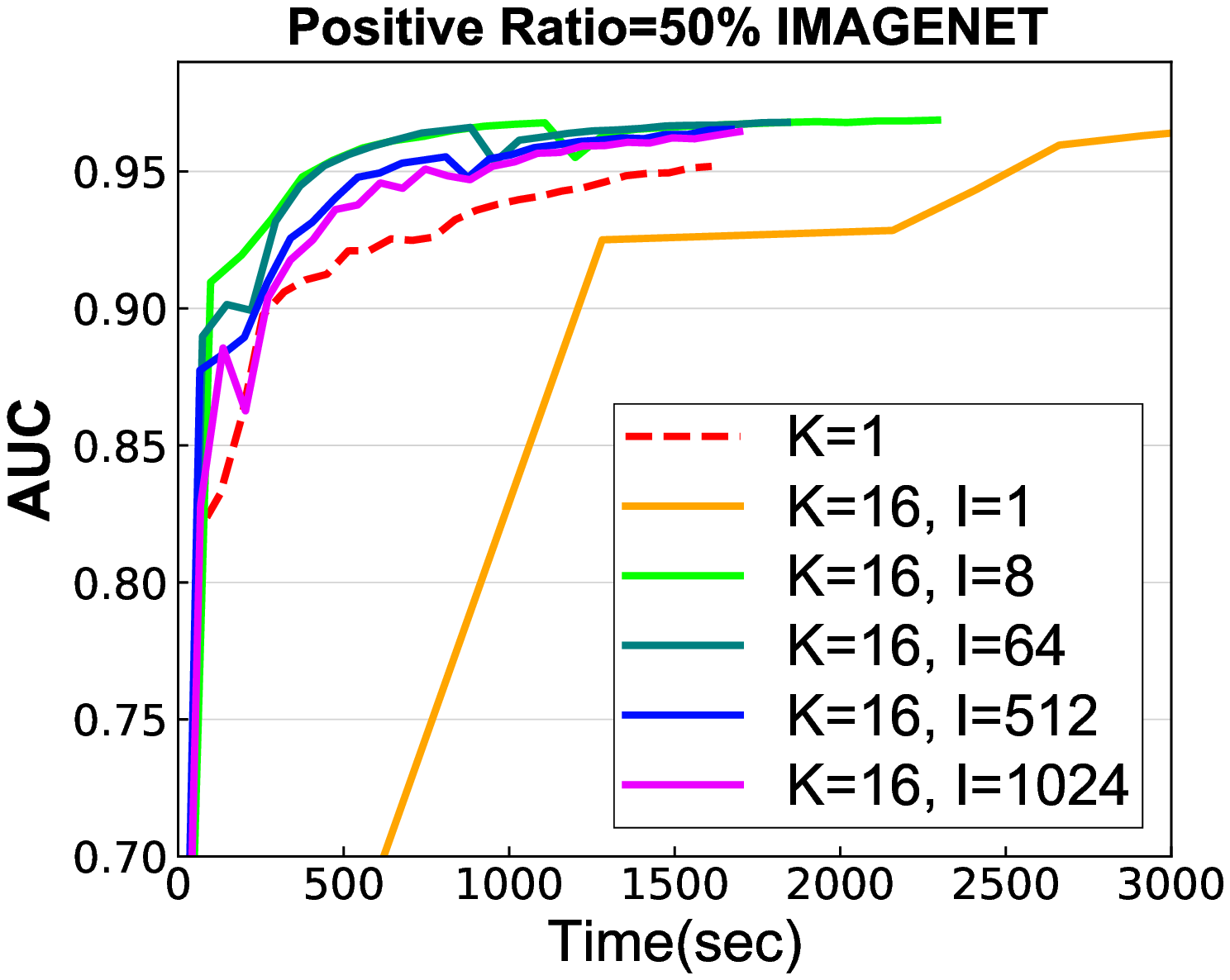}}
    \vspace{-0.4cm}
    \caption{ImageNet, positive ratio = 50\%.}
        \label{sup:fig:imagenet_71}
     \centering
    \subfigure[Fix $I$, vary $K$]
    {\includegraphics[scale=0.25]{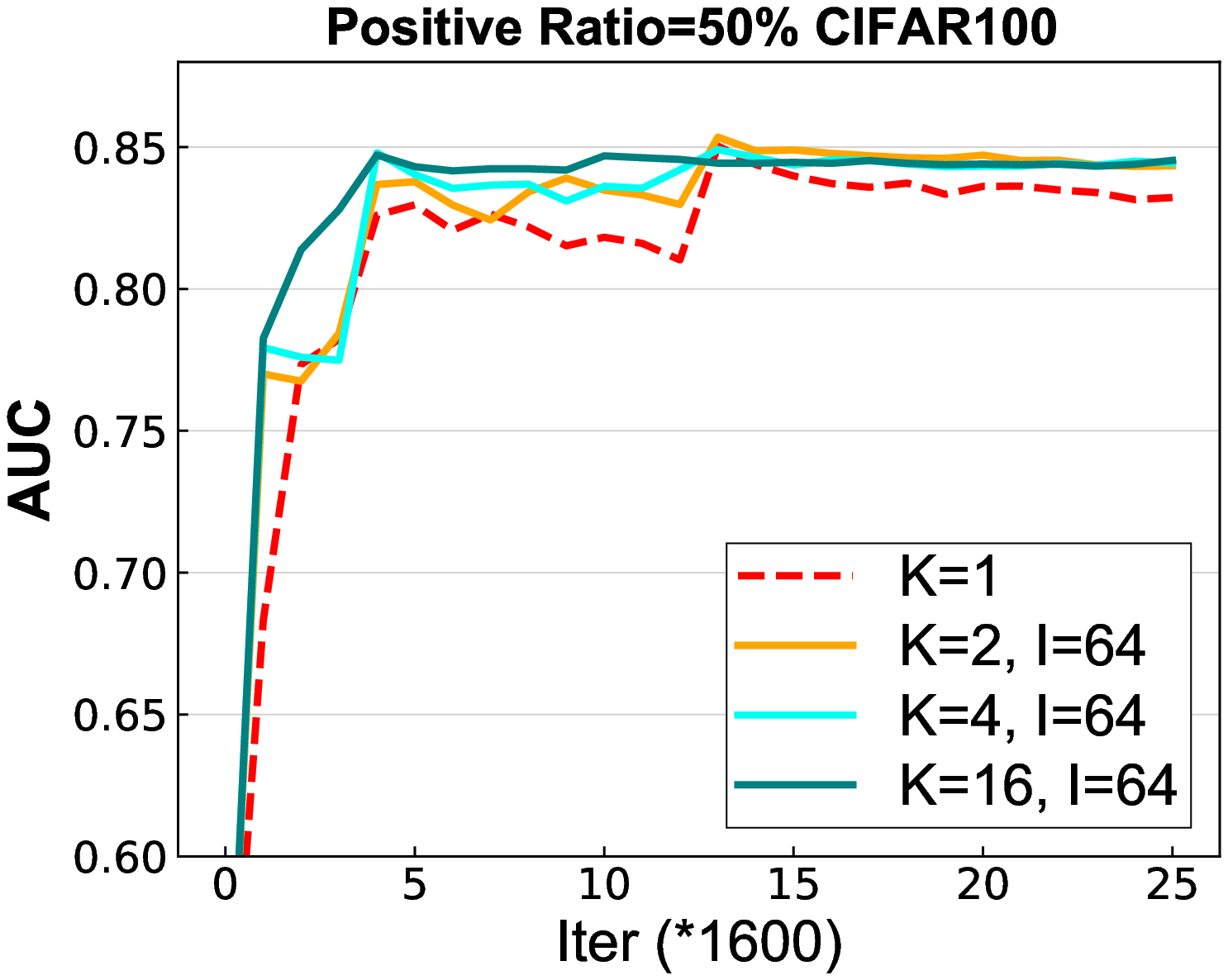}
    \includegraphics[scale=0.25]{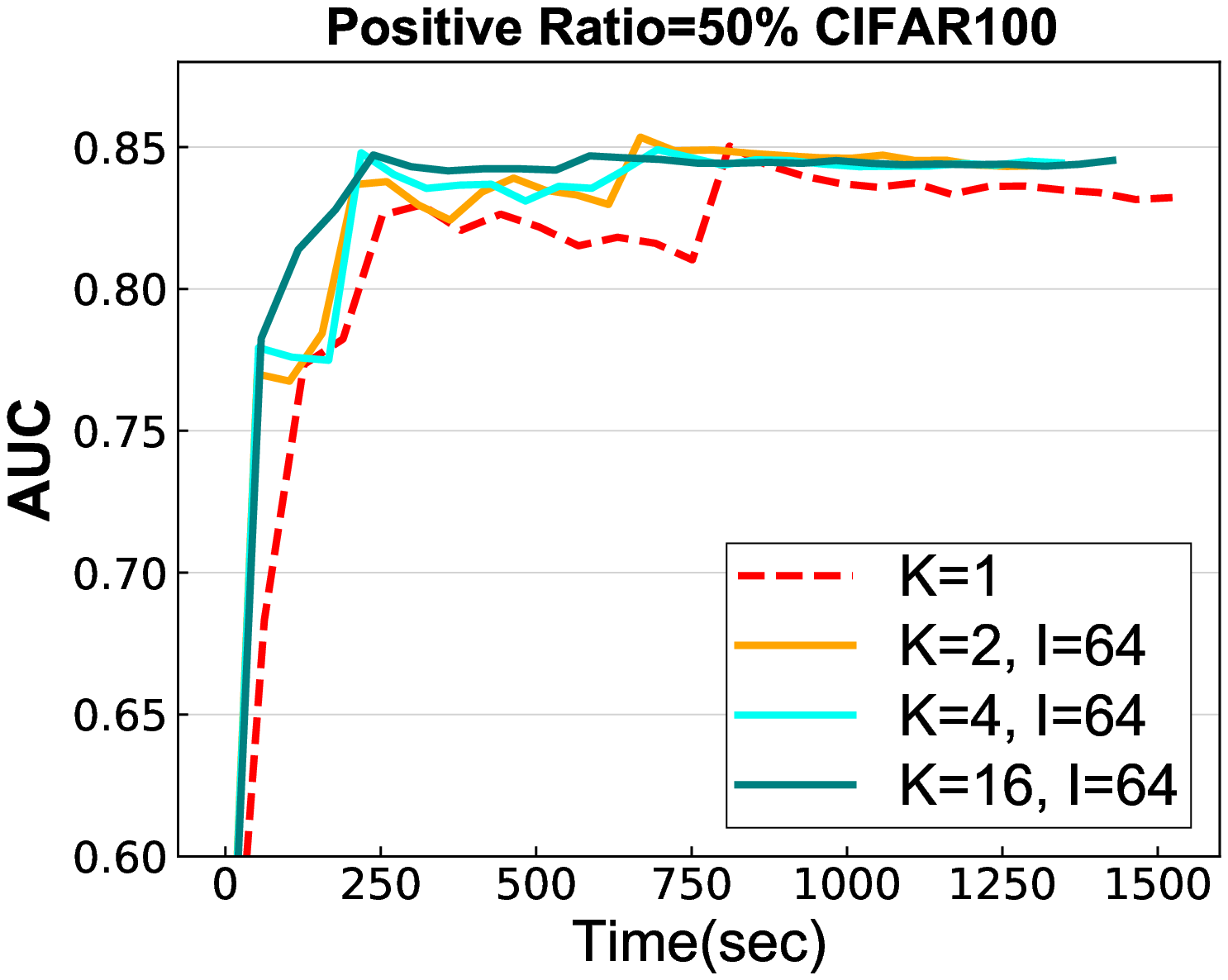}}
    \subfigure[Fix $K$, vary $I$]
    {\includegraphics[scale=0.25]{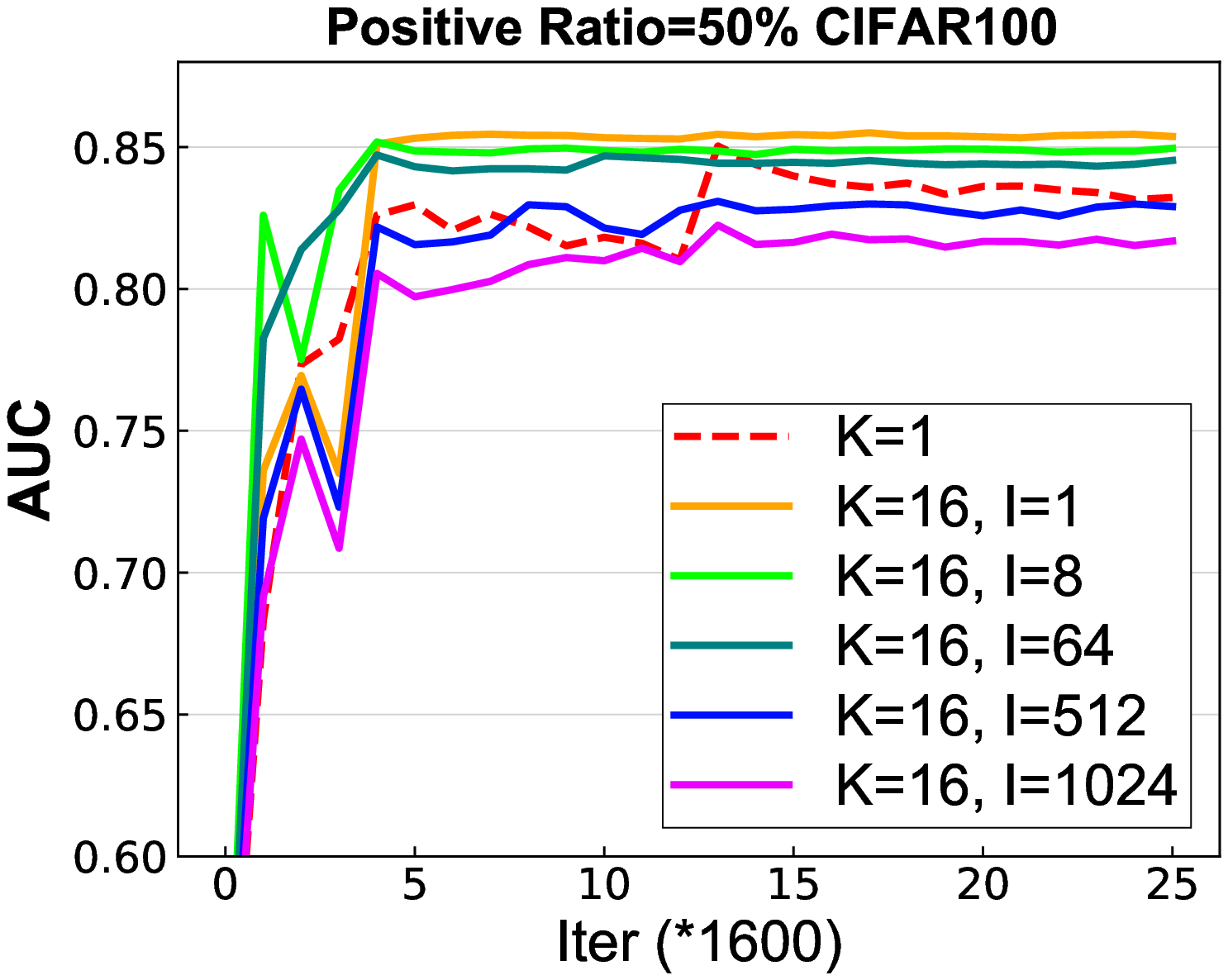}
    \includegraphics[scale=0.25]{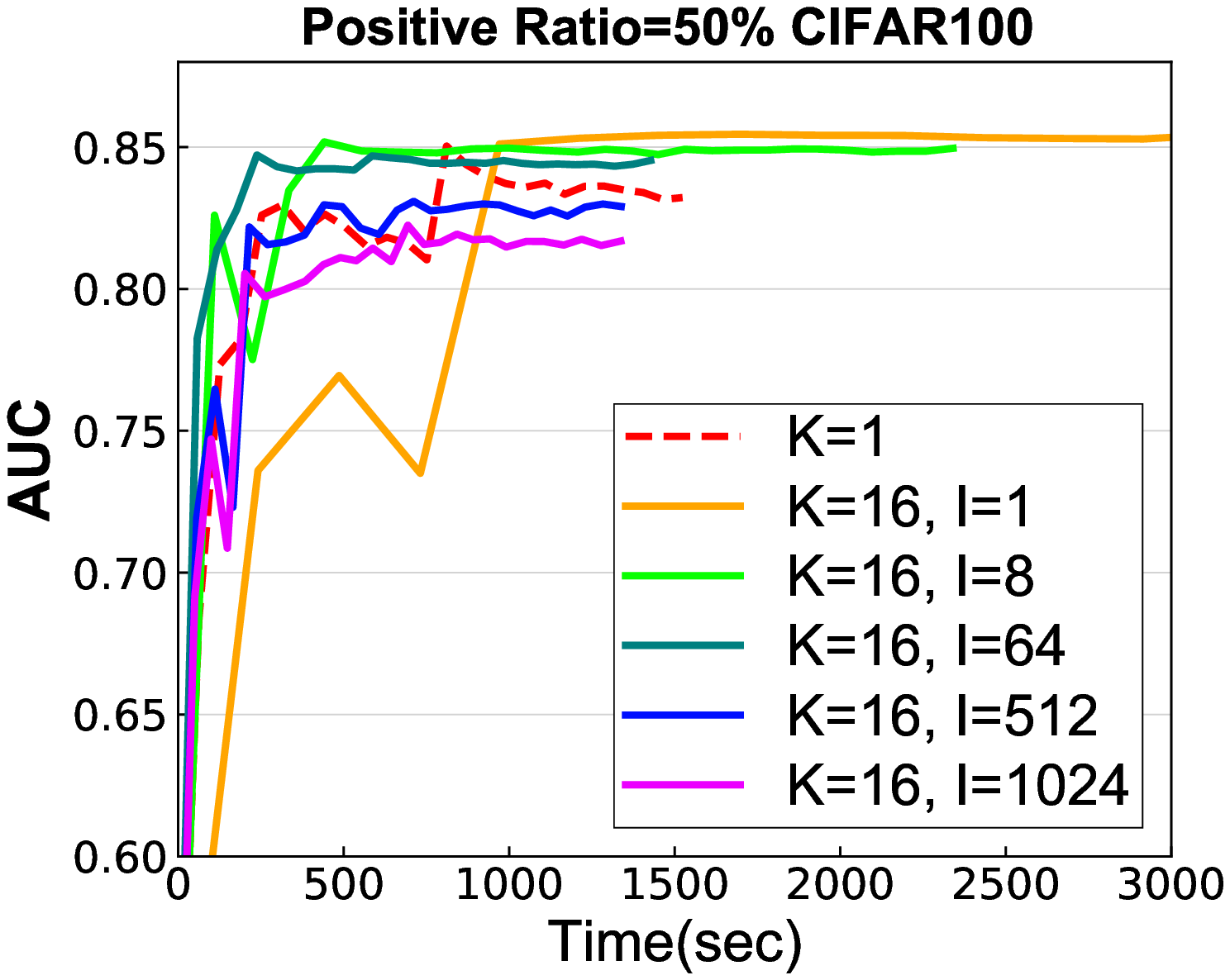}}
    \vspace{-0.4cm}
    \caption{Cifar100, positive ratio = 50\%.}
    \label{sup:fig:cifar100_71}
    \subfigure[Fix $I$, vary $K$] {\includegraphics[scale=0.25]{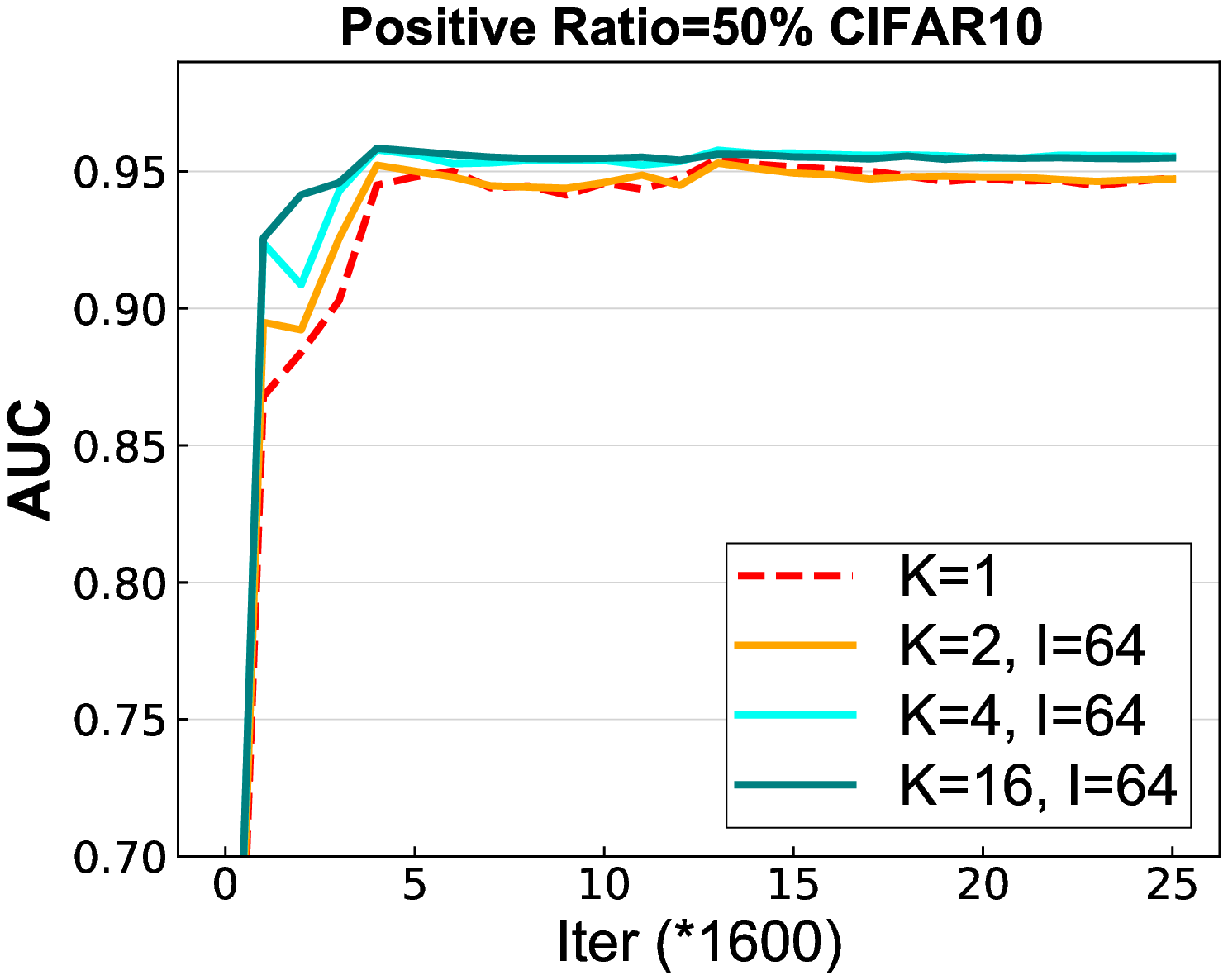}
    \includegraphics[scale=0.25]{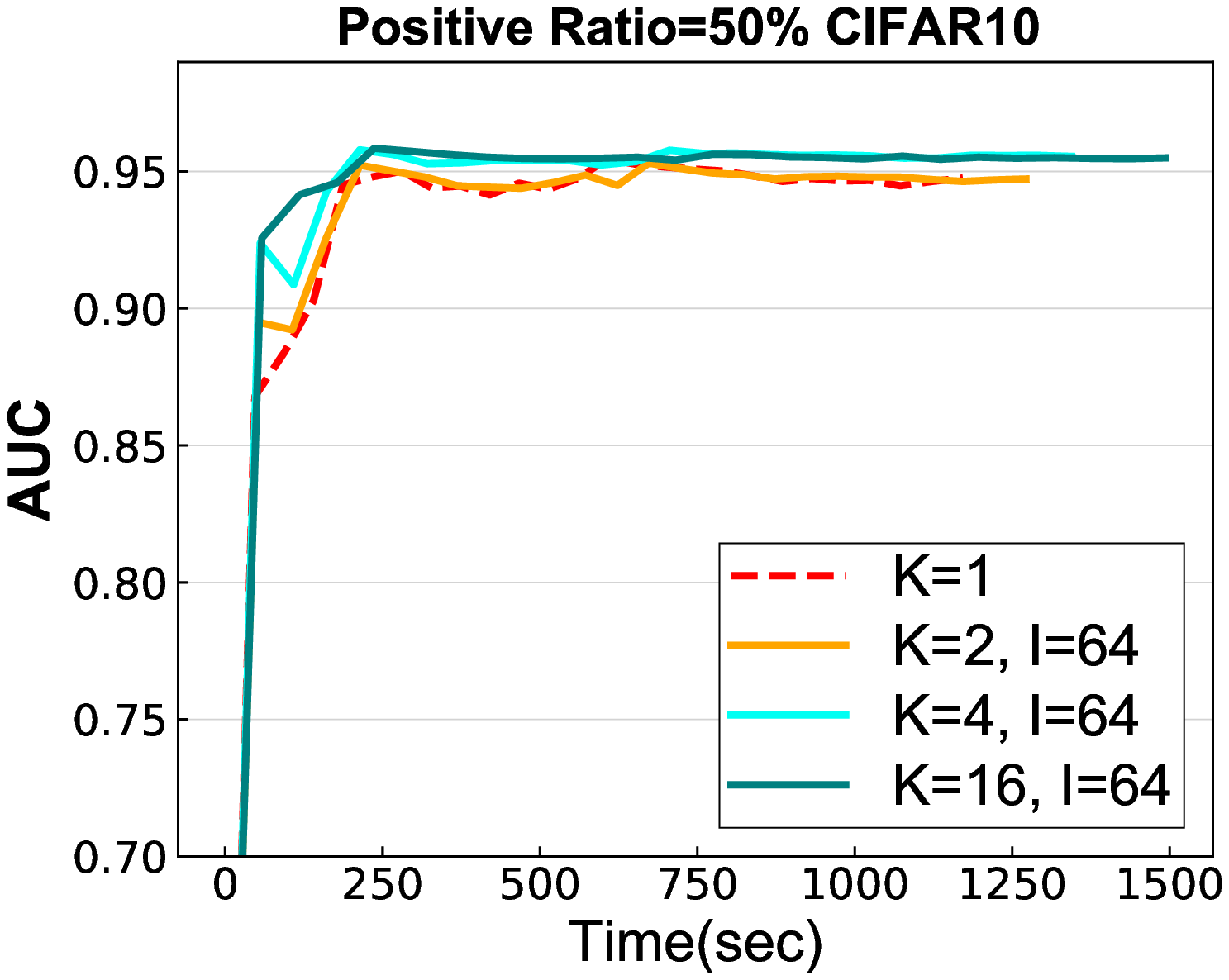}}
    \subfigure[Fix $K$, vary $I$]
    {\includegraphics[scale=0.25]{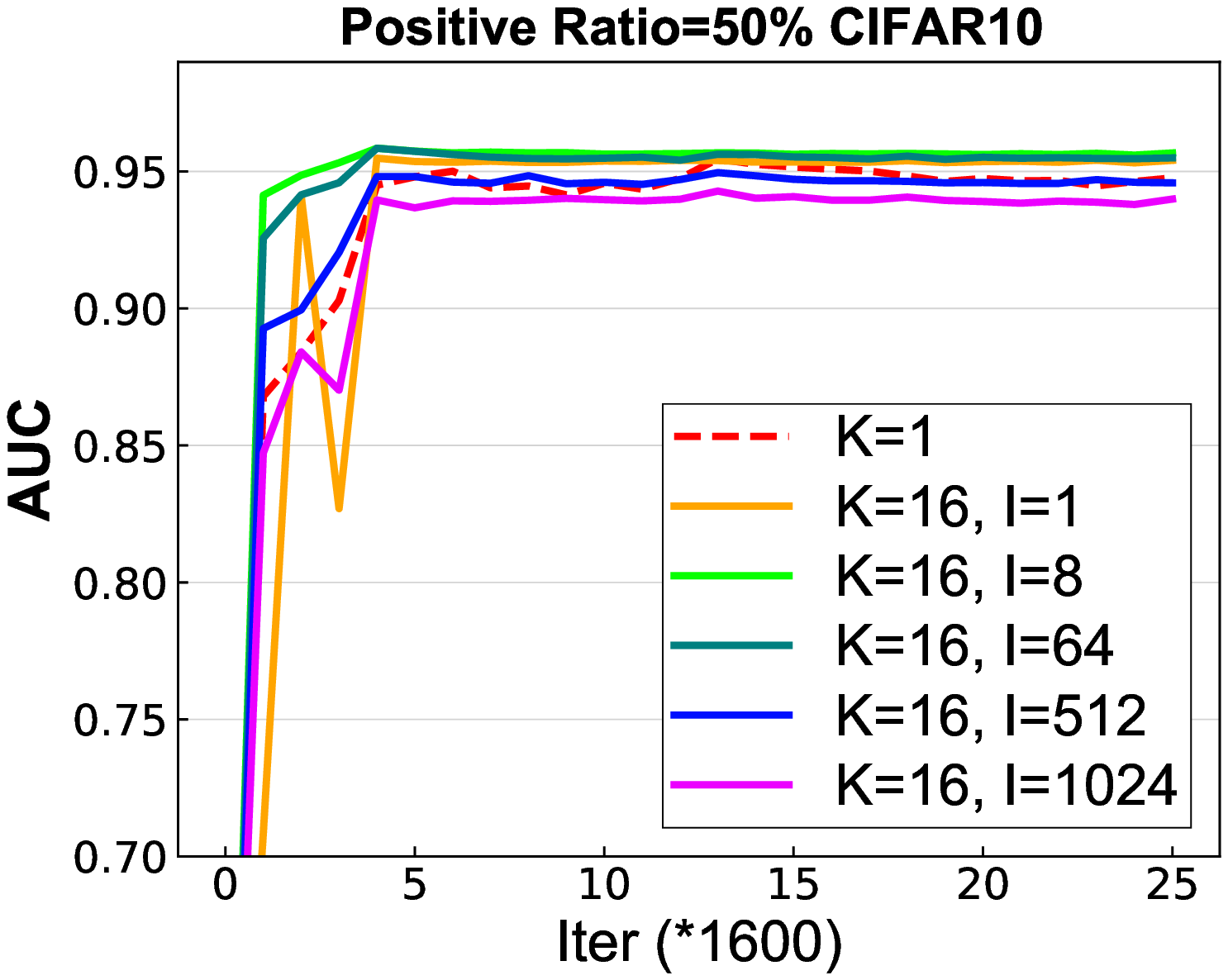}
    \includegraphics[scale=0.25]{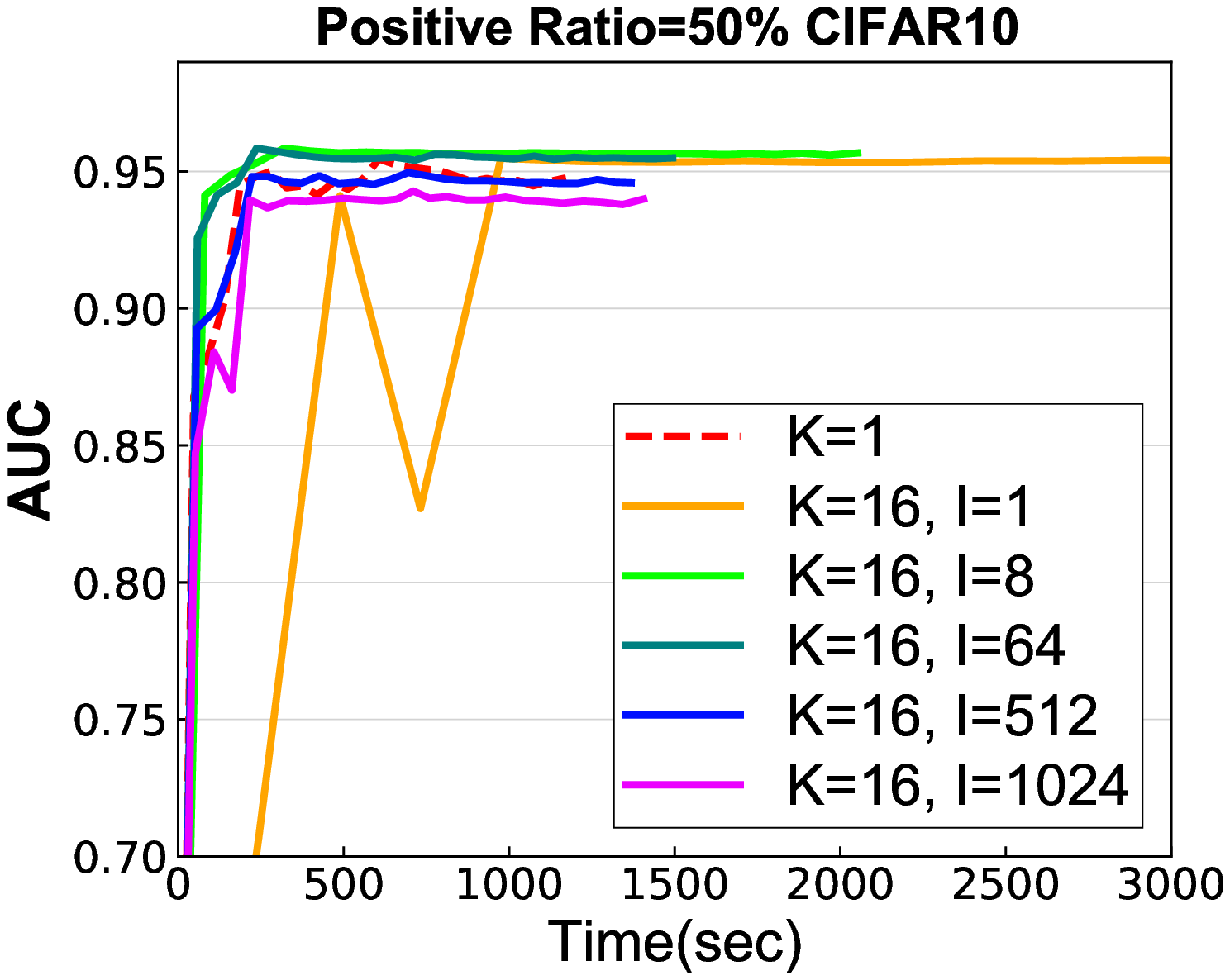}
    }
    \vspace{-0.4cm}
    \caption{Cifar10, positive ratio = 50\%.}
    \label{sub:fig:cifar10_71}
    \vspace{-0.15in} 
\end{figure*}

\begin{figure}[htbp]
    \centering
    \includegraphics[scale=0.23]{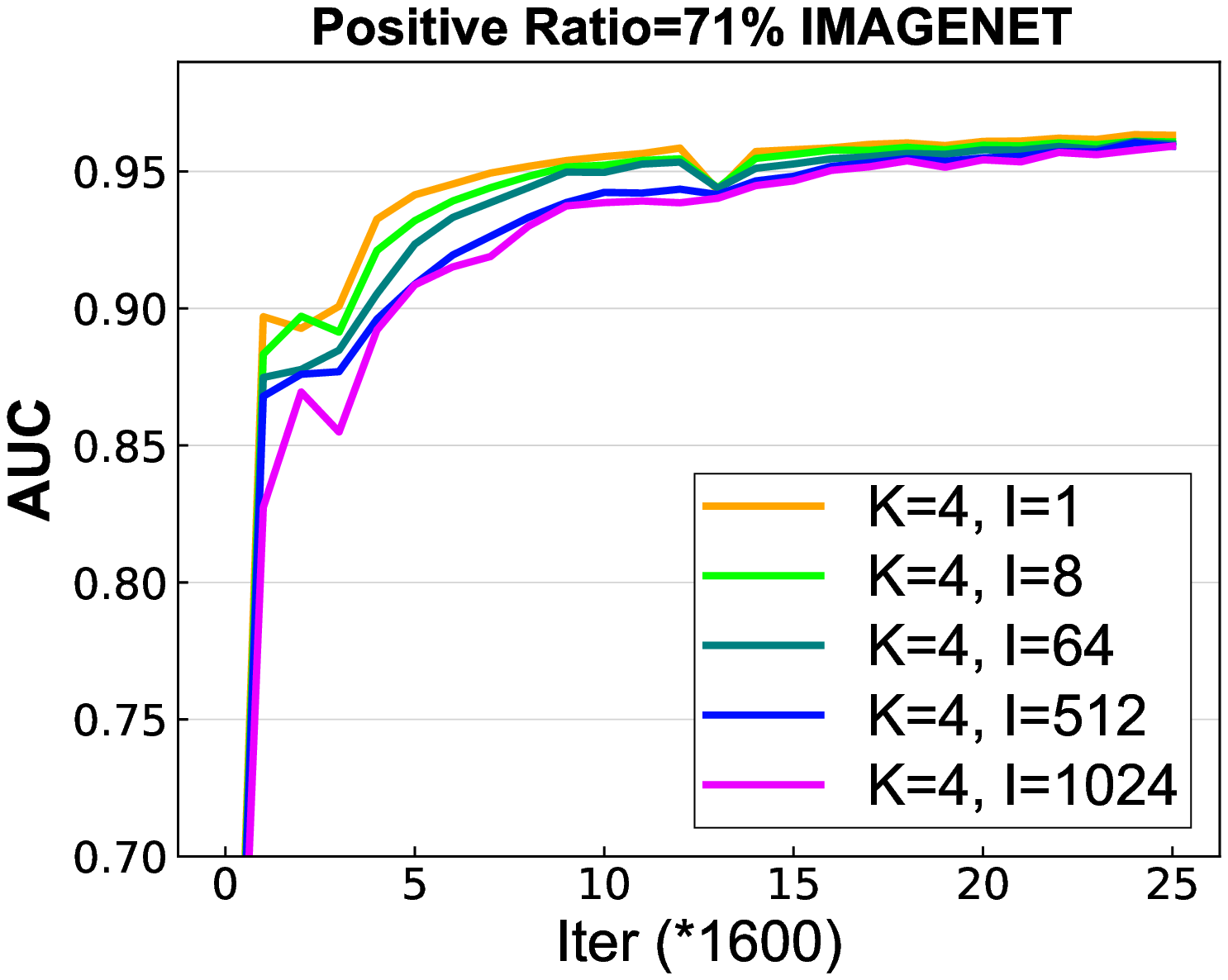}
    \includegraphics[scale=0.23]{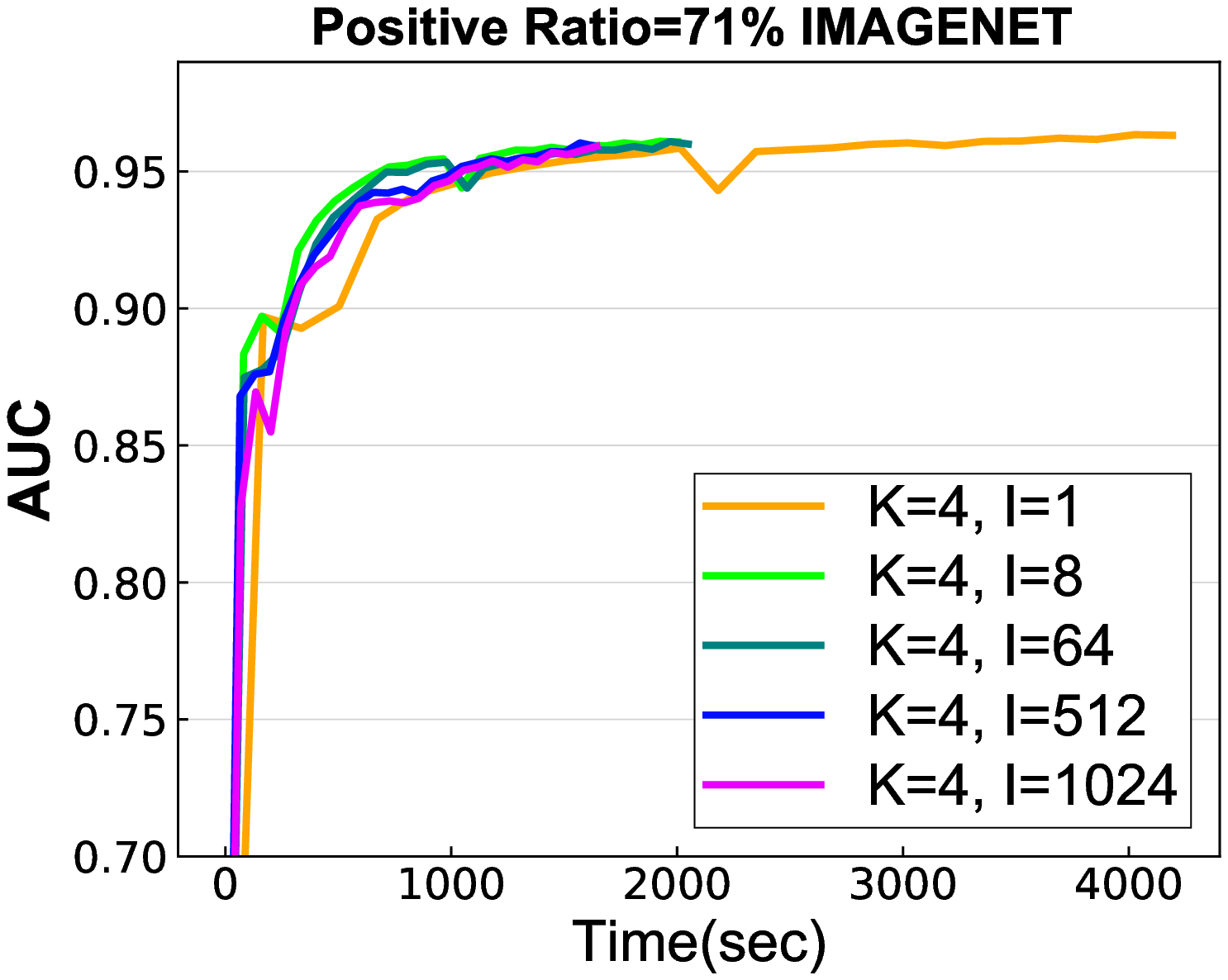}
    \vspace{-0.4cm}
    \caption{ImageNet, postive ratio=71\%, $K$=4.}
    \label{fig:ablation_imagenet_K_4_P_71}
\end{figure}

\end{document}